\documentclass[12pt,]{article}
\usepackage[utf8]{inputenc}
\usepackage[T1]{fontenc}

\usepackage{fullpage}
\usepackage{amssymb,amsmath}
\usepackage{mathtools}
\usepackage{bm}
\usepackage{siunitx}
\usepackage{csquotes}

\usepackage{xspace}
\newcommand\ie{i.\,e.\xspace}
\newcommand\eg{e.\,g.\xspace}

\DeclareFontFamily{U}{mathx}{\hyphenchar\font45}
\DeclareFontShape{U}{mathx}{m}{n}{<-> mathx10}{}
\DeclareSymbolFont{mathx}{U}{mathx}{m}{n}
\DeclareMathAccent{\widebar}{0}{mathx}{"73}

\newcommand{\figletter}[1]{{{\fontfamily{\sfdefault}\selectfont \textbf{#1}}}}

\usepackage{booktabs}
\usepackage{longtable}
\usepackage{multirow}
\usepackage{placeins}

\usepackage[dvipsnames]{xcolor}

\definecolor{darkgreen}{rgb}{0.0, 0.5, 0.0}

\usepackage{setspace}
\usepackage{times}

\usepackage{float}

\usepackage{graphicx}
\usepackage{subcaption}
\usepackage{rotating}
\usepackage{tabularx}
\usepackage{dcolumn}
\usepackage{pdflscape}
\usepackage{rotating}

\usepackage{tikz}
  \usetikzlibrary{positioning}
  \usetikzlibrary{arrows}
  \usetikzlibrary{fit}

\usepackage{array}

\newcommand{\specialcell}[2][c]{%
  \begin{tabular}[#1]{@{}l@{}}#2\end{tabular}}

\usepackage[colorlinks=true, allcolors=blue]{hyperref}

\usepackage[%
  sort&compress
]{cleveref}

\usepackage[superscript,biblabel,nomove]{cite}
\usepackage[resetlabels,labeled]{multibib}
\newcites{Supp}{References}
\crefname{Suppsec}{References}{References}

\makeatletter
\renewcommand{\fps@figure}{H}         
\renewcommand{\fps@table}{H}          
\makeatother 

\allowdisplaybreaks

\title{\centering\LARGE\singlespacing{Monitoring the COVID-19 epidemic\\ with nationwide telecommunication data}}

\date{}

\author{Joel Persson\thanks{ETH Zurich (Swiss Federal Institute of Technology), 8092 Zurich, Switzerland} \thanks{To whom correspondence may be addressed. Email: jpersson@ethz.ch}
        \and
        Jurriaan F. Parie\footnotemark[1]
        \and
        Stefan Feuerriegel\footnotemark[1]
        }

\begin{document}
\maketitle

\begin{abstract}\normalfont
\noindent
In response to the novel coronavirus disease (COVID-19), governments have introduced severe policy measures with substantial effects on human behavior. Here, we perform a large-scale, spatio-temporal analysis of human mobility during the COVID-19 epidemic. We derive human mobility from anonymized, aggregated telecommunication data in a nationwide setting (Switzerland; February~10--April~26, 2020), consisting of $\sim$1.5 billion trips. In comparison to the same time period from 2019, human movement in Switzerland dropped by \SI{49.1}{\percent}. The strongest reduction is linked to bans on gatherings of more than 5 people, which is estimated to have decreased mobility by \SI{24.9}{\percent}, followed by venue closures (stores, restaurants, and bars) and school closures. As such, human mobility at a given day predicts reported cases 7--13 days ahead. A \SI{1}{\percent} reduction in human mobility predicts a 0.88--1.11\,\% reduction in daily reported COVID-19 cases. When managing epidemics, monitoring human mobility via telecommunication data can support public decision-makers in two ways. First, it helps in assessing policy impact; second, it provides a scalable tool for near real-time epidemic surveillance, thereby enabling evidence-based policies.
\end{abstract}

\flushbottom
\maketitle
\thispagestyle{empty}

\begin{center}
\begin{tabular}{p{14.5cm}}
\small
\noindent\textbf{Keywords}: \mbox{COVID-19}, epidemiology, human mobility, telecommunication data, Bayesian modeling
\end{tabular}
\end{center}

\newpage 
\sloppy
\raggedbottom

\section{Introduction}

The novel coronavirus disease (COVID-19) has evolved into a global pandemic, which, as of December~15, 2020, has been responsible for more than 70 million reported cases \cite{WHO}. In response, governments around the world have put policy measures into effect with the aim of reducing transmission rates \cite{Flaxman.2020,Hsiang.2020,Ruktanonchai.2020,Lai.2020,Unwin.2020,Banholzer.2021}. Examples of policy measures are border closures, school closures, venue closures, and bans on gatherings. 

Prior literature has suggested the use of human mobility data to model the COVID-19 epidemic \cite{Grantz.2020}. Mobility patterns have been inferred from point-of-interest (POI) check-ins \cite{Benzel.2020,Gao.2020,Chang.2020,Dave.2020,Gupta.2020} and from location logs of smartphone apps \cite{Adiga.2020,Chinazzi.2020,Galeazzi.2020,Fang.2020,Kraemer.2020a,Li.2020,Tian.2020,Bonaccorsi.2020,Nouvellet.2020,Kang.2020,Kogan.2020,Huang.2020,Xiong.2020}. Other works have used telecommunication data to model spreading patterns \cite{Badr.2020,Jia.2020}, for exploratory analysis of mobility patterns \cite{Jeffrey.2020,Pullano.2020,Vinceti.2020}, for network analysis of structural changes in mobility \cite{Schlosser.2020}, and for modeling the spatio-temporal distribution of COVID-19 \cite{Jia.2020}, but none have yet empirically explored the link between telecommunication data and policy measures. Establishing such a link would provide a scalable tool for near real-time disease surveillance under policy measures and, in particular, enable evidence-based policies. Previously, the value of telecommunication data for disease surveillance has been studied in the context of malaria \cite{Ruktanonchai.2016,Wesolowski.2012}, influenza \cite{Viboud.2019}, and other infectious diseases \cite{Bengtsson.2015,Wesolowski.2015a,Wesolowski.2015b}, where the objective was to make spatio-temporal forecasts. In contrast, this paper demonstrates the utility of telecommunication data for near real-time assessments of COVID-19 policies. In fact, nationwide data from mobile telecommunication networks has been used by governments during the first wave of COVID-19 \cite{Reuters}. However, to the best of our knowledge, empirical evidence regarding the effectiveness of telecommunication data for epidemic surveillance in the context of COVID-19  is absent. 

In this paper, we analyze human mobility during the COVID-19 epidemic. Our analysis is based on large-scale, granular data of human movements (anonymized and aggregated) consisting of $\sim$1.5 billion trips in Switzerland during the first COVID-19 wave (February~10--April~26, 2020) derived from telecommunication data. Using regression models, we estimate the (1)~impact of policy measures on human mobility, and (2)~how mobility predicts the growth in reported COVID-19 cases. By establishing that policy measures reduce mobility and that mobility predicts reported cases, mobility insights can be used to inform when to implement policy measures. The findings are therefore of direct value to public decision-makers: monitoring human mobility through telecommunication data provides an effective and scalable tool for near real-time epidemiology and thus, management of the COVID-19 epidemic.

To establish the ability of telecommunication data for near real-time monitoring of the COVID-19 epidemic, we follow a two-stage approach (see \nameref{sec:methods}). We first study the reduction in mobility due to 5 different policy measures (bans on gatherings of more than 100 people, bans on gatherings of more than 5 people, school closures, venue closures, and border closures). We then estimate to the extent to which reduction in mobility predicts decreases in reported case growth. Here, we compare the predictive ability over a forecast window from 7 to 13 days. Taken together, the results confirm the effectiveness of policy measures for reducing human mobility and, in turn, human mobility as a predictor of reported cases by a lead time of approx. 7--13 days. The two-stage approach is repeated for total trips, 3 different modes of mobility (train, road, highway), and 2 different purposes for mobility (commuters vs. non-commuters). In an extended analysis, we further perform a mediation analysis. Here, we decompose the reduction in new reported cases due to the policy measures into (a)~the part that is only explained by reductions in mobility and (b)~the part that is explained by other behavioral adaptations.

\section*{Results}

\subsection*{Human mobility derived from nationwide telecommunication data}

We analyze large-scale data on human mobility during February~10--April~26, 2020 from Switzerland. For this, human movements derived from telecommunication data were obtained from a major telecommunications provider in Switzerland (see \nameref{sec:methods}). Telecommunication data provide more reliable and extensive information on mobility compared to alternative data sources (check-ins or location logs from smartphone apps) \cite{Buckee.2020,Kishore.2020,Grantz.2020}. In particular, our telecommunication data represents routine signal exchanges (``pings'') exchanged between mobile devices and network antennas. These were recorded for all mobile devices in Switzerland regardless of the mobile service provider. Based on the telecommunication data, granular locations (longitude, latitude) of individuals carrying a mobile device were inferred. This yields data on micro-level movements from all mobile devices in a nationwide setting. Altogether, the nationwide mobility for a population of $\sim$8.6 million people was estimated.

For the analysis, the telecommunication data were then processed to count the number of trips between posts codes per day and canton. All trips were further classified according to the mode of mobility (``train'', ``highway'', and other ``road'' movements) and purpose (``commuting'' vs. ``non-commuting''). For the time period of this study (February~10--April~26, 2020), our data include a total of $\sim$1.5 billion trips. Details are reported in \nameref{sec:methods}.

We further collected data on the use of policy measures in Switzerland. Switzerland comprises 26 member states at the sub-national level (called ``cantons''), each with a large degree of sovereignty. As a result, the use of policy measures varies across cantons with respect to their order and timing in a way that is similar to the variation among other European countries. Some cantons (\eg, Ticino at the border to Italy and Geneva at the border to France) showed epidemiological dynamics with large numbers of reported cases and responded with comparatively stringent policy measures during the first wave of COVID-19. Other cantons had lower case numbers and put policy measures in effect during a later phase of the epidemic. We followed a systematic procedure (see Supplement~\ref{supp:data}) based on which we encoded policy measures according to five categories: bans on gatherings ($>$\,100 people), bans on gatherings ($>$\,5 people), school closures, venue closures (stores, restaurants, bars), and border closures. The implementation dates of the chosen policy measures varied greatly across cantons (but then remained in effect for the complete study period, \ie, until April 10). 

\begin{figure}[H]
\thispagestyle{empty}
\vspace{-1cm}
\begin{minipage}[t]{0.3\linewidth}
    \figletter{a} \\
    \begin{subfigure}{\linewidth}
        \includegraphics[width=\linewidth]{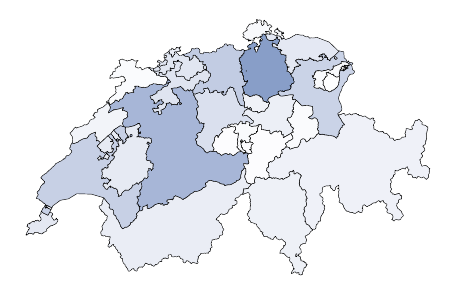}
        \label{subfig:SUI_heatmap1}
    \end{subfigure}
\end{minipage}
\hspace{1cm}
\begin{minipage}[t]{0.6\linewidth}
    \figletter{d} \\
    \begin{subfigure}{\linewidth}
        \includegraphics[width=\linewidth]{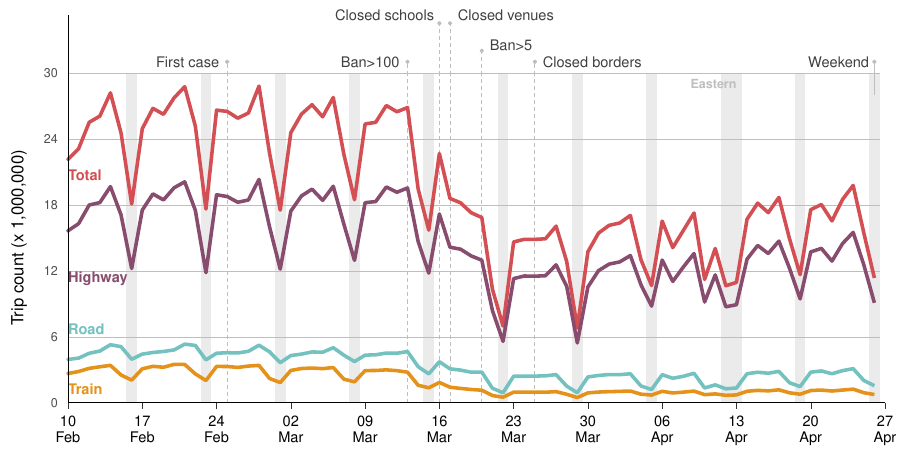}
        \label{subfig:diff_mobility_by_type}
    \end{subfigure}
\end{minipage}

\begin{minipage}[t]{0.3\linewidth}
    \figletter{b} \\
    \begin{subfigure}{\linewidth}
        \includegraphics[width=\linewidth]{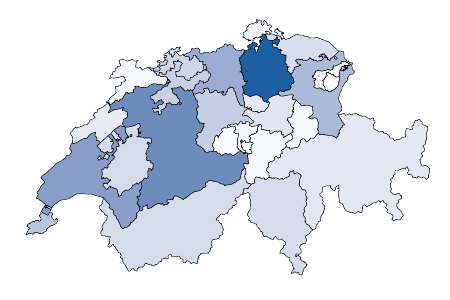}
        \label{subfig:SUI_heatmap2}
    \end{subfigure}
\end{minipage}
\hspace{1cm}
\begin{minipage}[t]{0.6\linewidth}
    \figletter{e} \\
    \begin{subfigure}{\linewidth}
        \includegraphics[width=\linewidth]{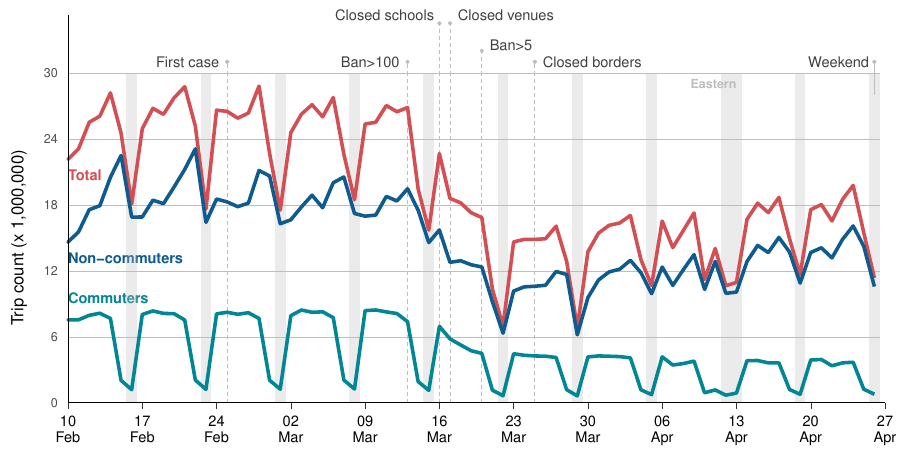}
        \label{subfig:diff_mobility_by_reason}
    \end{subfigure}
\end{minipage}

\vspace{-4pc}

\begin{minipage}[t]{0.3\textwidth}
    \begin{subfigure}{\linewidth}
        \includegraphics[width=\linewidth]{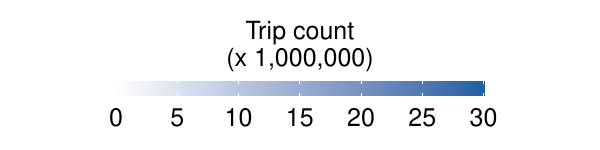}
        \label{subfig:colorbar}
    \end{subfigure}
\end{minipage}
\begin{minipage}[t]{0.6\textwidth}
    \hspace*{\fill}
\end{minipage} 

\begin{minipage}[t]{0.14\textwidth}
    \hspace*{\fill}
\end{minipage}
\begin{minipage}[t]{0.6\textwidth}
    \figletter{c}\\
    \begin{subfigure}{\linewidth}
        \includegraphics[width=\linewidth]{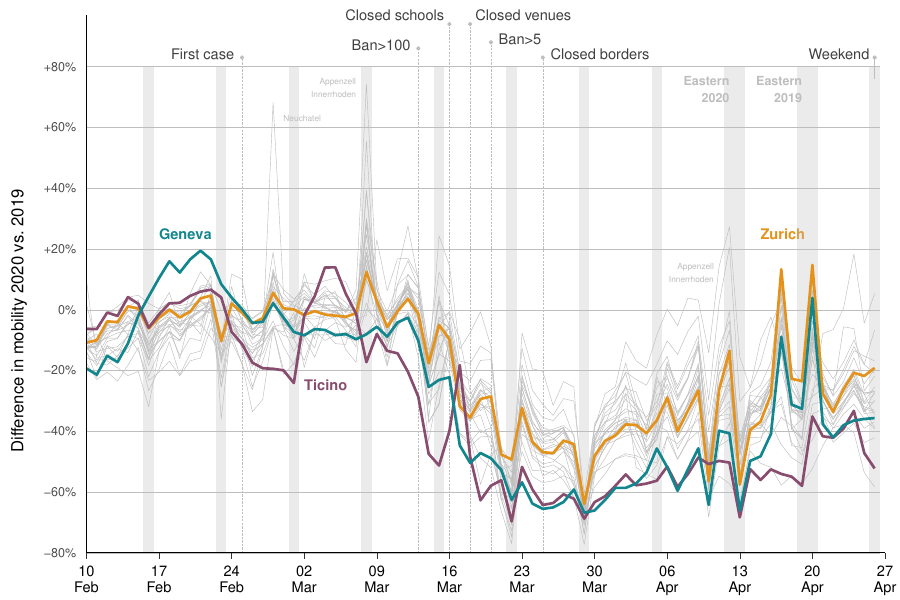}
        \label{subfig:diff_mobility_by_canton}
    \end{subfigure} 
\end{minipage}
\begin{minipage}[t]{0.14\textwidth}
    \hspace*{\fill}
\end{minipage}
\vspace{-0.5cm}

\caption{\footnotesize Nationwide human mobility in Switzerland during the first wave of the COVID-19 epidemic. Mobility is quantified by movements (``trips'') between different postcode areas. \textbf{(a)}~Total number of trips per canton for the first week after all policy measures were put in effect in all cantons (March 25--April 1, 2020). (\textbf{b})~Total number of trips for the same week in 2019 (\ie, as reference year). For this week, the total number of trips dropped from $\sim$186 million (in 2019) to $\sim$95 million (in 2020), \ie, a reduction of \SI{49.1}\,\%. \textbf{(c)}~Percentage change in total trips across 26 sub-national levels (cantons) for 2020 vs. 2019 (when aligned for day-of-week patterns). The reason for the comparison to 2019 is to show the reduction in mobility relative to a reference year, while accounting for seasonal changes in mobility. A higher reduction in mobility is observed for cantons that also reported a high number of COVID-19 cases (\ie, Ticino and Geneva). \textbf{(d)}~Reduction in trip count by mode of mobility. (\textbf{e})~Reduction in trip count by purpose of mobility. Annotations show nationwide implementation dates of policy measures (implementation dates at cantonal level are reported in Supplement~\ref{supp:data}).}
\label{fig:1}
\end{figure}

The spatio-temporal patterns of human movements in our data are as follows. Overall,  $\sim$95 million trips were recorded in the first week (February~24--March~1, 2020), during which all five policy measures were in effect in all 26 cantons. In comparison, the same time period in 2019 (as a reference period) registered $\sim$186 million trips. This amounts to a reduction of \SI{49.1}\,\%. The reduction occurred in all cantons (\Cref{fig:1}a,b). The highest decline was observed in Ticino and Geneva, which are located at the borders with Italy and France, respectively. Both cantons also reported the highest number of COVID-19 cases. 

The largest mobility reduction compared to 2019 occurred on Sunday, March~22, 2020 (\Cref{fig:1}c). In comparison to Sunday, March~24, 2019, the reduction in trip counts ranged between \SI{49.3}\,\% and \SI{77.0}\,\% across the 26 cantons (mean: \SI{61.6}\,\% reduction per canton). Overall, the reduction in mobility is of similar magnitude for both rural (\eg, canton of Valais) and urban regions (\eg, cantons of Basel-City and Zurich). Furthermore, movements declined for all modes of mobility (\Cref{fig:1}d) and for all purposes (\Cref{fig:1}e). After the implementation of the policy measures, trips by train remained low for the rest of the study period, while highway traffic was on an upward trend (\Cref{fig:1}d). Similarly, trips by commuters remained at a low level after the implementation of the policy measures, whereas trips not for commuting (\ie, personal purposes) started increasing in early April (\Cref{fig:1}e).

\newpage 

\thispagestyle{empty}

\FloatBarrier
\newpage
\thispagestyle{plain}

\subsection*{Estimating the reduction in human mobility due to policy measures}

We estimate the reduction in mobility due to the policy measures with a regression model. The estimates are identified via a difference-in-difference analysis and may thus be given a causal interpretation under certain assumptions (see Supplement~\ref{supp:identification_strategy_mobility}).

\begin{figure}[H]
    \begin{minipage}[t]{\linewidth}
            \hspace{1.5cm}
        \figletter{a}
            \hspace{4.7cm}
        \figletter{b}
            \hspace{4.6cm}
        \figletter{c}
    \end{minipage}
    \begin{minipage}[t]{\linewidth}
        \centering
        \includegraphics[width=\linewidth]{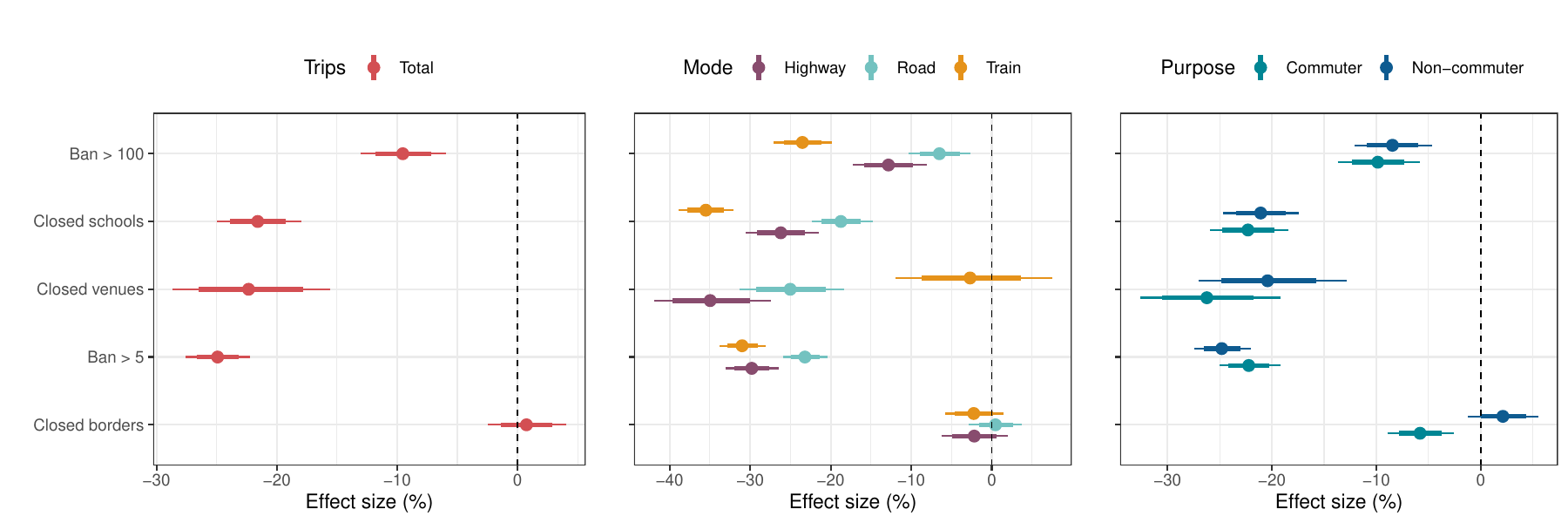}
    \end{minipage}
\caption{The estimated effects of policy measures on mobility. Shown are the estimated effects on (\textbf{a}) the total number of trips, (\textbf{b}) trips by mode, and (\textbf{c}) trips by purpose. The dots in \textbf{(a)}--\textbf{(c)} show posterior means; the thick and thin bars represent the 80\,\% and 95\,\% credible intervals, respectively. Policy measures are arranged from top to bottom in the order in which they were implemented (cf. Supplement~\ref{supp:data}).}
\label{fig:DiD_estimates}
\end{figure}

The most effective policies for reducing trip counts are as follows (\Cref{fig:DiD_estimates}a). Based on our model, bans on gatherings of more than 5 people reduced total trips by 24.9\,\% (95\,\% credible interval [CrI]: 22.1--27.6\,\%), venue closures reduced total trips by 22.3\,\% (95\,\% CrI: 15.6--29.0\,\%), and school closures reduced total trips by 21.6\,\% (95\,\% CrI: 17.9--25.0\,\%). For a precise ranking, the width of the credible intervals must be considered. Here, the aforementioned policy measures appear more effective at reducing total trips than the other two policy measures (\ie, bans on gatherings of more than 100 and border closures). In particular, bans on gatherings of more than 100 people are linked to a comparatively smaller change in total trips than bans on gatherings of more than 5 people (\ie, the 95\,\% CrIs of the estimates are disjoint). For border closures, the credible interval includes zero. Overall, policy measures are important determinants of mobility reductions during the COVID-19 epidemic.

The estimated mobility reduction depends on the underlying mode (\Cref{fig:DiD_estimates}b). Across all policy measures, the mobility reduction is more pronounced for highways than for road movements. This observation is to be expected. Highways are often used for long-distance travel, which is more likely to be suspended during an epidemic, while roads also include movements within close proximity and are more likely to correspond to routine or essential activities (\eg, grocery shopping). For the ban on gatherings and school closures, the largest reduction is seen in trips by train, which can be explained by the widespread use of public transportation in Switzerland. Finally, we observe a wide credible interval for the estimated effect of venue closures on trips by train. A potential reason for this is that the use of trains (\eg, for visiting stores) varies across cantons, as some cantons (\eg, Zurich) have a high population density with extensive shopping infrastructure, while others (\eg, Appenzell Innerrhoden) have only a few stores due to their low population density, resulting in the need for travel to visit stores.

The estimated effect sizes are fairly similar for trips made for commuting versus non-commuting (\Cref{fig:DiD_estimates}c). This is interesting considering that no policy measure in Switzerland directly prohibited movement to and from work. The efficacy of border closures is uncertain since the credible interval for its estimated effect includes zero. In contrast, a negative effect is observed for commuting. Here, one potential reason is that border closures have reduced the number of cross-border commuters. A validation analysis supports this explanation (see Supplements~\ref{supp:validating_border_closures} for details).

The findings are robust to alternative model specifications (see the robustness checks in Supplement~\ref{supp:robustness_checks}). Specifically, changing the specification of time-related control variables still gives parameter estimates for the policy measures that imply decreases in mobility.

\subsection*{Estimating the relationship between mobility and COVID-19 cases}

\begin{figure}[H]
    \begin{minipage}[t]{0.4\textwidth}
        {\fontfamily{\sfdefault}\selectfont \textbf{a}} \\
        \begin{subfigure}{\linewidth}
            \includegraphics[height=6cm]{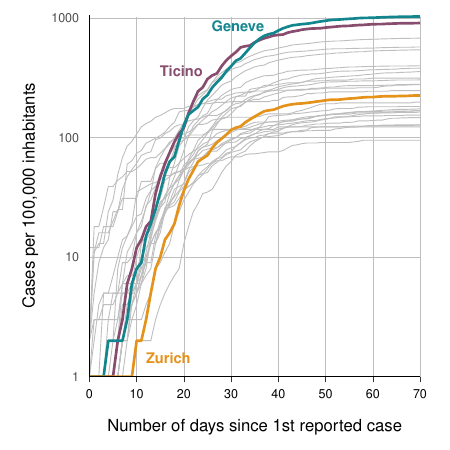}
        \end{subfigure}
    \end{minipage}
    \begin{minipage}[t]{0.59\textwidth}
        {\fontfamily{\sfdefault}\selectfont \textbf{b}} \\
        \begin{subfigure}{\linewidth}
            \includegraphics[width=\textwidth]{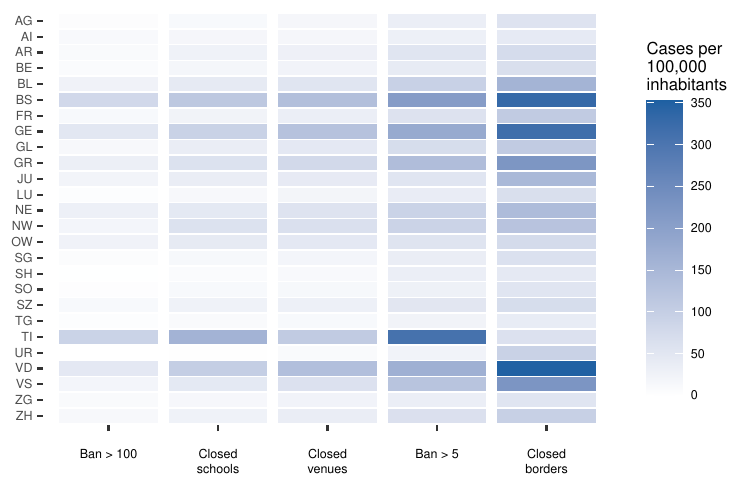}
        \end{subfigure}
    \end{minipage}
    \\
    \\
    \begin{minipage}[t]{\linewidth}
            \hspace{0.8cm}
        \figletter{c}
            \hspace{4.9cm}
        \figletter{d}
            \hspace{4.9cm}
        \figletter{e}
    \end{minipage}
    
    \begin{minipage}[t]{\linewidth}
        \centering
        \includegraphics[width=\linewidth]{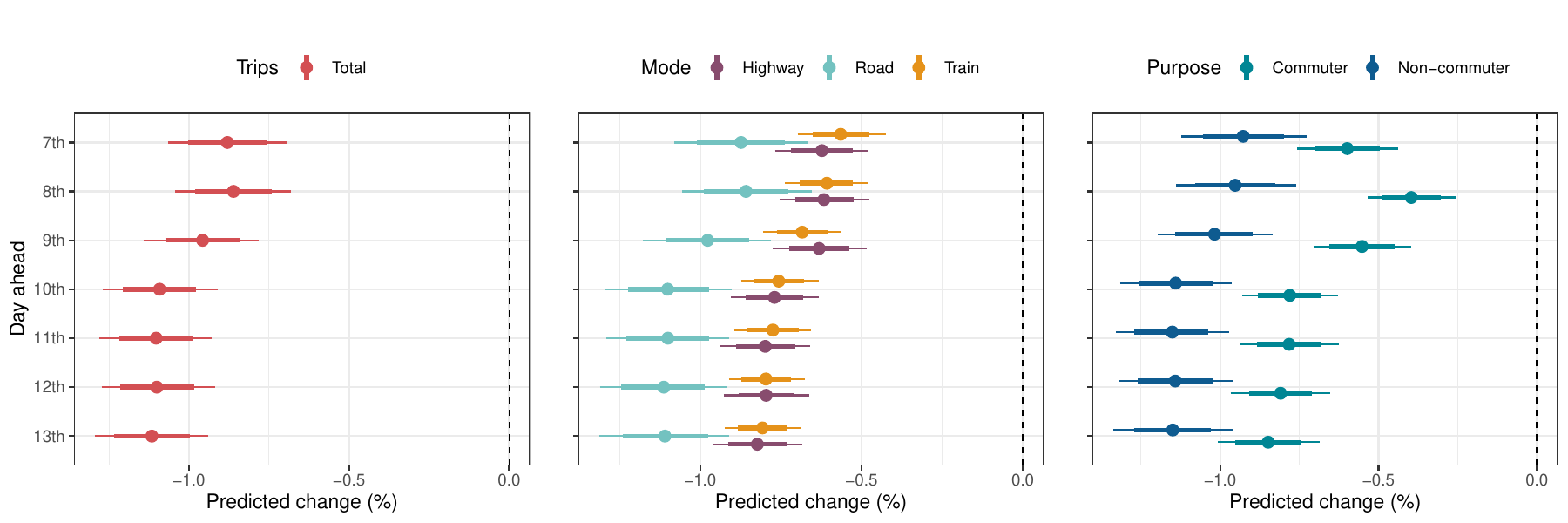}
    \end{minipage}
    
\caption{Decreases in human mobility predicts reductions in reported cases of COVID-19 over a forecast window of 7--13 days. \textbf{(a)}~COVID-19 case growth since the 1st reported case for 26 Swiss cantons. \textbf{(b)}~Number of COVID-19 cases when the policy measures were implemented for all 26 Swiss cantons. Shown are abbreviated names. The predicted change in reported new COVID-19 cases at a given day is based on mobility lagged by 7 days to 13 days. The predicted change is reported given a 1\,\% decrease in \textbf{(c)}~total trips, \textbf{(d)}~mode, and \textbf{(e)}~purpose. Posterior means are shown as dots, while 80\,\% and 95\,\% credible intervals are shown as thick and thin bars, respectively.}
\label{fig:cases_estimates}
\end{figure}

The epidemic dynamics during the first wave of COVID-19 in Switzerland are as follows. The initial exponential growth rates exhibit considerable heterogeneity across cantons (\Cref{fig:cases_estimates}a). The strongest initial growth is observed for the cantons Ticino and Geneva, resulting in the largest number of cases towards the end of the sample. Moreover, the number of reported cases at the dates that policy measures were implemented varies greatly across cantons (\Cref{fig:cases_estimates}b). This reflects different responses among cantons to local infection dynamics.

We use regression models to estimate the extent to which decreases in mobility predict future reductions in the reported number of new cases (\Cref{fig:cases_estimates}c). The predicted decrease is studied with a forecast window over 7--13 days. The forecast window is set analogous to previous research \cite{Xiong.2020}, and so that it covers variations in incubation time combined with reporting delay. 

We find that decreases in mobility at a given day predict decreases in reported new cases 7 to 13 days later. For a 7-day ahead forecast, we find that a 1\,\% decrease in the total number of trips predicts a 0.88\,\% (95\,\% CrI: 0.7--1.1\,\%) reduction in the reported number of new cases. For a 13-day ahead forecast, a 1\,\% decrease in the total number of trips predicts a 1.11\,\% (95\,\% CrI: 0.9--1.6\,\%) reduction in the reported number of new cases. Overall, mobility predicts decreases in the reported number of new cases over the whole forecast horizon. The predicted decrease is larger for longer forecasts. This result is to be expected, as a longer time window accommodates the full distribution of incubation periods (plus reporting delays). Altogether, the regression analysis provides evidence of that mobility predicts epidemic dynamics.

Our analysis also shows that the predicted change in the reported number of new cases varies across the mode and purpose of trips. In terms of mode, decreases in trips by highway and train predict reductions in the reported number of new cases of similar magnitude (\Cref{fig:cases_estimates}d). Their estimates have comparatively narrow credible intervals, reflecting a higher degree of certainty. Trips are also categorized according to their purpose, namely commuting vs. non-commuting. The results show that decreases in trips for commuting predict smaller reductions in the number of reported new cases compared to decreases in non-commuting trips (\Cref{fig:cases_estimates}e). Predicted reductions are nonetheless found for both modes of mobility (\ie, commuting vs. non-commuting), all categories of purpose (\ie, highway, road, and train), and for the whole 7--13 day forecast window. Again, a larger reduction is predicted for longer forecasts.

The predictive ability of mobility for reported new cases holds with alternative model specifications. For most of the 7--13 day forecasts, changing how we control for time-related factors still results in a predicted decrease in reported cases given decreases in mobility. Moreover, changing the dependent variable to daily hospitalizations or deaths attributed to COVID-19 leads to qualitatively the same results over a forecast horizon of 10--20 days. Details on these robustness checks are provided in Supplement~\ref{supp:robustness_checks}.

\subsection*{Estimating the mediating role of mobility}

In an extended analysis, we study how decreases in reported case growth is explained by reductions in mobility due to policy measures versus other behavioral changes due to policy measures. The estimates are obtained from a mediation analysis that decomposes the total effects of the policy measures on reported case growth into (1)~their direct effects not explained by changes in mobility and (2)~their indirect effects through mobility. The mediation analysis is performed by combining our two regression models into a structural equation model (see Supplement~\ref{supp:mediation_analysis} for details).  Results from mediation analysis are reported for the total number of trips.

\begin{figure}[H]
    \begin{minipage}[t]{\linewidth}
        \hspace{0.6cm}
        \figletter{a} \\
        \begin{subfigure}{\linewidth}
            \includegraphics[width=\linewidth]{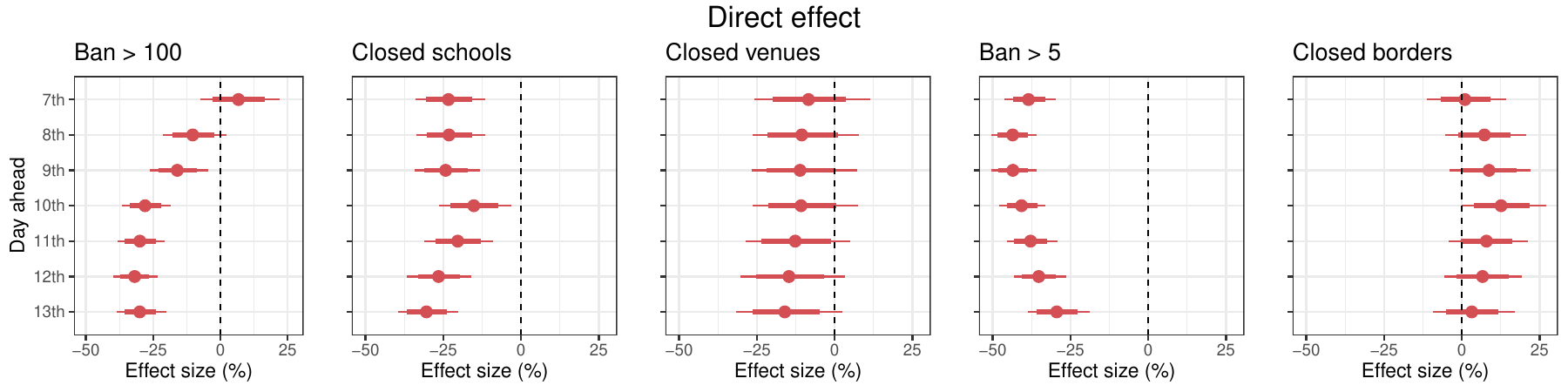}
        \end{subfigure}
    \end{minipage}
    \\
    \begin{minipage}[t]{\linewidth}
        \hspace{0.6cm}
        \figletter{b} \\
        \begin{subfigure}{\linewidth}
            \includegraphics[width=\linewidth]{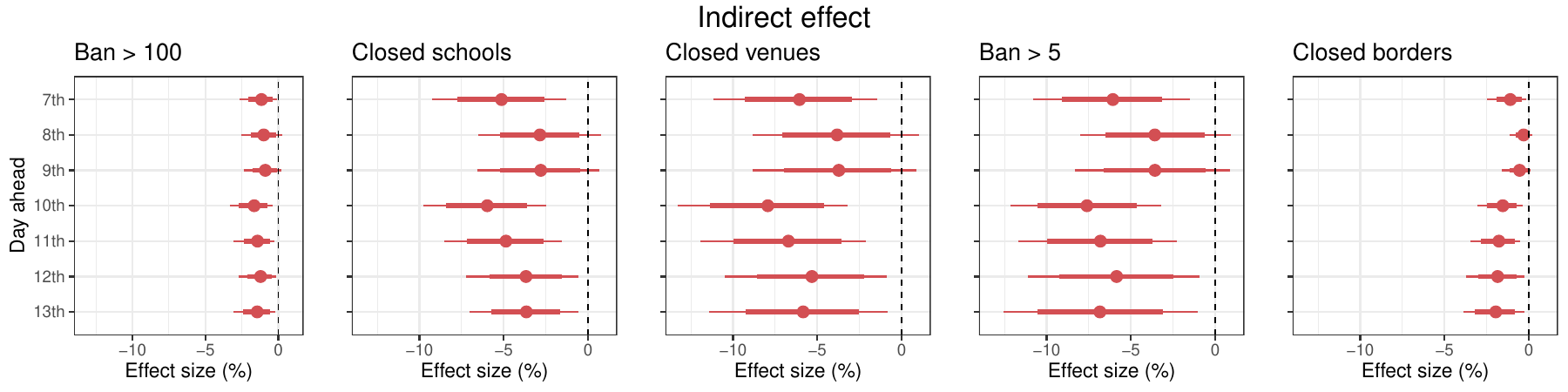}
        \end{subfigure}
    \end{minipage}
    \\
    \begin{minipage}[t]{\linewidth}
        \hspace{0.6cm}
        \figletter{c} \\
        \begin{subfigure}{\linewidth}
            \includegraphics[width=\linewidth]{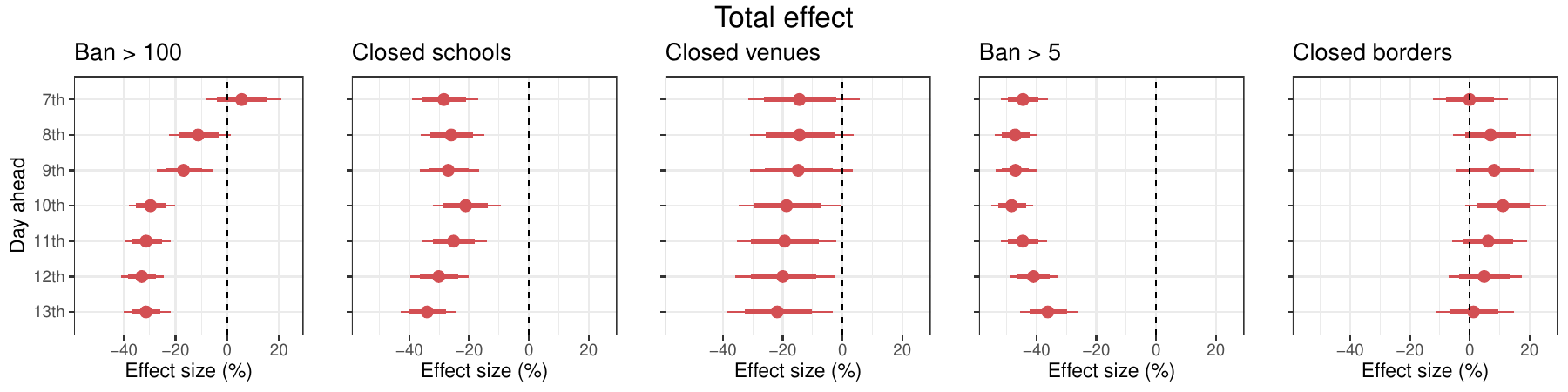}
        \end{subfigure}
    \end{minipage}
    \caption{Mobility mediates the effect of policy measures on the reported number of new cases. Estimated \textbf{(a)}~direct effect of policy measures, \textbf{(b)}~indirect effect of policy measures via total trips, and \textbf{(c)}~total effect of policy measures on the 7th to 13th day ahead. Posterior means are shown as dots, while 80\,\% and 95\,\% credible intervals are shown as thick and thin bars, respectively. Policy measures are arranged from top to bottom in the order in which they were implemented (cf. Supplement~\ref{supp:data}).}
\label{fig:mediation_results_1}
\end{figure}

The mediation analysis shows a large direct effect for bans on gatherings of more than 5 people, bans on gatherings of more than 100 people, and school closures (\ref{fig:mediation_results_1}). Pronounced indirect effects are found for all policy measures. In particular, the indirect effect of venue closures makes up about a third of their total effect at several lags. Moreover, border closures are estimated to only have reduced the reported number of new cases indirectly through mobility. The results are discussed in further detail in Supplement~\ref{supp:mediation_results}. In summary, the results show that mobility is an important mediator: the studied policy measures operate -- to a large degree -- through mobility. Thus, policy measures aimed at reducing mobility appear to be effective for reducing COVID-19 case growth.

\section*{Discussion}

This study shows the ability of telecommunication data for near real-time monitoring of the COVID-19 epidemic. Our analysis is based on nationwide telecommunication data during February~10--April~26, 2020 from Switzerland, which were used to infer nationwide mobility patterns. This supports monitoring of the COVID-19 epidemic as follows: (1)~We first studied the link between policy measures and human mobility. In particular, we performed a difference-in-difference analysis quantifying how mobility was reduced due to 5 different policy measures (bans on gatherings, school closures, venue closures, and border closures). The largest reduction in total trips was linked to bans on gatherings of more than 5 people, followed by venue closures and school closures. Overall, the policy measures resulted in substantial reductions of human mobility. (2)~We then studied the link between human mobility and reported COVID-19 cases. Reductions in mobility predicted decreases in the number of reported new cases. Specifically, a reduction in human movement by 1\,\% predicted a 0.88--1.11\,\% reduction in the daily number of new cases of COVID-19 over a forecast horizon of 7 days to 13 days. Our modeling approach with telecommunication data therefore provides near real-time insights for disease surveillance. Taken together, the findings enable quantitative comparisons of the extent to which policy measures reduce mobility and, subsequently, reduce reported cases of COVID-19.

The use of telecommunication data for nationwide monitoring has several benefits \cite{Desai.2019,Grantz.2020}. First, telecommunication data from mobile networks provide comprehensive coverage. Specifically, such data capture all movements of individuals carrying mobile devices without explicit user interaction, including those from non-residential and even foreign individuals. Mobile devices routinely exchange information when searching for signals from adjacent antennas; hence, metadata are retrieved regardless of the underlying mobile service provider. Second, such metadata can be collected in an anonymized manner that is compatible with data privacy laws. Third, movements at a micro-level (\eg, trips to other households, school, and work) can be inferred. Thus, compared to alternative sources of mobility information such as check-ins or smartphone apps, telecommunication data are considered to be more complete \cite{Buckee.2020,Kishore.2020,Grantz.2020}. Fourth, unlike smartphone apps, telecommunication data are also available in low-income countries \cite{Blumenstock.2015}. Finally, telecommunication data are measured with high frequency (\eg, daily), thereby enabling regularly updated monitoring as needed by decision-makers. Based on these benefits, telecommunication data appear to be highly effective for policy monitoring during the COVID-19 epidemic. 

This work is subject to the typical limitations of observational studies. First, the findings depend on the accuracy of the data on COVID-19 cases. Second, our models are informed by recommendations for COVID-19 modeling \cite{Bertozzi.2020} and, therefore, follow parsimonious specifications to isolate features of the epidemic for policy-relevant insights. We cannot, however, rule out the possibility that there exist external factors beyond those that are captured by the spatially and temporally varying variables in the models. To address this, we use flexible models and conduct extensive robustness checks (Supplement~\ref{supp:robustness_checks}). Third, the model linking policy measures to mobility estimates effects, while the model linking mobility to cases is predictive. The different objectives of the models address the needs of public decision-makers: the former serves policy assessments and the latter epidemiological forecasting, respectively. Therefore, the estimates from the former are identified with a difference-in-difference analysis and may thus warrant causal interpretations under certain assumptions (see Supplement~\ref{supp:identification_strategy_mobility}). On the other hand, the estimates from the latter are conditional associations since the model does not control for that policy measures reduce both mobility and cases. Therefore, the latter model predicts reductions in reported cases from reductions in mobility when both reductions are driven by policy measures (see Supplement~\ref{supp:identification_strategy_cases} for a discussion of this approach). Fourth, our findings are limited to our study setting, that is, the first wave in Switzerland. Future research may confirm the external validity of our findings by analyzing other countries or time periods.

Inferring mobility patterns from telecommunication data is inherently coupled to the coverage of such data and our definition of trips. Only movements for individuals who carry a mobile device with a SIM card are included. In particular, trips are not included for individuals who do not carry SIM cards. Similarly, trips may be counted several times if an individual carries several SIM cards (\eg, when carrying both a phone and a SIM-based tablet). It is also possible that trips by children, elderly, or other groups of the population with less phone usage are underrepresented in the data. Furthermore, micro-level movements are not observed but inferred via triangulation between antennas through the use of a positioning algorithm achieving state-of-the-art accuracy. In spite of these limitations, telecommunication data are considered to be scalable and, in particular, more complete for inferring mobility patterns compared to alternative data sources \cite{Buckee.2020,Kishore.2020,Grantz.2020}. Moreover, our objective is not to obtain accurate estimates of mobility in itself, but to evaluate the predictive ability of telecommunication data for reported case growth. Our analysis confirms telecommunication data as such a monitoring tool.  

Our findings are of direct value for public decision-makers. Nationwide mobility data from mobile telecommunication networks can be leveraged for the management of epidemics. Thereby, we fill a previously noted void in the case of COVID-19 \cite{Buckee.2020,Kishore.2020,Oliver.2020}. Specifically, monitoring mobility supports public decision-makers when managing the COVID-19 epidemic in two ways. First, it helps public decision-makers in assessing the impact of policy measures targeted at mobility behavior. Second, by predicting epidemic growth, it provides a scalable tool for near real-time epidemic surveillance. Such tools are relevant for evidence-based policy-making of public authorities in the current COVID-19 epidemic.

\newpage
\section*{Materials and methods}
\label{sec:methods}

The aim of this study is to make population inference of mobility from nationwide telecommunication data. In our study, mobility estimates derived from telecommunications data were obtained from Swisscom Mobility Insights, a commercial data platform of a major telecommunications provider in Switzerland, and then further processed for analysis.

\begin{figure}
\centering
\includegraphics[width=.95\textwidth]{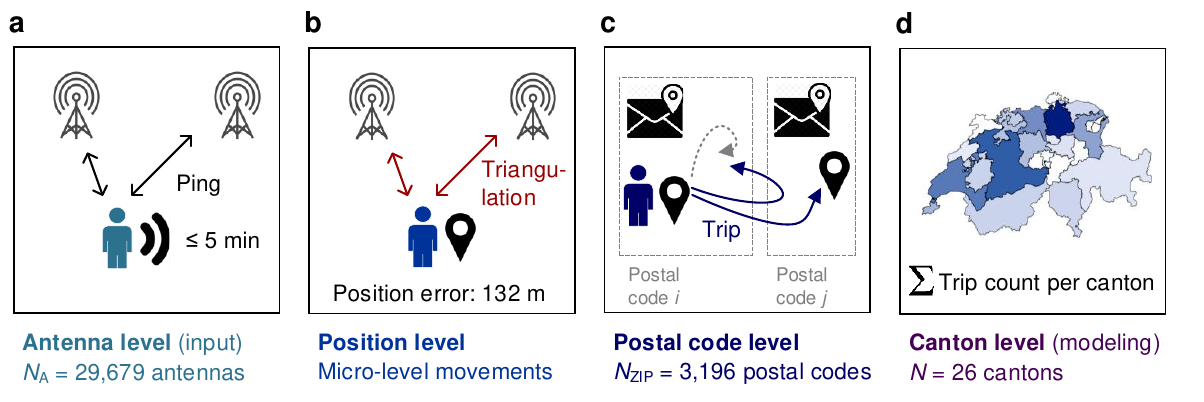}
\caption{Deriving mobility estimates from nationwide telecommunication data for monitoring the COVID-19 epidemic. (\textbf{a})~Telecommunication data are collected at the level of antenna. (\textbf{b})~Telecommunication data at antenna level are used to infer micro-level movements of individuals via triangulation. (\textbf{c})~Data on micro-level movements are used to count movements between postal codes (named ``trips'') over time. (\textbf{d})~The data are aggregated at the cantonal level per day in order to link them to policy measures and COVID-19 case numbers.}
\label{fig:mapping_mobility}
\end{figure}

Swisscom collects telecommunications data from routine signal exchanges (\ie, pings) with antennas, regardless of the actual service provider. Based on the telecommunication data, mobility estimates are inferred as follows (Fig.~\ref{fig:mapping_mobility}): (a)~Telecommunication data are collected at the level of antenna. (b)~Telecommunication data at antenna level are used to infer micro-level movements of individuals via triangulation. (c)~Data on micro-level movements are used to count movements between postal codes (named ``trips'') over time. This procedure is performed to capture mobility levels in the population. (d)~The data are further aggregated at the cantonal level per day in order to link them to policy measures and COVID-19 case numbers. The following sections explains Swisscoms procedure for obtaining nationwide mobility estimates from telecommunication data and how we further processed the data for analysis.

\subsection*{Nationwide telecommunication data}

Telecommunication data are routinely collected from signal exchanges (\ie, pings) between mobile devices and adjacent antennas. Such signal exchanges occur for all SIM-based mobile devices (\eg, mobile phones, smartphones), regardless of the actual service provider. In particular, our data also include movements of people with a foreign SIM card and, hence, represents nationwide telecommunication. A network event between a mobile device and mobile network compromises metadata as follows: the IMSI number of the SIM card, the date and time the SIM card connected to the mobile network, and the ID of the mobile antenna to which the SIM card was connected. The IMSI number is available for all SIM cards and thus represents a unique identifier, independent of the actual service provider. The events from mobile networks are extracted from the mobile communications systems every night, and thus the mobile data is available the following day. In our analysis, we use telecommunication metadata collected by Swisscom according to the above description \cite{MIP.2020}. Swisscom also ensured that IMSI numbers are stored in an anonymized format (see Ref. \cite{MIP.2020} for details).

Telecommunication data hold advantages over alternative data sources for the purpose of measuring human mobility. The advantages become especially clear in comparison to location data from check-ins \cite{Gao.2020,Benzel.2020,Chang.2020,Dave.2020,Gupta.2020} or location logs from smartphone apps \cite{Adiga.2020,Kraemer.2020b,Chinazzi.2020,Galeazzi.2020,Fang.2020,Kraemer.2020a,Li.2020,Tian.2020,Bonaccorsi.2020}. First, compared to smartphone applications, SIM-based devices are fairly ubiquitous. This holds for both high-income countries (such as Switzerland) and low-income countries. Second, the use of telecommunication data ensures coverage for large parts of society. Specifically, it reduces the risk of an age bias (\eg, check-ins are known to be more frequent among younger, technology-savvy people). Third, telecommunication data avoid the need for user interaction with a device. Hence, many micro-level movements are captured (\eg, school visits, commuting to work, grocery shopping) that would otherwise not be subject to monitoring. 

The telecommunication infrastructure operated by Swisscom has wide coverage \cite{MIP.2020}. Specifically, it covers 99.9\% of the geographic area in Switzerland. The infrastructure records telecommunication metadata via almost 30,000 antennas across the country. Out of these, approx. 7,000 are of the GSM type (2G), 11,000 of the UMTS type (3G), and 12,000 of the LTE type (4G) \cite{GSM.2020, UMTS.2020, LTE.2020}. As Switzerland has a total of 3,196 postcode areas, there are on average approx. 9 antennas per postcode. Additional details on antennas are provided in Supplement~\ref{supp:antenna}.

The frequency of pings is determined by how often a mobile device connects to a new antenna or, if in between two antennas, every 5 minutes. Hence, if a mobile device is stationary and does not connect to a new antenna, the ping rate is once every 5 minutes. The rate will momentarily increase if a stationary device connects to a new antenna or if a mobile device connects to a new antenna during a trip. Importantly, the variation in ping rates across mobile devices has no influence in our analysis as the lowest possible ping rate (every 5 minutes) produces data of considerably higher temporal resolution than the daily aggregated data we use for the analysis. An internal algorithm by Swisscom ensures that bouncing (\ie, when a phone bounces between antennas) is correctly addressed and attributed to a single antenna. The telecommunication metadata are then used by Swisscom to infer actual locations over time via triangulation (see next section).

\subsection*{Inferring positions of SIM cards via triangulation}

The locations of SIM cards within antenna areas are determined via triangulation between antennas through the use of a positioning algorithm \cite{Kafsi.2019}. A high-level description of the algorithm is as follows. Every signal from a SIM card in the telecommunication data is associated with a probability distribution over locations that represents the uncertainty of its actual location in a given antenna area at the time. The location is estimated from two random variables: the radius $R$, given by the distance of the signal from the origin of the antenna area, and its angle $\Theta$ to the antenna azimuth. Here, $R$ is Gaussian distributed with empirical mean and variance estimated via maximum likelihood, and $\Theta$ follows a multinomial distribution depending only on the antenna azimuth and its bandwidth. The inferred locations are subject to a delay between signals between antennas and SIM cards. To address this, the location at a given point in time is estimated by marginalizing the probability distribution of the radius over the empirical distribution of signal delays estimated from all observations. Details are available in \cite{Kafsi.2019}. In sum, by tracking the location of SIM cards over time, we are able to capture micro-level movements of individuals.

The accuracy of the positioning algorithm has been empirically validated \cite{Kafsi.2019}. The median positioning error is 132 meters, making it highly accurate compared to state-of-the art methods \cite{Leontiadis.2014}. The accuracy was determined by comparing the algorithm's predicted positions to self-reported actual positions for more than 6,000 trips with over 12,000 end-points \cite{Kafsi.2019}.

\subsection*{Deriving mobility estimates from telecommunication data}
\label{sec:deriving_mobility}

Mobility has been frequently found to be helpful for understanding urban phenomena \cite{Bassolas.2019,Gonzalez.2008}. In this study, we use mobility estimates derived from nationwide telecommunication data.

Trips are computed as follows. A single trip is defined as the movement of a SIM card between two different postcode areas after the location has been static for 20 minutes \cite{MIP.2020}. The trip is then counted for both postcodes. Similarly, trips that cross midnight are counted for both days. Trips from neighboring countries into Switzerland are counted for the arrival postcode.

Swisscom define trips as movements between postal code areas as they represent the smallest spatial unit that is officially defined by the federal government. Switzerland has 3,196 postcode areas with high spatial granularity. The exact size varies between urban and rural regions, but, on average, a postcode area in Switzerland covers merely \SI{12.9}{\kilo\meter\squared}. Moreover, 71\,\% of the Swiss working population commute between different postcodes for work and oftentimes even between cantons (\url{https://www.bfs.admin.ch/bfs/en/home/statistics/mobility-transport/passenger-transport/commuting.html}). The average (one-way) travel distance to work is \SI{15.0}{km} (see previous URL), and, the average daily travel distance for leisure activities is \SI{36.8}{km} (\url{https://www.bfs.admin.ch/bfs/en/home/statistics/mobility-transport/passenger-transport/travel-behaviour.html}). For both work and leisure activities, travel routinely spans several postal code areas. Hence, the use of nationwide telecommunication data combined with our definition of trips provides comparatively large-scale estimates of aggregate mobility.

For the analysis, daily trips from Swisscom Mobility Insights were further aggregated as follows. First, each trip between postcodes (including trips with departure location in a neighboring country) was mapped onto cantons at the sub-national level. Here, we used cantonal shape files from the Swiss government and aggregated all daily trips within each canton. The attribution of trips to both the departure and arrival postcodes enables us to capture the number of trips between cantons as well as the number of trips between border cantons and neighbouring countries. The result of the aggregation is a panel (longitudinal) data set of trip counts in all cantons. 

The reason for the aggregation to the cantonal level per day is twofold. First, policy measures are implemented within cantons, and, second, COVID-19 case data are only published per day at the cantonal level. Therefore, data on policy measures and case number are not available on a more granular level.

Swisscom further labels trips according to both mode and purpose. The mode of trips was differentiated based on estimating the location of SIM cards with the positioning algorithm and the position of antennas along train, highway, and road networks. If several modes of mobility were used in the same trip, the mode with the longest leg was chosen. For comparison, public transport through train is an important mode of transportation in Switzerland  (\url{https://www.bfs.admin.ch/bfs/en/home/statistics/mobility-transport/passenger-transport/travel-behaviour.html}) that is relevant for explaining the results. The purpose of mobility was classified based on trips to/from work (called ``commuting'') and all other trips (called ``non-commuting''). This differentiation was based on the home and work location of individuals (in terms of postal code area). These locations were derived from the most frequent geographic location of individuals between 8\,pm--8\,am for home locations and 8\,am--5\,pm for work locations. Afterwards, both home and work location were matched against the departure and arrival (postal code area) of a trip to determine whether the trip was to/from work, and hence labeled ``commuting''. The classification or trips into mode of transport is highly accurate. Specifically, a validation against self-reported data showed that \SI{90}\,\% of all trips were correctly classified \cite{Kafsi.2019}.

\subsection{Merging of data for analysis}

For analysis, the mobility data was merged with data on (1)~policy measures, (2)~the reported number of cases and hospitalizations and deaths attributed to COVID-19, (3)~the number of tests conducted and their share of positive results, and (4)~population sizes. All data were at the daily and cantonal level except the data on testing, which due to lack of availability, were at the daily and country level. Data (1)--(4) are either publicly available online in the form used in the analysis or constructed from other publicly available sources. Supplement \cref{supp:data} describes how the data on policy measures were collected and validated. See \nameref{sec:data} for details on data availability and how the remaining data were collected.

\subsection*{Modeling overview}
\label{sec:models}

In this section, we present the regression models used to estimate the relationship between (1)~policy measures and mobility and (2)~mobility and reported cases. Here, the first model estimates the reduction in mobility due to policy measures. The estimates are identified with a difference-in-difference analysis and may therefore be given a causal interpretation under certain assumptions (see Supplement~\ref{supp:identification_strategy_mobility}). The second model, in turn, estimates the extent to which reductions in mobility predict decreases in the reported number of new cases as policy measures are being implemented. 

The models have parsimonious specifications recommended for isolating policy-relevant insights \cite{Bertozzi.2020} and are informed by epidemiology. In particular, they are formalized as Bayesian hierarchical negative binomial regression models. Rather than modeling the disease dynamics themselves (as with a compartment model), our focus is on estimating the relative effect of other determinants, namely, policy measures and mobility. The use of negative binomial distributions is common in epidemiological modeling, as it allows for overdispersion in dependent variable (\ie, the number of trips and the reported number of new cases). Furthermore, each model uses a log-link between the dependent and explanatory variables. For the model of the reported number of new cases, it enables us to capture the exponential growth in cases during the initial stages of an epidemic, also observed in our data. For the model of mobility, it makes the estimates relative to the observed levels of mobility. Both dependent variables were found to follow negative binomial distributions (with overdispersion). See Supplement~\ref{supp:model_diagnostics_overdispersion} for an analysis of overdispersion.

The models include further controls for (1)~population size per canton, (2)~unobserved heterogeneity between cantons, and (3)~time effects as follows. (1)~We control for differences in population size among cantons with an offset term. This is motivated by the fact that the magnitude of the estimated effects depend on the population size. Hence, the model estimates are relative to the number of inhabitants per canton. (2)~Unobserved heterogeneity is estimated with a canton random effect. We thereby account for unobserved factors that affect both policy measures and mobility (for the former model) and mobility and cases (for the latter model). (3)~Time effects are modeled in two ways. On the one hand, we include weekday fixed effects to control for variations in the implementation of policy measures, levels of mobility, and reporting/testing across weekdays (\eg, mobility is higher on weekdays, whereas testing is lower on weekends; thus, reported cases are lower on weekends and tests conducted on weekends may be reported on Mondays or Tuesdays). On the other hand, we incorporate a trend variable that controls for changes in case dynamics or behavioral adaptations towards social distancing that occur over time since a canton first reported a case. This could for instance occur due to unobserved changes in adherence to other policy measures (\eg, wearing masks and keeping physical distance of at least 1.5 meters) over time. Here, we model the variation in when cantons reported their first case as potentially dependent on the unobserved canton heterogeneity.

The results with additional controls (\eg, testing, spatial correlation between cantons, and dependence between the different trip variables) and alternative dependent variables (\eg, hospitalizations and deaths attributed to COVID-19 instead of reported cases) are part of the robustness checks. 

\subsection*{Model for estimating the reduction in human mobility due to policy measures}
\label{sec:model_mobility}

A multiple time period, multiple group difference-in-difference (DiD) analysis \cite{Goodman.2018} is conducted to estimate the effect of each policy measure on mobility. We restrict the analysis to the time period between February~24 and April~5, 2020; that is, starting before the first reported COVID-19 case in Switzerland and ending prior to Easter holidays. With this time period, the initial observations act as a control group in which mobility is at the baseline level (as individuals may not yet have voluntarily reduced their mobility as a response to reported cases). Furthermore, by ending at April~5, there can be no confounding of the effects of policy measures due to Easter holidays. Such confounding would be caused by that during holidays, mobility generally changes from regular levels and, as a consequence, policy measures are more or less likely to be implemented relative non-holidays.

Let $M_{itk}$ denote the trip count on mobility variable $k =1,2,\ldots,K$ (\ie, total trips, road trips, train trips, etc.) in canton $i = 1,2,\ldots,N$ on day $t =1,2,\ldots,T$. The variable $M_{itk}$ is derived as explained in the previous section and represents the dependent variable in regression model $k$ for the DiD analysis. The values of the model parameters depend on which mobility variable is the dependent variable of the regression; hence, we index all model parameters with $k$. 

We model $M_{itk}$ to follow a negative binomial distribution with conditional mean function
\begin{align}
    \label{eq:mobility_mean}
    \mathbb E[M_{itk}\,|\,\eta^{(k)}_{it},E_i]
    = \mu(M_{itk})
    = E_i \exp \eta^{(k)}_{it}
\end{align}
where $E_i$ denotes the population size of canton $i$. Then, $\exp \eta^{(k)}_{it} = (\mu(M_{itk}))/E_i$ is the expected number of daily trips per inhabitant in canton $i$. The estimates of this model are therefore adjusted according the variation in canton population sizes. The term $\eta^{(k)}_{it}$ is the linear predictor, specified in hierarchical form as
\begin{align}
    \label{eq:mobility_linear_predictor}
    \eta^{(k)}_{it}
    &=
    \alpha^{(k)}_i + 
    \delta^{(k)}_{w(t)} +
    \gamma^{(k)} \log z_{it} +
    \sum^L_{l=1} \beta^{(k)}_l
    d_{itl}, \\
    \label{eq:mobility_level_2}
    \alpha^{(k)}_i
    &=
    \alpha^{(k)} + 
    \theta^{(k)}_i + 
    \gamma^{(k)}_B \widebar{\log z_i},
\end{align}
whose variables and parameters are explained in the following. 

The first term $\alpha^{(k)}_i$ is a time-invariant effect specific to canton $i$. We model $\alpha^{(k)}_i$ as a function of several variables that vary across cantons; see Equation \eqref{eq:mobility_level_2}. Here, $\alpha^{(k)}$ is the intercept among all cantons, which represents the overall baseline relative mobility on Mondays before any policy measure was implemented and any COVID-19 cases were reported. The term $\theta^{(k)}_i$ is a random effect that captures unobserved time-invariant factors for canton $i$ (\eg, population density) that confound the effect of policy measures on mobility. The final variable $\widebar{\log z_i}$ is discussed in detail below. The subscript $B$ on the associated parameter denotes that it measures between-canton effects; that is, the parameter only measures the effect of increases in the variable across cantons.

The variable $d_{itl}$ is a dummy variable that takes a value of $1$ if policy measure $l \in \{1,2,\ldots,L\}$ is implemented by canton $i$ at day $t$ and $0$ otherwise. Hence, $\exp(\beta^{(k)}_l)$ measures the multiplicative effect of policy measure $l$ on the expected number of daily trips on mobility variable $k$ per canton inhabitant. Note that all policy measure variables are included in \eqref{eq:mobility_linear_predictor}. Hence, the effect of each policy measure is conditional on the other policy measures being held fixed. \Cref{fig:1}c shows that the reduction in mobility is similar across cantons; therefore, we do not estimate the heterogeneity in the effect of the policy measures on mobility across cantons.

The term $\delta^{(k)}_{w(t)}$ represents the fixed effect of weekday $w$ on the relative (log-transformed) mobility compared to the reference weekday (here: Monday). The term controls for the confounding factor that aggregate mobility and the probability of implementing a policy measure is likely higher on, \eg, Mondays than Sundays. 

The variable $z_{it}$ captures other sources of time-related confounding and is derived as follows. Let $q_{it}$ be the number of days since the first reported COVID-19 case in canton $i$. The variable is calculated as
\begin{align}
    q_{it}
    =
    \begin{cases}
         t-t'_i, & \text{ if } t > t'_i, \\
         0, &  \text{ if } t \leq t'_i ,
    \end{cases}
\end{align}
where $t'_i$ is the date the first case was reported in canton $i$. Since the logarithm of zero is undefined, we then set $z_{it} = q_{it}+1$ and include the logarithm of $z_{it}$ in the model. The associated parameter $\gamma^{(k)}$ is therefore interpreted as the percentage increase in relative mobility given a 1\,\% increase in the number of days since a canton first reported a case. The rationale for including $z_{it}$ is that individuals may adapt their mobility behavior over time irrespective of social distancing policies. Therefore, the variable captures how mobility would trend over time even if the policy measures were not implemented.

The variable $\widebar{ \log z_i} = T^{-1} \sum_{t=1}^T \log z_{it}$ is the time average of $\log z_{it}$ in canton $i$ (the bar over the expression denotes an average value). The variable is included in the model to allow the canton-specific effect $\theta^{(k)}_i$ to be correlated with $\log z_{it}$ over the cantons. Such correlation would, for instance, arise if the date that the first COVID-19 case is reported in each canton depends on the unobserved canton-specific factors. As an example, the date the first case is reported in a canton could depend on the (unobserved) adherence of inhabitants to social distancing recommendations. If such correlation exists but is ignored, it would instead enter the error term of the model, leading to a violation of the exogeneity assumption and incorrect parameter estimates. By including $\widebar{\log z_i}$, we essentially make $\theta^{(k)}_i$ a Mundlak-type correlated random effect \cite{Mundlak.1978}. The benefit of correlated random effects over fixed effects or standard random effects is that they use only the within-unit variation to estimate parameters (and, hence, give identical estimates to those of fixed effects models) while also having the random effects property of estimating the variation in the unobserved heterogeneity via partial pooling. Note that time averages of the policy measure variables or weekday effects are not included in the model since those are conditionally exogenously determined and, therefore, uncorrelated with the model errors. We refer to \cite{Bell.2015} for a detailed discussion of the underlying benefits of this approach relative to fixed effects.

By substituting the linear predictor \eqref{eq:mobility_linear_predictor} into the conditional mean function \eqref{eq:mobility_mean} and expanding $\alpha^{(k)}_i$, the full model of mobility variable $k$ becomes
\begin{align}\label{eq:mobility_model}
    \log \mathbb E[M_{itk}\,|\,\eta^{(k)}_{it}, E_i]
    =
    \log E_i +
    \alpha^{(k)} + 
    \theta^{(k)}_i + 
    \delta^{(k)}_{w(t)} +
    \gamma^{(k)} \log z_{it} +
    \gamma^{(k)}_B \widebar{\log z_i} +
    \sum^L_{l=1} \beta^{(k)}_l
    d_{itl}.
\end{align}
The conditional variance of $M_{itk}$ is given by
\begin{align}\label{eq:mobility_model_variance}
    \mathbb V[M_{itk}\,|\,\eta^{(k)}_{it}, E_i, \zeta^{(k,M)}]
    =
    \mu(M_{itk})
    \left(
        1 +
        \frac{ \mu(M_{itk}) }{\zeta^{(k,M)}}
    \right),
\end{align}
where $\zeta^{(k,M)}$ is the overdispersion parameter (the superscript $M$ distinguishes the overdispersion parameter of the mobility model from the model of reported cases).

We specify one regression equation in the form of~\eqref{eq:mobility_model} for each mobility variable and estimate them separately. Each regression has the same explanatory variables but a different mobility variable as the dependent variable (\ie, total trips or one of the mobility variables based on mode or purpose).

\subsection*{Model for estimating the relationship between mobility and COVID-19 cases}
\label{sec:model_cases}

The model for estimating the relationship between mobility and reported COVID-19 cases is similar in structure to the model used to link policy measures to mobility. To accommodate the forecast horizon, we lag the mobility variables to estimate how a decrease in mobility at a given day predicts reductions in the reported number of new cases at a later day. This enables forecasting of future reported case growth by evaluating the model at daily observed mobility levels.

Let $C_{it}$ denote the cumulative number of reported cases in canton $i=1,2,\ldots,N$ until and including day $t=1,2,\ldots,T_i$. Then, $Y_{it} = C_{it} - C_{i,t-1}$ is the number of new cases that are reported in canton $i$ on day $t$. Note that the time period (\ie, total number of days $T_i$) varies across cantons $i = 1,\ldots,N$. The reason for this is that the dependent variable of this regression model is the reported number of new cases, and as such we restrict the data to start at the date of each cantons first reported case. Hence, the data for this model starts between February~24 and March~16 (depending on the canton) and ends at April~5 (for all cantons). Our modeling approach accounts for the resulting unbalanced number of observations per canton.

We model $Y_{it}$ as following a negative binomial distribution with conditional mean function
\begin{align}\label{eq:covid_mean}
    \mathbb E[Y_{it}\,|\,\eta^{(k,s)}_{it},E_i]
    = \mu^{(k,s)}(Y_{it})
    = E_i \exp \eta^{(k,s)}_{it},
\end{align}
where
\begin{align}
    \label{eq:cases_linear_predictor}
    \eta^{(k,s)}_{it}
    &= 
    \alpha^{(k,s)}_i +
    \delta^{(k,s)}_{w(t)} +
    \gamma^{(k,s)} \log z_{it} +
    \xi_{ks} \log m_{i,t-s,k}, \\
    \alpha^{(k,s)}_i
    &=
    \alpha^{(k,s)} + 
    \theta^{(k,s)}_i + 
    \gamma^{(k,s)}_B \widebar{\log z_i} + 
    \xi_{ks,B} \widebar{\log m_{ik}}
\end{align}
is the hierarchical linear predictor. In this model, $\exp \eta^{(k,s)}_{it} = (\mu^{(k,s)}(Y_{it}))/E_i$ is the expected number of reported positive cases in canton $i$ on day $t$ relative to the canton population size (sometimes called the relative risk in spatial epidemiology \cite{Riebler.2016, Asmarian.2019}). The model for the daily growth in reported cases can then be written as
\begin{align}\label{eq:cases_model}
    \log \mathbb E[Y_{it}\,|\,\eta^{(k,s)}_{it}, E_i]
    =
    &\log E_i +
    \alpha^{(k,s)} + 
    \theta^{(k,s)}_i + 
    \delta^{(k,s)}_{w(t)} + \nonumber \\
    &\gamma^{(k,s)} \log z_{it} +
    \gamma^{(k,s)}_B \widebar{\log z_i} + 
    \xi_{ks} \log m_{i,t-s,k} +
    \xi_{ks,B} \widebar{\log m_{ik}}.
\end{align}
The conditional variance of $Y_{it}$ given mobility variable $k$ lagged by $s$ days is given by
\begin{align}\label{eq:cases_model_variance}
    \mathbb V[Y_{it}\,|\,\eta^{(k,s)}_{it}, E_i, \zeta^{(k,s,Y)}]
    =
    \mu^{(k,s)}(Y_{it})
    \left(
        1 +
        \frac{ \mu^{(k,s)}(Y_{it}) }{\zeta^{(k,s,Y)}}
    \right)
\end{align}
where $\zeta^{(k,s,Y)}$ is the overdispersion parameter. 

For simplicity, we use the same notation for variables and parameters in this model as in the model used for the effect of policy measures on mobility (but the estimated parameters have, of course, a different interpretation and values). The superscript $(k,s)$ is attached to parameters to indicate that their values depend on the choice of mobility variable $k$ and its lag $s$.

The parameter of interest is $\xi_{ks}$. It measures the expected percentage change in the reported number of new cases per canton inhabitant $s$ days after a 1\,\% increase in mobility variable $k$. Hence, the parameter shows how the relative growth rate in reported cases changes as a function of lagged mobility, after adjusting for relevant factors but where mobility varies according to which policy measures are implemented. Note that we intentionally include only a single lag $s$ (and then refit the model for different lags) rather than including multiple lags at the same time. We follow this approach to avoid issues related to multicollinearity between the mobility variables and because one cannot condition on mobility in future days when predicting future reported cases from current mobility. Supplement~\ref{supp:identification_strategy_cases} further explains our rationale.

The intercept $\alpha^{(k,s)}$ gives the baseline number of reported cases relative to the canton population for Mondays. 

The parameter $\delta^{(k,s)}_{w(t)}$ is the effect of weekday $w$ relative to the Monday effect. By including weekday effects in the model, we control for confounding differences in the number of trips and the number of reported COVID-19 cases between weekdays that would bias the parameter estimate on the lagged mobility variable. For instance, people travel to schools and work primarily on weekdays and, similarly, there are fewer COVID-19 tests on weekends and thus fewer reported cases on Monday/Tuesday (due to reporting delays). 

The variable $\log z_{it}$ is also included in this model. It now controls for the fact that mobility and the reported case growth both depend on when the first case was reported in a canton.

The Mundlak-style random effects $\theta^{(k,s)}_i$ estimate the impact of unobserved canton-specific factors that may be correlated with both the variation in mobility across cantons and the logarithm of the number of days since the first case was reported in each canton. We achieve this by including variables of the time averages of $\log z_{it}$ and lagged mobility, that is, $\widebar{\log m_{ik}} = T^{-1}\sum_{t-s} \log m_{i,t-s,k}$. Then, any potential cross-canton correlation between the random effect and $\log m_{i,t-s,k}$ or $\log z_{it}$ via the models error term is controlled for.

The above regression model is fitted separately for each $(k,s)$, that is, each pair of mobility variable and lag. This allows us to investigate to what extent different lags of each mobility variable predict the number of reported cases.

\subsection*{Estimation details}
\label{sec:estimation}

We estimate our models in a fully Bayesian framework. We run 4 Markov chains for every model, each with 2000 warm-up samples and another 2000 samples from the posterior distributions. Since our models are fitted with a log-link, we transform the posterior parameter samples so that they give estimates for the original scale of the dependent variable. For each parameter, we report in our plots the posterior mean and the associated 80\,\% and 95\,\% credible intervals (CrI) of the transformed posterior distribution.

The software used for estimation is the \textsf{R} package \texttt{brms} \cite{Burkner.2017, Burkner.2018} version 2.11.1 built upon the statistical modeling platform Stan \cite{Carpenter.2017}. Parameter estimates are obtained by Markov chain Monte Carlo sampling in Stan version 2.19.2 using the Hamiltonian Monte Carlo algorithm \cite{Neal.2011, Duane.1987} and the No-U-Turn sampler (NUTS) \cite{Hoffman.2014}.

\Cref{tbl:priors} presents our choices of priors for the variables in the models. We use weakly informative priors to stabilize the computations and provide some regularization. Our prior on each $\beta_l$ reflects that we expect each policy measure to reduce the logarithm of expected mobility with 25\,\%, on average, but that effects between 0--50\,\% are relatively probable. The prior on each $\xi_{ks}$ implies that we expect that a 1\,\% decrease in the lagged mobility variable predicts a 1\,\% decrease in reported cases for each of the considered lags, with negative effect sizes or effect sizes exceeding 2\,\% being unlikely. The prior on $\gamma$ implies that we expect the relative outcome to increase with 1\,\% for each 1\,\% increase in the number of days since the first reported case. The intercept, overdispersion parameter, and standard deviation of the canton random effects are given weakly informative priors. The prior on $\delta_{w(t)}$ states that the effect of a given weekday that is not Monday should fall within ~50--150\,\% of the Monday effect. The parameters for the variables of between-canton averages are assigned vague priors since we have no a~priori belief of their effects.

\begin{table}[H]
\begin{center}
\footnotesize
\begin{tabular}{l l l l}
\toprule
\text{Parameter}  & \text{Description} & \text{Prior} & \text{Model} \\
\midrule
$\beta_l$         & \text{Effect of policy measure $l$} &  $\mathsf N(-0.25,0.25)$ & \eqref{eq:mobility_model} \\
$\xi_{ks}$        & \text{Effect of log mobility variable $k$ with a lag of $s$} & $\mathsf N(1,1)$  & \eqref{eq:cases_model} \\
$\alpha$          & \text{Intercept}                             & $\mathsf{Half\text{-}t}(3, 1.8, 2.5)$        & \eqref{eq:mobility_model}, \eqref{eq:cases_model} \\
$\theta_i$       & \text{Canton random effect}              & $\mathsf{N}(0, \sigma_\theta)$        & \eqref{eq:mobility_model}, \eqref{eq:cases_model} \\
$\sigma_\theta$  & \text{Standard deviation for canton random effect}              & $\mathsf {Half\text{-}t}(3, 0, 2.5)$       & \eqref{eq:mobility_model}, \eqref{eq:cases_model} \\
$\delta_{w(t)}$   & \text{Effect of weekday $w$ (compared to Monday)}                 & $\mathsf N(0,0.5)$        & \eqref{eq:mobility_model}, \eqref{eq:cases_model} \\
$\gamma$          & \text{Effect of log no. of days since 1st reported case}    & $\mathsf N(1,1)$       & \eqref{eq:mobility_model}, \eqref{eq:cases_model} \\
$\gamma_B$        & \text{Effect of between-canton average of log no. of days since 1st reported case}    & $\mathsf N(0,5)$       & \eqref{eq:mobility_model}, \eqref{eq:cases_model} \\
$\xi_{ks,B}$      & \text{Effect of Between-canton average of log mobility with a lag of $s$} & $\mathsf N(0,5)$  & \eqref{eq:cases_model} \\
$\zeta$           & \text{Overdispersion in dependent variable}        & $\mathsf {Gamma}(0.01, 0.01)$   & \eqref{eq:mobility_model}, \eqref{eq:cases_model} \\
\bottomrule
\multicolumn{4}{p{16cm}}{\emph{Note:} The superscripts $(k)$ and $(k,s)$ are omitted as the same priors are assigned to each model. The column ``Description'' states what effect the associated parameter represent (except for the overdispersion parameter).}
\end{tabular}
\caption{Choice of priors}
\label{tbl:priors}
\end{center}
\end{table}

\subsection*{Model diagnostics}

We followed common practice for model diagnostics of Bayesian models \cite{Gelman.2013}. For each of the models, we inspected (1)~posterior predictive checks, (2)~divergent transitions, (3)~effective sample size and convergence of the Markov chains, (4)~overdispersion in the dependent variables, (5)~influential observations, and (6)~correlation between the policy parameters. All model diagnostics indicate a good fit. Details are provided in Supplement~\ref{supp:model_diagnostics}.

\subsection*{Robustness checks}

First, we checked the robustness of the model estimates against alternative specifications of time effects: (a)~A model specified as in the main paper but where the logarithmic trend is replaced with corresponding linear and quadratic trends (of the number of days since the first reported case in each canton) to capture nonlinearities in both the reported case dynamics and general behavior towards social distancing. (b)~A model with additional week fixed effects (\ie, a weekday fixed effect, a week fixed effect, and a trend variable in logarithmic form). This model allows us to control for weekly exogenous shocks (\eg, media reports about the shortage of critical care in Italy) but acknowledge that such fixed effects would be unknown at the time of forecasting and, therefore, cannot be used to predict reported case growth at a future date. All models yield similar estimates and hence confirm the predictive ability of the mobility variables (see \Cref{supp:robustness_checks_time}). 

Second, the number of reported cases could potentially depend on the number of conducted tests per canton and day. When controlling for this, we obtain similar estimates (\Cref{supp:robustness_checks_tests}). 

Third, we extend the models by including a spatial random effect, as commonly used in the spatial epidemiology and disease mapping literature \cite{Besag.1991, Wakefield.2000}. This approach allows us to account for the spatial dependence in mobility between neighboring cantons. We find that the spatial dependence is low and retrieve estimated effects of policy measures that are practically identical to those of the main analysis (see \Cref{supp:extension_spatial}).  

Fourth, we account for potential dependence between different mobility variables by estimating their models jointly (that is, by modeling the covariance of the canton random effects for the different mobility variables). This model yields slightly narrower credible intervals but almost identical point estimates for the effects of the policy measures (see \Cref{supp:extension_multivariate}).   

\newpage 
\section*{Data availability}
\label{sec:data}

Human mobility data presented in this work are available from the Swisscom Mobility Insights platform (\url{https://mip.swisscom.ch}). Cantonal geographic boundaries can be found as shape files at the Federal Office of Topography; swisstopo (\url{https://shop.swisstopo.admin.ch/en/products/landscape/boundaries3D}). Data on reported COVID-19 cases per canton and relative to the cantonal population size (\ie, cases per 100,000 inhabitants) come from the Federal Office of Public Health of the Swiss Confederation; BAG (\url{https://www.covid19.admin.ch/en/overview}). We also use this source to obtain data on deaths and hospitalizations attributed to COVID-19 per canton, and data on testing conducted in Switzerland. Additional information on the Swiss population come from the Swiss Federal Statistical Office; BFS (\url{https://www.bfs.admin.ch/bfs/en/home/statistics/population.html}). Data on antenna locations are obtained from a web tool at the Swiss federal geoportal (\url{https://www.geo.admin.ch/}, tool: \url{https://map.geo.admin.ch/}), which is based on data provided by the Federal Office of Topography; swisstopo (\url{https://www.swisstopo.admin.ch/en/home.html}), and the Federal Office of Communications; OFCOM (\url{https://www.bakom.admin.ch/bakom/en/homepage.html}) (direct links to the maps: \url{https://s.geo.admin.ch/8f7aa435b8}, \url{https://s.geo.admin.ch/8f7aa5536b}, and \url{https://s.geo.admin.ch/8f7aa69498}). Details on data collection for policy measures are provided in \Cref{supp:data}. When referring to cantons, we use abbreviations instead of full canton names (\url{https://www.admin.ch/ch/d/gg/pc/documents/1336/Abkuerzungsverzeichnis.pdf}).

Code and data for reproducing our results are available at our GitHub page (\url{https://github.com/jopersson/covid19-mobility}).

\section*{Ethics declarations} 

\textbf{Competing interests.} S.F. declares membership in a \emph{COVID-19 Working Group} by the \emph{World Health Organization} (WHO) but without competing interests. J.P. and S.F. acknowledge funding from the \emph{Swiss National Science Foundation} (SNSF) on data-driven health management (grant 186932).

\vspace{0.4cm}
\noindent
\textbf{Ethics approval.} Ethics approval \mbox{(2020-N-41)} was obtained from the institutional review board at (name anonymized for blind peer-review). 

\vspace{0.4cm}
\noindent
\textbf{Author contributions.} J.P. and S.F. contributed to conceptualization, data collection, data analysis (modeling), results interpretation, and manuscript writing. J.F.P. contributed to conceptualization, data collection, data analysis (exploratory), results interpretation, and manuscript writing.

\vspace{0.4cm}
\noindent
\textbf{Acknowledgments.} We thank Dominik Hangartner and Achim Ahrens for the invaluable feedback. We also thank Swisscom for their extensive support.

\vspace{0.4cm}
\noindent
\textbf{Correspondence.} Joel Persson (\url{jpersson@ethz.ch})

\newpage

\bibliography{literature}

\newpage 

\appendix
\begin{center}
\Large Supplemental Information
\end{center}

\section{Data}
\label{supp:data}

\subsection{Policy measures}

We collected data on policy measures implemented at both national level directly from the official resources of the federal government in Switzerland (\url{www.admin.ch}). Implementations of policy measures at sub-national (cantonal) level were collected from official resources of the cantonal authorities (\eg, \url{www4.ti.ch}, \url{www.ge.ch}). These policy measures were implemented throughout the complete canton (\ie, there is no ``partial'' implementation). We then checked our data on policy measures against benchmark datasets, namely the government response event dataset CoronaNet \cite{Cheng.2020}, the Government Response Tracker \cite{Hale.2020} and the Swiss National COVID-19 Science Task Force (\url{https://ncs-tf.ch/en/situation-report}). As our goal is to study the response and impact of mobility, we excluded policy measures that were primarily used to target physical distance and not mobility. Examples of such policy measures is the requirement to wear a mask or the recommendation to keep a physical distance of at least 1.5 meters between people.

The policy measures were encoded as follows. We encoded ``school closures'' such that the closure falls on a weekday. That is, when school closures were put into effect on a Saturday, we encoded ``school closures'' as being closed from Monday onwards. The reason is that both primary and secondary schools are, by default, closed on weekends and, hence, movements from/to school can only be in effect on the next weekday. Different from other countries, schools were closed not ``partially'', \eg, for specific age groups, but for all. We encoded ``closed border'' such that this policy measure is in effect when any side of the border is closed. As an example, when Italy closed its border, we encoded the borders for the adjacent cantons as closed. The rationale is that trips will be cancelled if people cannot return. For all cantons without borders to other countries, we encoded ``border closures'' as implemented when the national government decided put travel restrictions in place. Consider Zurich as an example. The cantonal government of Zurich did not put travel restrictions into effect, and, hence, we set ``border closures'' for the canton of Zurich to March~25, 2020, which is the date when the national government enforced travel restrictions. Before that date, travel into the canton was possible through Zurich airport and it was only restricted from March~25, 2020 onward due to the national (and not the cantonal) government. 

The resulting list of policy measures is shown in \Cref{tbl:NPI_dscr}. All of these policy measures remained in effect from the day of their implementation until the end of our study period (\ie, through April~26, 2020). Furthermore, note the cross-canton variation in implementation dates for some of the policy measures. The difference in timing for a given policy measure is one of the aspects that enables us identify the effects of the policy measures with the difference-in-difference analysis. These comparisons better approximate differences between a treatment and control group when the cantons are comparable by having similar values on unobservable and included control variables in the regression model (see \nameref{sec:model_mobility}). An example of a comparable difference is that between St. Gallen~(SG) and Lucerne~(LU). These cantons are similar in geographical size, population size, and population density, but whereas SG implemented border closures on March~14, LU was not affected by border closures until March~25 (\Cref{tbl:NPI_dscr}). Hence, between these dates, LU provide control observations for SG that enable us to identify the effect of borders closures in the whole country.

\begin{table}[H]
\singlespacing
\centering
\footnotesize
\begin{tabular}{l l m{5cm} m{4.75cm}}
\toprule
\specialcell{\footnotesize{Implementation}\\\footnotesize{date}} & Policy measure & Description & Cantons \\
\midrule
10-03-2020 & Border closures & Italian border closed & {\footnotesize GR, TI, VS}\\
\\
\vspace{1cm}
13-03-2020 & Ban $>$ 100 & Ban on gatherings with $>$ 100 people & \specialcell{\footnotesize AG,AR,AI,BL,BS,BE,FR,GE,GL,\\\footnotesize GR,JU,LU,NE,NW,OW,SH,SZ,SO,\\ \footnotesize SG,TI,TG,UR,VD,VS,ZG,ZH}\\
\vspace{0.6cm}
14-03-2020 & Venue closures &  Closure of venues & \footnotesize{TI}\\ 
\\
14-03-2020  &  Border closures &  Austrian border closed & \footnotesize{GR, SG}\\ 
\\
16-03-2020 & School closures & Closure of primary and high schools &  \specialcell{\footnotesize AG,AR,AI,BL,BS,BE,FR,GE,GL,\\\footnotesize GR,JU,LU,NE,NW,OW,SH,SZ,SO,\\ \footnotesize SG,TI,TG,UR,VD,VS,ZG,ZH} \\
\\
17-03-2020 & Venue closures & Closure of non-essential stores (all businesses except supermarkets, food suppliers and pharmacies), museums, zoos, hairdressers, garden centers, restaurants, nightclubs and bars &   \specialcell{\footnotesize AG,AR,AI,BL,BS,BE,FR,GE,GL,\\\footnotesize GR,JU,LU,NE,NW,OW,SH,SZ,SO,\\ \footnotesize SG,TI,TG,UR,VD,VS,ZG,ZH} \\ 
\\
17-03-2020 &  Border closures & German border closed & {\footnotesize AG, BL, BS, SH, TG, ZH} \\ 
\\
18-03-2020 & Ban $>$ 5 &  Ban on gatherings with $>$ 5 people & {\footnotesize JU, VD} \\ 
\\
20-03-2020 & Ban $>$ 5 & Ban on gatherings with $>$ 5 people &  \specialcell{\footnotesize AG,AR,AI,BL,BS,BE,FR,GE,GL,\\\footnotesize GR,JU,LU,NE,NW,OW,SH,SZ,SO,\\ \footnotesize SG,TI,TG,UR,VD,VS,ZG,ZH}\\
\\
18-03-2020 & Border closures & French border closed & {\footnotesize GE, JU, NE, VD}\\ 
\\
25-03-2020 & Border closures & Swiss border closed &  \specialcell{\footnotesize AG,AR,AI,BL,BS,BE,FR,GE,GL,\\\footnotesize GR,JU,LU,NE,NW,OW,SH,SZ,SO,\\ \footnotesize SG,TI,TG,UR,VD,VS,ZG,ZH}\\ 
\bottomrule
\end{tabular}
\caption{Timeline of policy measure implementations across sub-national levels (cantons). When referring to cantons, we use abbreviations from \url{https://www.admin.ch/ch/d/gg/pc/documents/1336/Abkuerzungsverzeichnis.pdf}.}
\label{tbl:NPI_dscr}
\end{table}

\subsection{Variation in timing of policy measures across cantons}
 
We summarize the spatio-temporal heterogeneity across cantons of when policy measures were put into effect. For this, we first study the difference in days between cantons of when any pair of two policy measures were implemented. We take the absolute value of the difference in days as the difference should not depend on which canton implemented a policy measure first. We then calculate the mean and median of the absolute differences to obtain easily interpretable summary statistics of the variation in timing. \Cref{tbl:NPI_diff} shows that the policy measures are rarely implemented close in time; instead, there is a substantial delay between implementation dates across cantons. This variation among policy measures between cantons enables us to disentangle the effects of the different policy measures and affects the precision of their estimated effects. See Supplement~\ref{supp:param_corr} and \ref{fig:traffic_bivardist_total} for a further discussion.

\begin{table}[H]
    \begin{subtable}[t]{.45\textwidth}
        \raggedright
        \footnotesize
        \begin{tabular}{@{} l *{10}{c} @{}}
            \toprule
            &
            \rotatebox[origin=l]{90}{Ban $>$ 100} &
            \rotatebox[origin=l]{90}{School closures} &
            \rotatebox[origin=l]{90}{Venue closures} &
            \rotatebox[origin=l]{90}{Ban $>$ 5} &
            \rotatebox[origin=l]{90}{Border closures} \\
            \midrule
                 Ban $>$ 100 & {---} & 3.0 & 3.9 & 6.9 & 7.8 \\ 
                 School closures & 3.0 & {---} & 1.0 & 3.9 & 5.6 \\ 
                 Venue closures & 3.9 & 1.0 & {---} & 3.0 & 4.8 \\ 
                 Ban $>$ 5 & 6.9 & 3.9 & 3.0 & {---} & 4.6 \\ 
                 Border closures & 7.8 & 5.6 & 4.8 & 4.6 & {---} \\ 
            \bottomrule
        \end{tabular}
    \end{subtable}%
    \hfill
   \begin{subtable}[t]{.45\textwidth}
        \raggedleft
        \footnotesize
        \begin{tabular}{@{} l *{10}{c} @{}}
            \toprule
            &
            \rotatebox[origin=l]{90}{Ban $>$ 100} &
            \rotatebox[origin=l]{90}{School closures} &
            \rotatebox[origin=l]{90}{Venue closures} &
            \rotatebox[origin=l]{90}{Ban $>$ 5} &
            \rotatebox[origin=l]{90}{Border closures}\\
            \midrule
                 Ban $>$ 100 & {---} & 3.0 & 4.0 & 7.0 & 5.0 \\ 
                 School closures & 3.0 & {---} & 1.0 & 4.0 & 6.0 \\ 
                 Venue closures & 4.0 & 1.0 & {---} & 3.0 & 7.0 \\ 
                 Ban $>$ 5 & 7.0 & 4.0 & 3.0 & {---} & 5.0 \\ 
                 Border closures & 5.0 & 6.0 & 7.0 & 5.0 & {---} \\ 
            \bottomrule
        \end{tabular}
    \end{subtable}
    \caption{Variation in implementation dates of policy measures across cantons (absolute time difference in days): mean (left) and median (right).}
    \label{tbl:NPI_diff}
\end{table}

\clearpage

\section{Validation of mobility data and border closures}

In this paper, we measure mobility using variables of trip counts derived from telecommunication metadata. Here, we validate the mobility variables against alternative measurements. We furthermore validate the estimated effect of border closures, the only policy measure whose credible interval covered an effect of zero.

\subsection{Validating mobility data}

We validate our mobility data against alternative measures of mobility derived from (1)~the telecommunication service provider (Swisscom Mobility Insights) and (2)~a representative panel survey (Intervista AG). 

\begin{figure}[H]
    \centering
    \begin{minipage}[t]{0.75\textwidth}
        \figletter{a} \\
        \begin{subfigure}{\textwidth}
            \centering
            \includegraphics[width=\linewidth]{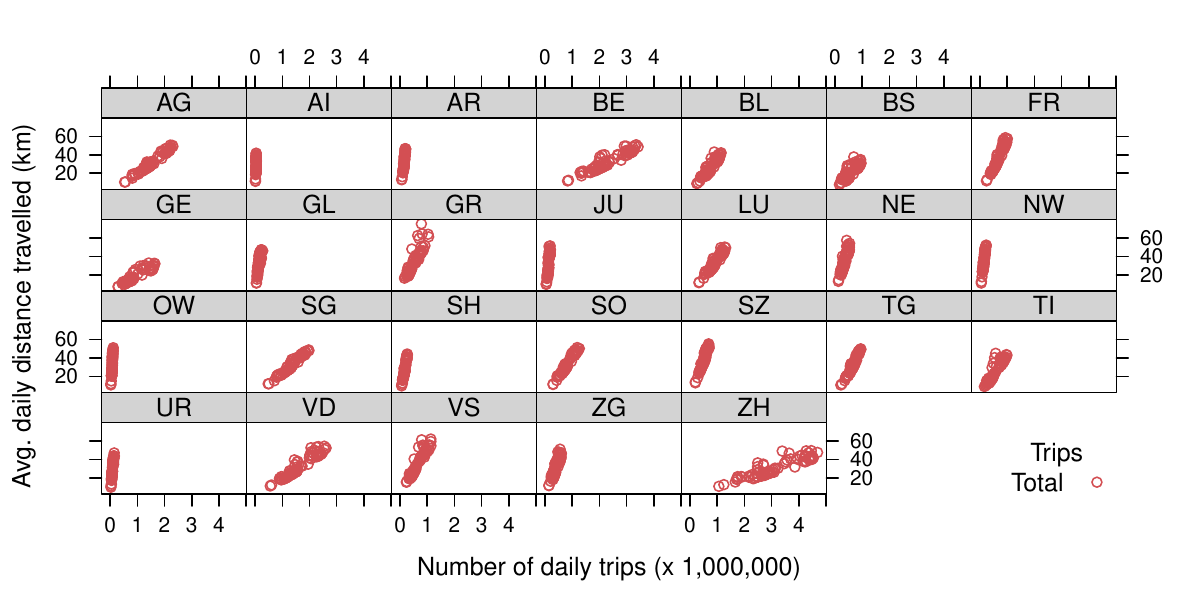}
        \end{subfigure}
    \end{minipage}
    \\
    \begin{minipage}[t]{0.75\textwidth}
        \figletter{b} \\
        \begin{subfigure}{\textwidth}
            \centering
            \includegraphics[width=\linewidth]{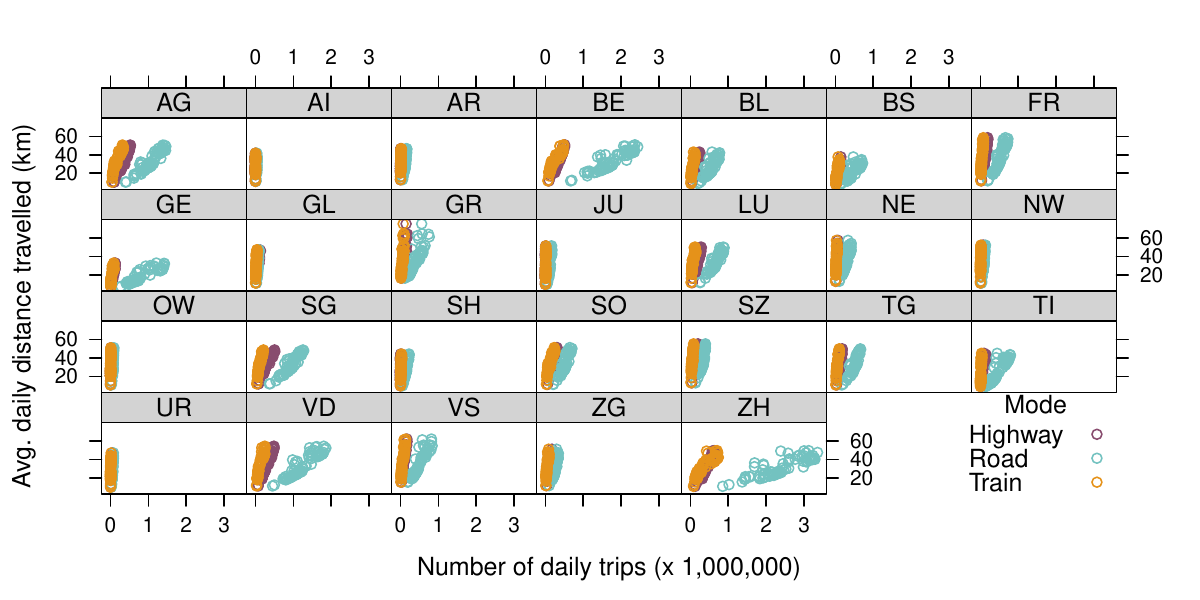}
        \end{subfigure}
    \end{minipage}
    \\
    \begin{minipage}[t]{0.75\textwidth}
        \figletter{c} \\
        \begin{subfigure}{\textwidth}
            \centering
            \includegraphics[width=\linewidth]{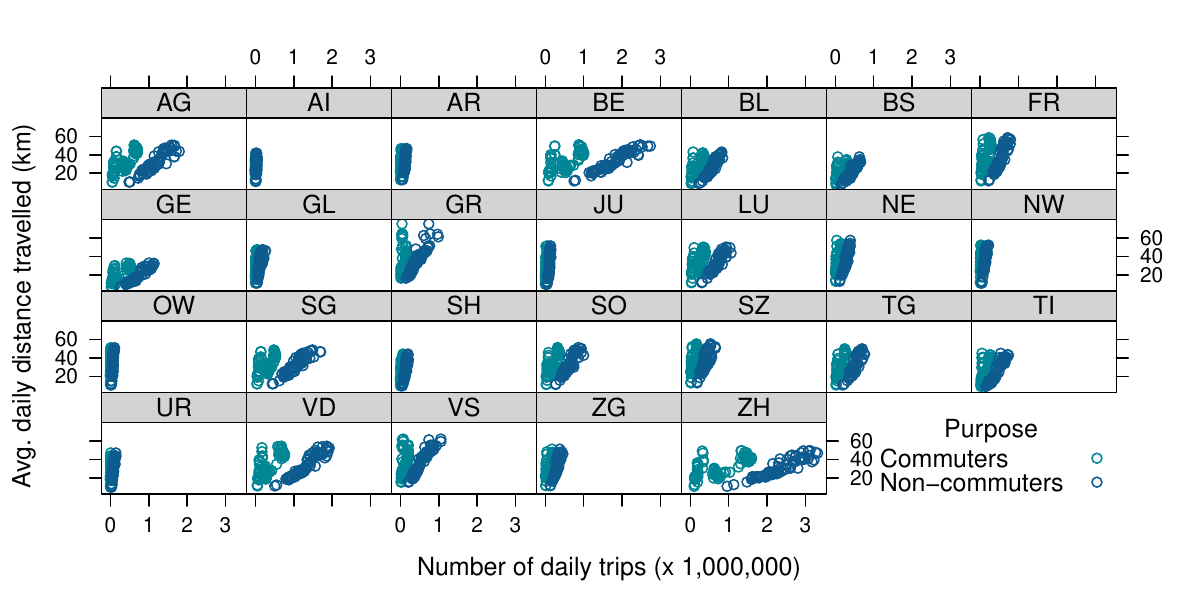}
        \end{subfigure}
    \end{minipage}
\caption{Scatterplots of data on the average distance travelled per person and day inferred from nationwide telecommunication data (Swisscom) against data on the number of \textbf{(a)} total trips, \textbf{(b)} trips by mode, and \textbf{(c)} trips by purpose, also inferred from nationwide telecommunication data (Swisscom).}
\label{fig:avg_distance_trips}
\end{figure}

For (1), we plot the relationship between the mobility variables of daily trip counts (by mode, by purpose, and in total) used in this study and the average daily distance travelled (km) per day and canton inhabitant, both obtained from Swisscom Mobility Insights. \Cref{fig:avg_distance_trips} shows that overall, the trip count variables are positively correlated with average distance travelled. This means that for days and cantons in which inhabitants made more trips, inhabitants also travelled longer, on average. 

\begin{figure}[H]
    \centering
            \includegraphics[width=0.9\textwidth]{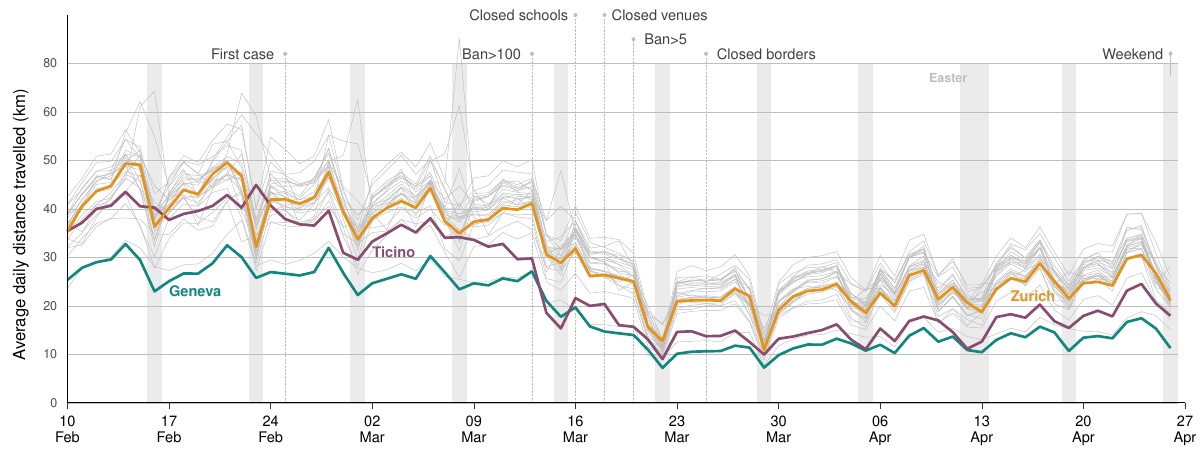}
\caption{Average daily distance travelled (km) per canton as inferred from nationwide telecommunication data (Swisscom).}
\label{fig:parallel_trends_total_distance}
\end{figure}

\begin{figure}[H]
    \centering
    \includegraphics[width=0.5\linewidth]{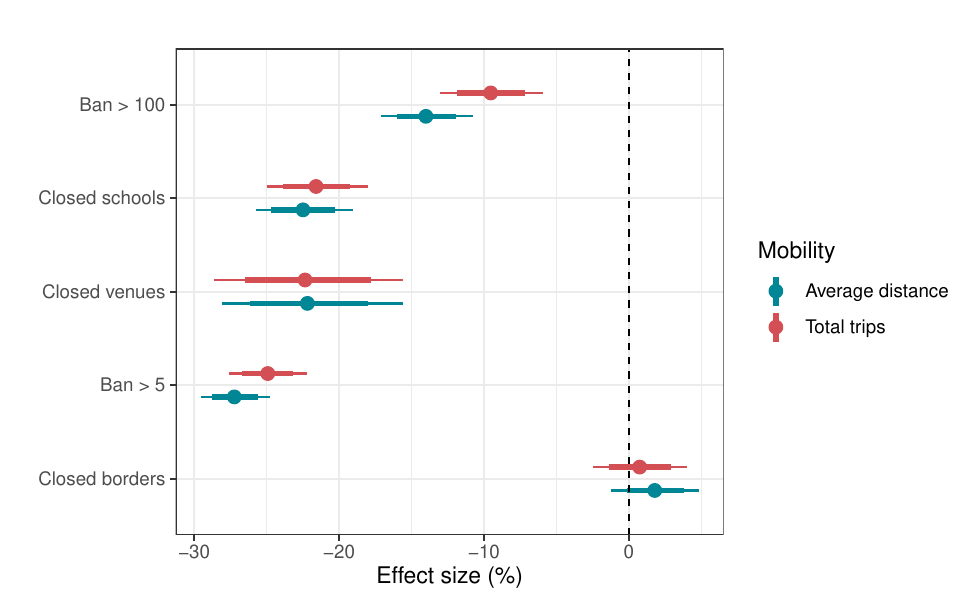}
\caption{Estimated effect of policy measures on total daily trip counts and average daily distance travelled (km) as inferred from nationwide telecommunication data (Swisscom). Posterior means are shown as dots, while 80\,\% and 95\,\% credible intervals are shown as thick and thin bars, respectively. Policy measures are arranged in the order in which they were implemented (cf. Supplement~\ref{supp:data}).}
\label{fig:did_dist}
\end{figure}

We now report the average distance travelled per canton as inferred from telecommunication data. This is shown in \Cref{fig:parallel_trends_total_distance}. Indeed, it follows the same general patterns over time as trip counts. The figure suggests that, not only did policy measures reduce the number of trips, they also reduced the average distance of trips. To check this, we re-estimate our regression model for the difference-in-difference analysis with average distance travelled as the dependent variable. \Cref{fig:did_dist} shows that the estimates for average daily distance travelled are consistent with the estimates for total trip counts. This validates our results against alternative measures of mobility.  

For (2), we draw upon a representative panel survey on mobility in Switzerland. The survey recorded the average distance travelled per day in Switzerland (at the country level and not at the canton level). The mobility survey is based on 2,561 participants that were representative of the population in Switzerland by age, gender, and region. See \url{https://www.experimental.bfs.admin.ch/expstat/de/home/innovative-methoden/mobil.html} and the links therein for details.

We plot the relationship between our variables interest (\ie, daily trip counts inferred from nationwide telecommunication data) and the variables of the average distance travelled per day as obtained through the panel survey. Because the trip counts are available at the canton level but the data from the panel survey are only available at the country level, they cannot be directly compared and, instead, we proceed as follows. The trip counts per canton from telecommunication data are first aggregated to the country level. We then match each variable of trip counts by mode, by purpose, or in total with the closest category of average distance travelled. Thus, the daily count of, \eg, commuter trips is plotted against the average daily distance travelled to work or school. We estimate a linear regression and report the estimated regression coefficients and the total variance explained ($R^2$). Here, we consider the new variables from the panel survey (Intervista AG) to be the dependent variables, while the trip count variables from telecommunication data (\ie, from our main paper) are the independent variables.

\begin{figure}[H]
    \centering
    \includegraphics[width=0.5\linewidth]{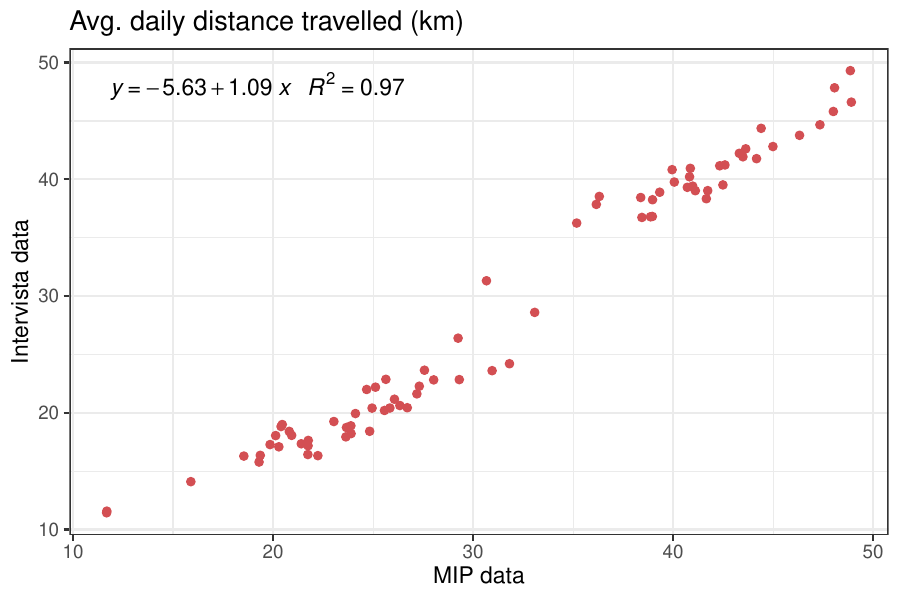}
\caption{Relationship between measures of mobility from two sources. The $x$-axis is average daily distance travelled (km) per inhabitant as inferred from nationwide telecommunication data (Swisscom). The $y$-axis is average daily distance travelled (km) per inhabitant from a representative panel survey on mobility (Intervista AG).}
\label{fig:avg_distance}
\end{figure}

\begin{figure}[H]
    \centering
    \includegraphics[width=\linewidth]{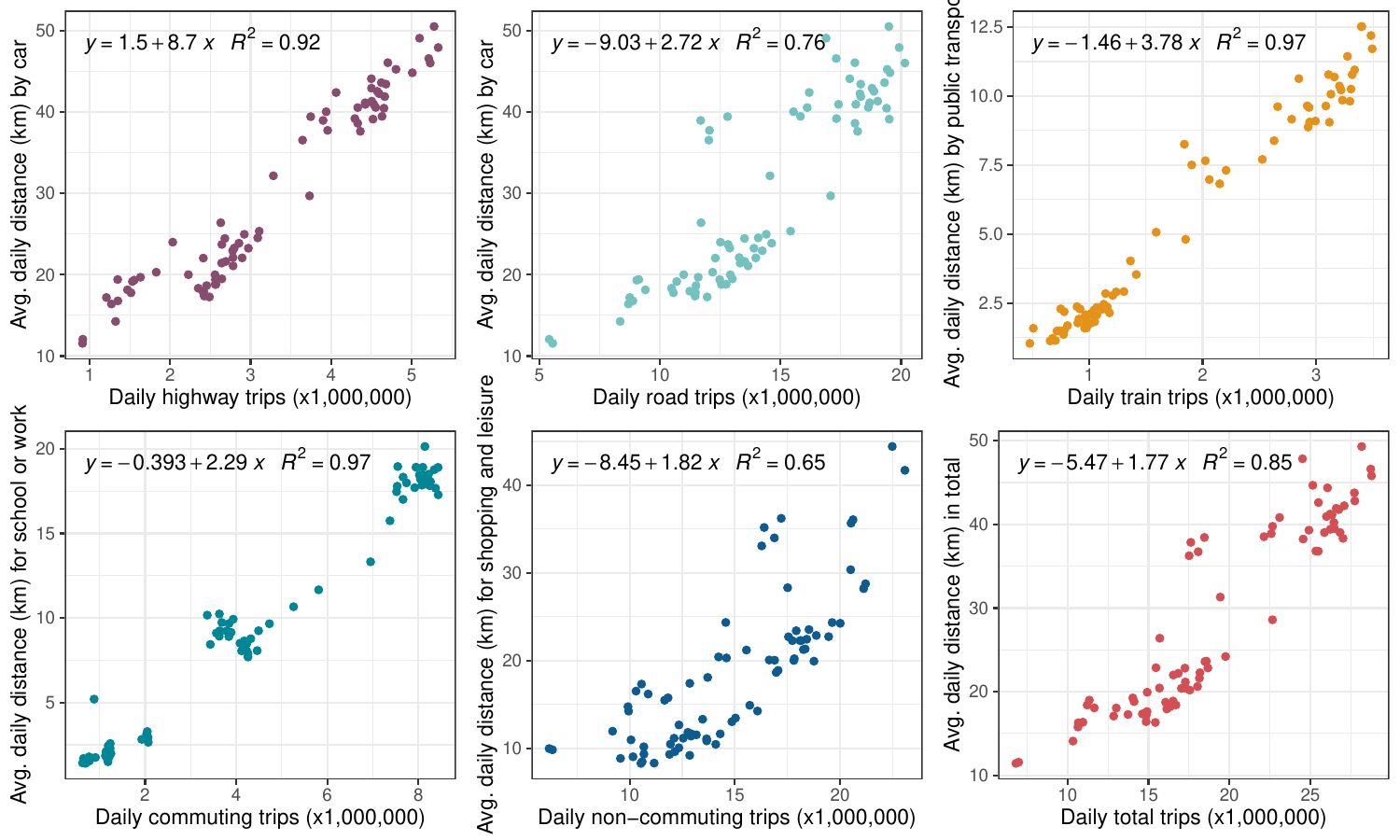}
\caption{Relationship between measures of mobility per category from two sources. The $x$-axis is total daily trip counts as inferred from nationwide telecommunication data (Swisscom). The $y$-axis is average daily distance travelled (km) per inhabitant from a representative panel survey on mobility (Intervista AG). Each category of trips (\ie, trips by mode or purpose) in the Swisscom data is matched with the closest category in the Intervista AG data.}
\label{fig:mip_intervista_category}
\end{figure}

Average distance travelled from the telecommunication data account \SI{97}\,\% of the variance in average distance travelled from the representative panel survey; see \Cref{fig:avg_distance}. Moreover, for each category of mobility, trip counts from the telecommunication data account for between \SI{65}--\SI{97}\,\% of the variance in average distance travelled from the representative panel survey (\Cref{fig:mip_intervista_category}). We conclude that the overall patterns of mobility captured by our approach are therefore consistent with other measures of mobility.

\subsection{Validating the effect of border closures}
\label{supp:validating_border_closures}

Border closures was the only policy measure whose credible interval covered zero (\Cref{fig:DiD_estimates}), meaning they may not have affected mobility. Here, we validate this finding.

First, we note that border closures will only directly limit mobility across borders to neighboring countries and not mobility within Switzerland. Second, salaries and living costs are generally higher in Switzerland than in neighboring countries. This suggest that cross-border mobility is expected to be coming from two types of people: (1)~People living in Switzerland and that travel across the border to buy groceries, do activities, etc. (2)~People living in a neighboring country and that work in Switzerland. It is reasonable to assume that mobility of type (2) is of greater magnitude than (1) as such people would typically cross the border more often. In our data, mobility by (1) is attributed to non-commuting trips and mobility by (2) to commuting trips, per Swisscoms classification procedure \cite{Kafsi.2019}. This follows from that trips with a mobile device connected to the Swisscom network will be captured irrespective of departure and arrival location as long as (a) the trip covers at least two Swiss postcodes, and (b) the mobile device is, at some point after departure, static for at least 20 minutes in Switzerland (including on arrival). This follows directly from the rule that Swisscom uses to define a trip. Hence, our data captures also cross-border trips. 

Given the above, border closures are expected to predominantly reduce commuting trips. This is confirmed by \Cref{fig:DiD_estimates}c, which shows that closed borders is estimated to have reduced commuting trips but not non-commuting trips. Furthermore, the time series in \Cref{fig:1}e shows that commuting trips remained low or even slightly decreased after borders were closed, whereas non-commuting trips gradually increased. Thus, border closures indeed appear to only have affected commuting trips.

Another implication from the above arguments it was mostly commuting trips across the border, and not commuting trips within Switzerland, that were responsible for the reduction in commuting trips following border closures. Assuming this is true, there can be two possible explanations for the fact: (a)~foreign cross-border commuters stayed in their neighboring country of residence, but because the border was closed, could not make their commute, or (b)~because the border was closed, foreign cross-border commuters could not work as previously and hence moved. Here, explanation (b) implies that the population of cross-border commuters changed, thus confounding the effect of border closures with population changes. Hence, only under explanation (a) is the reduction in mobility following border closures coming from a reduction in mobility and not people.

\begin{figure}[H]
        \centering
        \includegraphics[width=\linewidth]{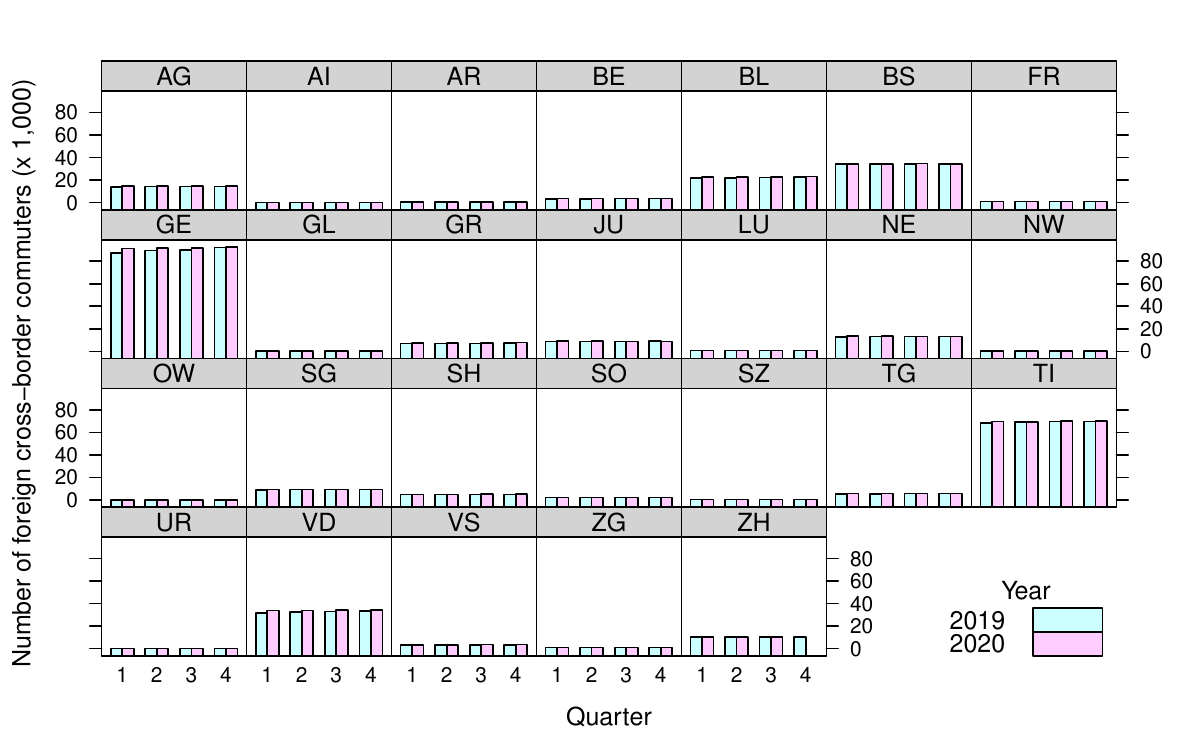}
\caption{Official statistics on the number of foreign cross-border commuters per canton by quarter in 2019 and 2020. The number of foreign cross-border commuters are stable across quarters within each year and, for any given quarter, practically equal between the two years.}
\label{fig:cross_border_commuters}
\end{figure}

\Cref{fig:cross_border_commuters} shows official statistics on foreign cross-border commuters per canton by quarter in 2019 and 2020. The number of foreign cross-border commuters are stable across quarters within each year and, for any given quarter, practically equal between the two years. Thus, the data is more consistent with that the reduction in commuting trips was due to people not being able to make their commutes than a reduction in people commuting across the border. The data is available at \url{https://www.bfs.admin.ch/bfs/de/home/statistiken/arbeit-erwerb/erhebungen/ggs.assetdetail.15864921.html}.

\section{Antenna coverage}
\label{supp:antenna}

\begin{figure}[H]
    \centering
    \begin{minipage}[t]{0.4\linewidth}
        \figletter{a} \\
        \begin{subfigure}{\linewidth}
            \includegraphics[width=\linewidth]{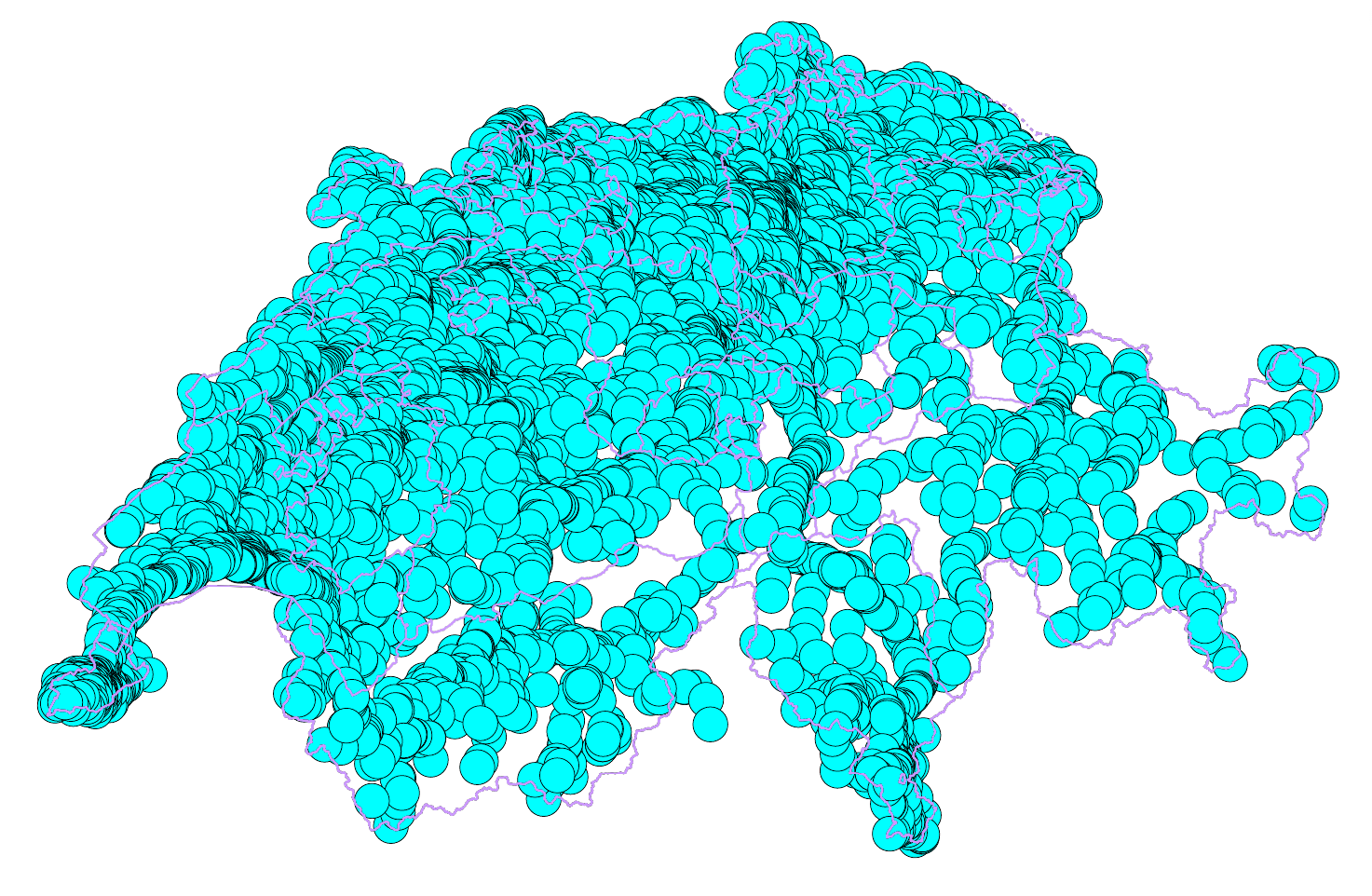}
        \end{subfigure}
    \end{minipage}
    \hspace{1cm}
    \begin{minipage}[t]{0.4\linewidth}
        \figletter{b} \\
        \begin{subfigure}{\linewidth}
            \includegraphics[width=\linewidth]{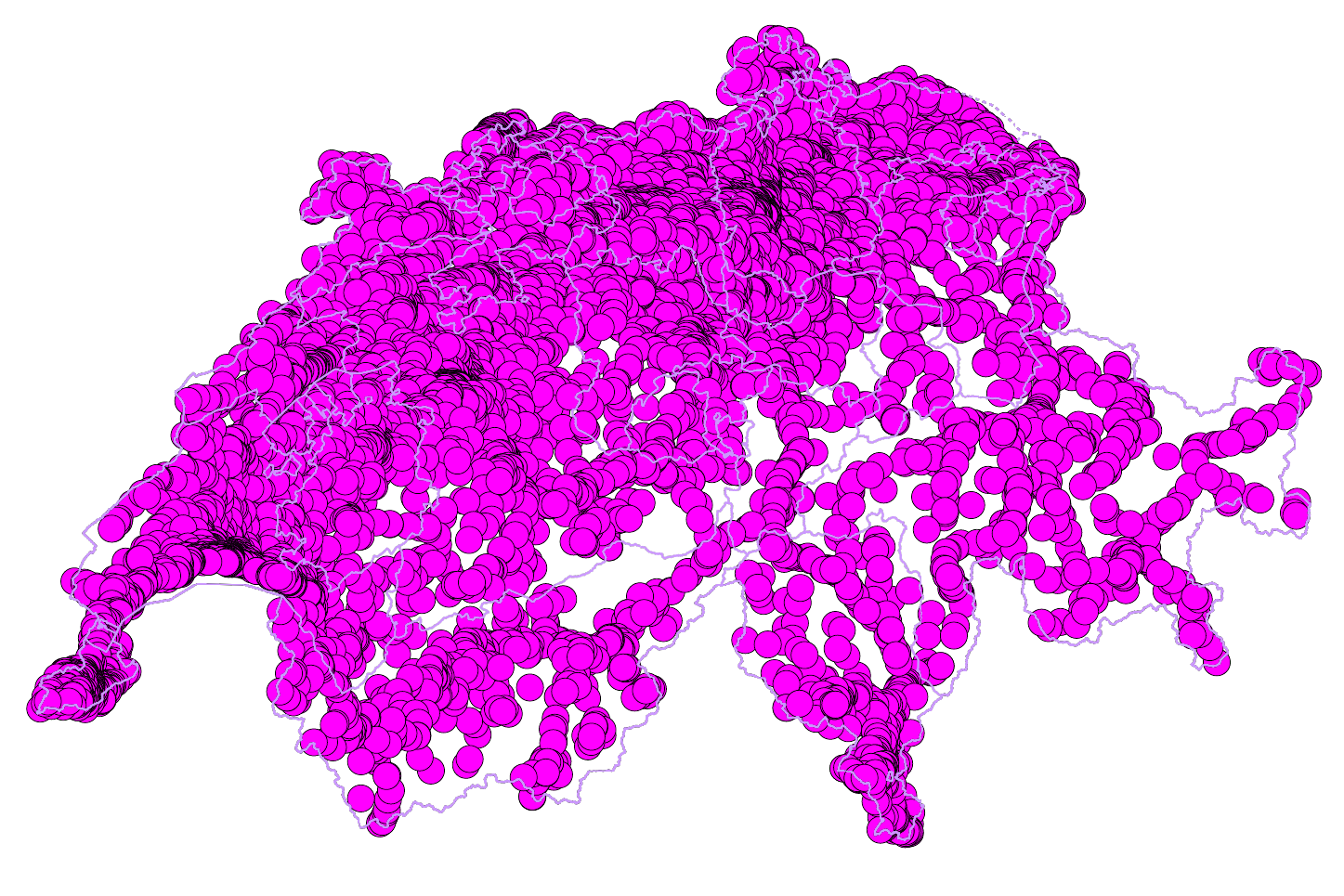}
        \end{subfigure}
    \end{minipage}
    \\
    \vspace{1cm}
    \begin{minipage}[t]{0.4\linewidth}
        \figletter{c} \\
        \begin{subfigure}{\linewidth}
            \includegraphics[width=\linewidth]{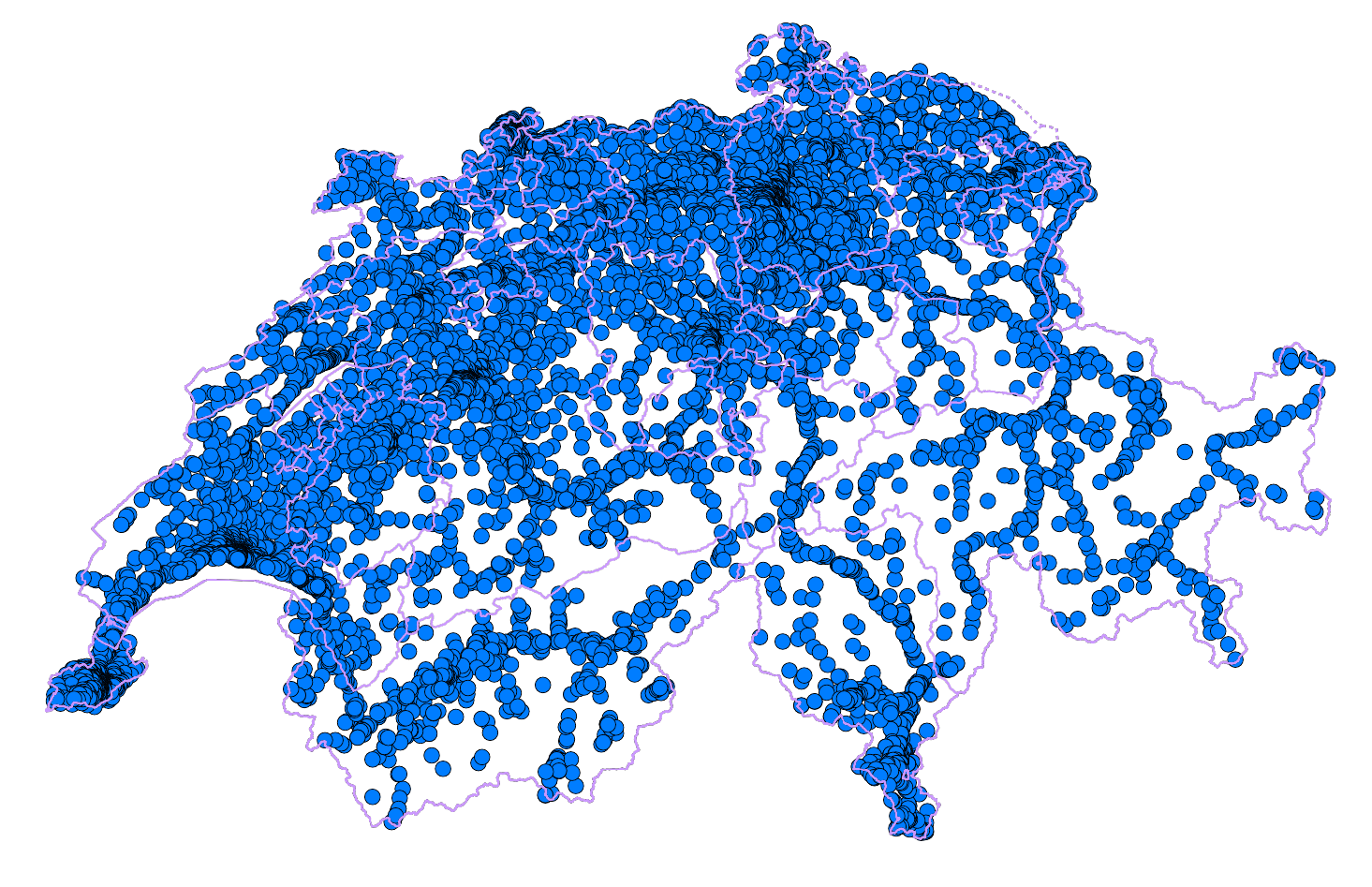}
        \end{subfigure}
    \end{minipage}
    \hspace{1cm}
    \begin{minipage}[t]{0.4\linewidth}
        \figletter{d} \\
        \begin{subfigure}{\linewidth}
            \includegraphics[width=0.95\linewidth]{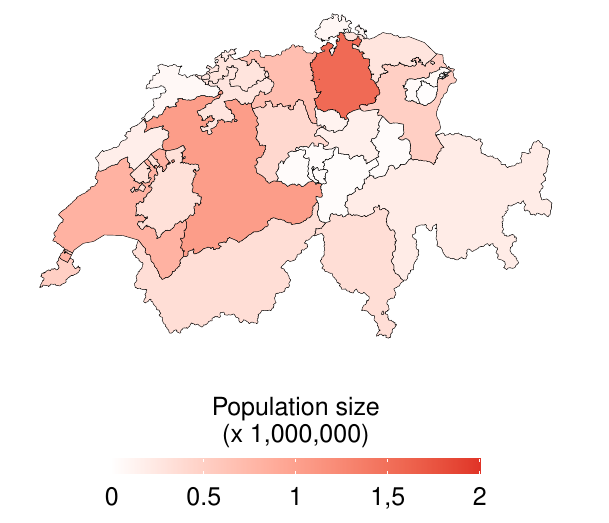}
        \end{subfigure}
    \end{minipage}
\caption{Geographical locations of mobile network antennas in Switzerland by type \textbf{(a)} GSM (2G), \textbf{(b)} UMTS (3G), \textbf{(c)} LTE (4G), and \textbf{(d)} canton population sizes as of February 2020.}
\label{fig:antenna}
\end{figure}

The number of antennas by type and size are summarized in \Cref{tbl:antennas}. \Cref{fig:antenna} further shows the geographical locations of antennas in Switzerland by type and the canton population sizes as of February, 2020 (the calculation of the population sizes is explained in Supplement~\ref{supp:check_assumptions}). More (less) densely populated cantons appear to have more (less) antennas. A comparison to \Cref{fig:1}a--b further shows that cantons with more (less) antennas also had a higher (lower) number of trips in 2019 and after all policy measures were implemented in 2020. Altogether, this ensures that trips can be captured in the whole country despite cantonal differences in population density, mobility, and antenna coverage. The spatial density of antennas can also be checked at the district and municipal level. The Swiss federal geoportal operates a web tool with open data for this purpose (\url{https://map.geo.admin.ch/}). 

\begin{table}[H]
\setlength\tabcolsep{10pt}
\singlespacing
\centering
\footnotesize
\begin{tabular}{l c c c c l}
\toprule
                    & \multicolumn{4}{c}{Antenna size}       \\
                    \cmidrule(lr){2-5}
Type                & Very small    & Small         & Mid       & Large & Sum   \\
\midrule
GSM (2G)            & 1,369          & 1,648        & 3,931     & 55    & 7,003 \\
UMTS (3G)           & 1,835          & 1,007        & 7,115     & 1,078 & 11,035   \\
LTE (4G)            & 1,939          & 567          & 5,720     & 3,415 & 11,641   \\
\midrule
Sum                 & 5,143          & 3,222        & 16,766    & 4,548 & 29,679 \\
\bottomrule
\end{tabular}
\caption{Number of mobile network antennas per size and type in Switzerland. The data are available at \url{https://opendata.swisscom.com/explore/?sort=modified}.}
\label{tbl:antennas}
\end{table}

\clearpage
\section{Identification strategy}

This section explains how the model estimates are identified from the data. In doing so we also explain the assumptions underlying the modeling.

\subsection{Estimating the reduction in human mobility due to policy measures through a difference-in-difference analysis}
\label{supp:identification_strategy_mobility}

The parameter estimates for the policy measures can be interpreted as causal effects under the following standard assumptions for difference-in-difference (DiD) analysis: (1a)~the policy measures were implemented ``as if'' independently of the level of mobility or, (1b)~in the absence of any policy measure, all cantons would have parallel trends in mobility; (2)~Differences between cantons were stable over time; and (3)~the implementation of each policy measure in any canton did not affect the mobility in any other canton. 

Assumption~(1a) implies that there are no omitted confounders of the relation between policy measures and mobility. This seems credible conditional on unobserved canton-specific and weekday-specific factors, the time since the a canton's first reported case, and a canton's population size. We therefore adjust for these factors in our model. In our context, assumption 1b implies that, if no policy measures were implemented, mobility would trend (or not trend) in the same way for all cantons throughout the whole study period. Then for a given day, the cantons that have not yet implemented a policy measure provide valid control observations for the cantons that have implemented the policy measure. The plausibility of the assumption can be investigated by inspecting mobility prior to policy measures. If cantons had parallel trends (or no trends) in mobility prior to policy measures, then this suggest that they would also have parallel trends (or no trends) throughout the whole study period if policy measures were not implemented. The credibility of assumptions (1a) and (1b) is graphically analyzed and further discussed in Supplement~\ref{supp:check_assumption_1}.

Assumptions (2) and (3) are together known as the stable unit treatment value assumption \cite{Rubin.1974, Holland.1986}, which is a standard assumption in the causal inference literature and for difference-in-difference analysis. Assumption 2 seems plausible given no substantial change in canton population sizes. This, in turn, appears to be likely given the official recommendations to limit long-distance travel, the lockdown of many European countries at the time, and that our data cover close to the whole Swiss population. In Supplement~\ref{supp:check_assumption_2}, we show that canton population sizes indeed were close to constant over time. The credibility of Assumption~(3) is debatable since some people may, for precautionary reasons, reduce their mobility when a policy measure that is not yet implemented in their canton is implemented in another canton. The assumption could be relaxed by estimating spillover effects of the policy measures between cantons, but this would increase the modeling complexity. 

Note that assumption 1a is stronger than assumption 1b but that both are sufficient for the causal interpretation together with assumption~(2) and (3). Hence if assumption~(1a) does not hold, the DiD estimates can still be interpreted as causal effects under assumptions (1b), (2), and (3) together. Importantly, even if the above assumptions required for causal interpretation do not hold, the estimates from the DiD analysis are still useful as they then give the conditional association between mobility and policy measures. 

The analysis also requires the variables to follow a specific measurement order in time. To measure the effect of policy measures on mobility within days, they must -- for each day -- be implemented earlier than mobility is measured. This holds true in our study. The policy measures were put into effect at midnight (\ie, start of day) of their date of implementation in Switzerland, while movements were measured at the end of each day. Recall that, for each mobility variable, we calculate the number of trips by aggregating the total number over the full day. Hence, the measurement order is such that the policy measures precede mobility.

\subsection{Estimating the extent to which mobility predict COVID-19 cases}
\label{supp:identification_strategy_cases}

We estimate the conditional relationship between lagged mobility and the reported number of COVID-19 cases using a regression model. We control for variables that confound the relationship between mobility and reported cases but let the policy measures vary. The regression thereby identifies the extent to which the level of mobility at a given day predicts the number of new cases reported in a later day when both depend on policy measures. By lagging mobility, we rule out the possibility that the parameter estimates for mobility are biased by simultaneity or reversed causality, meaning that there are feedback loops between mobility and reported cases or that the number of reported cases affect mobility. 

Two factors determine the delay of mobility in predicting the reported number of new cases: (1) the incubation period, and (2) the reporting delay. Previously, the mean incubation period was estimated to 5.1 days, while \SI{97.5}\,\% of infected people had symptoms within 11.5 days after exposure \cite{Lauer.2020}. Adding a 2 day reporting delay (cf. the case data from the Federal Office of Public Health of the Swiss Confederation; BAG) leads to the chosen lags of mobility of 7--13 days. 

We fit the regression model separately for each mobility variable and lag. We include only a single lag as we are interested in the extent to which mobility at a given day predicts reported cases in the future, and thus cannot condition on the mobility in days that are also in the future. The reason for including only one mobility variable is two-fold. First, the mobility variables within each category of trips (\eg, trips by mode or purpose) are different subsets of total trips, and thus by definition collinear. Hence, including all trip variables of either mode or purpose in the same model would result in biased point and interval estimates due to multicollinearity. Second, a trip by one mode is also a trip for one purpose, and vice versa. In addition, every trip by a mode and purpose contributes to total trips. Thus, by including only a single mobility variable, we avoid counting the same trips several times. Summarising, by refitting the model for each pair of mobility variable and lag we avoid introducing bias in its estimates for mobility.

\section{Credibility of identifying assumptions}
\label{supp:check_assumptions}

Supplement \ref{supp:identification_strategy_mobility} discusses the assumptions required for a causal interpretation of the DiD estimates of the effects of the policy measures from data. As is well known in the causal inference literature, these assumptions cannot be fully verified from data. Still, the credibility of the assumptions can be assessed. We now perform a series of checks that demonstrate their credibility in this study.

\subsection{Credibility of Assumption 1}
\label{supp:check_assumption_1}

We first assess the credibility of assumption (1a),~that the policy measures were implemented ``as if'' independently of the level of mobility, and (1b),~that in the absence of any policy measure, all cantons would have parallel trends in mobility. Here, (1a)~is a statement about how the public authorities decided to implement the policy measures, and (1b)~is a statement about how mobility would change in a counterfactual scenario in which no policy measures were implemented. Therefore neither assumption can be fully verified. Nonetheless, to check their credibility, we inspect the time series of trip counts per canton for all mobility variables. Now, it is well-known that the parallel trends assumption is scale dependent \cite{Athey.2006, Lechner.2011}. As the regression model used in the DiD analysis contains a log-link, this means that assumptions (1a) and (1b) should hold in log-levels of mobility. We therefore plot the time series of trip counts on the logarithmic scale.

\begin{figure}[H]
    \centering
    \begin{minipage}[t]{0.9\textwidth}
        \figletter{a} \\
        \begin{subfigure}{\linewidth}
            \includegraphics[width=\linewidth]{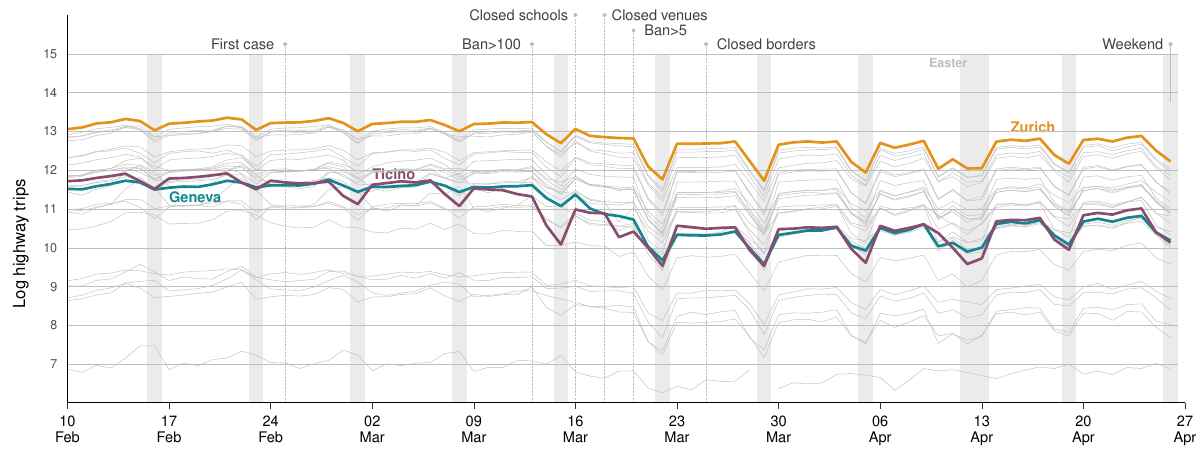}
        \end{subfigure}
    \end{minipage}
    \\
    \vspace{1cm}
    \begin{minipage}[t]{0.9\textwidth}
        \figletter{b} \\
        \begin{subfigure}{\linewidth}
            \includegraphics[width=\linewidth]{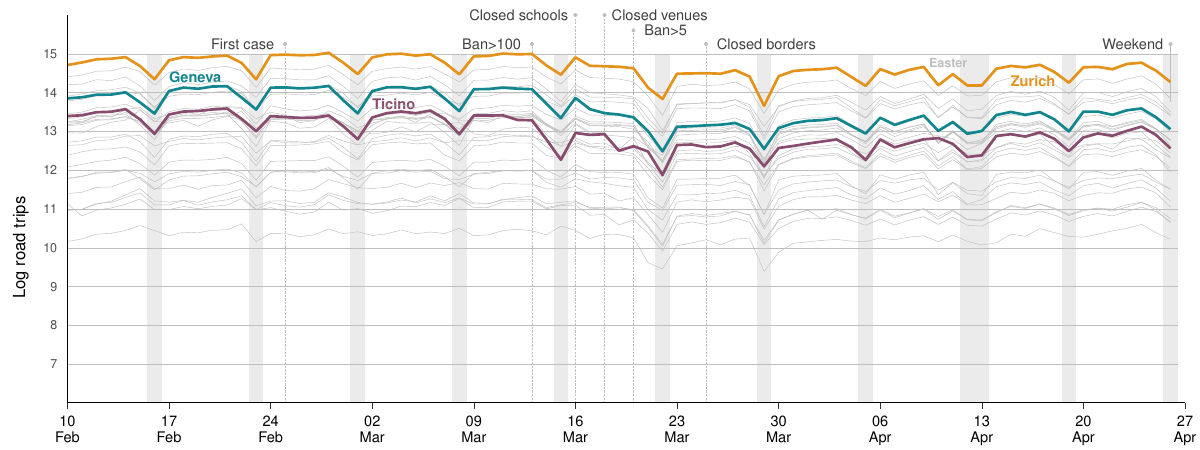}
        \end{subfigure}
    \end{minipage}
    \\
    \vspace{1cm}
    \begin{minipage}[t]{0.9\textwidth}
        \figletter{c} \\
        \begin{subfigure}{\linewidth}
            \includegraphics[width=\linewidth]{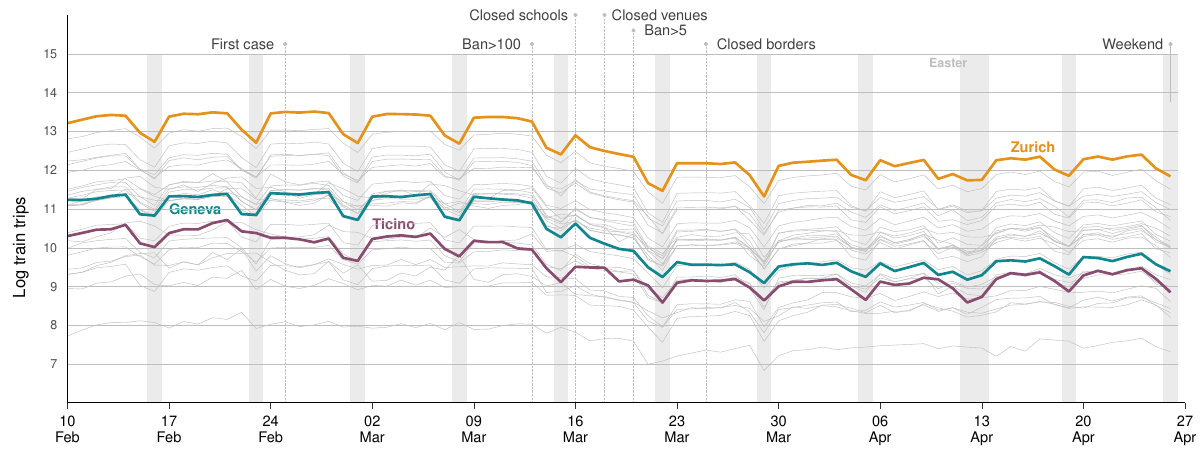}
        \end{subfigure}
    \end{minipage}
\caption{Parallel trends in log mobility by mode across cantons: \textbf{(a)} log highway trips, \textbf{(b)} log road trips, and \textbf{(c)} log train trips.}
\label{fig:parallel_trends_mode}
\end{figure}

\begin{figure}[H]
    \centering
    \begin{minipage}[t]{0.9\textwidth}
        \figletter{a} \\
        \begin{subfigure}{\linewidth}
            \includegraphics[width=\linewidth]{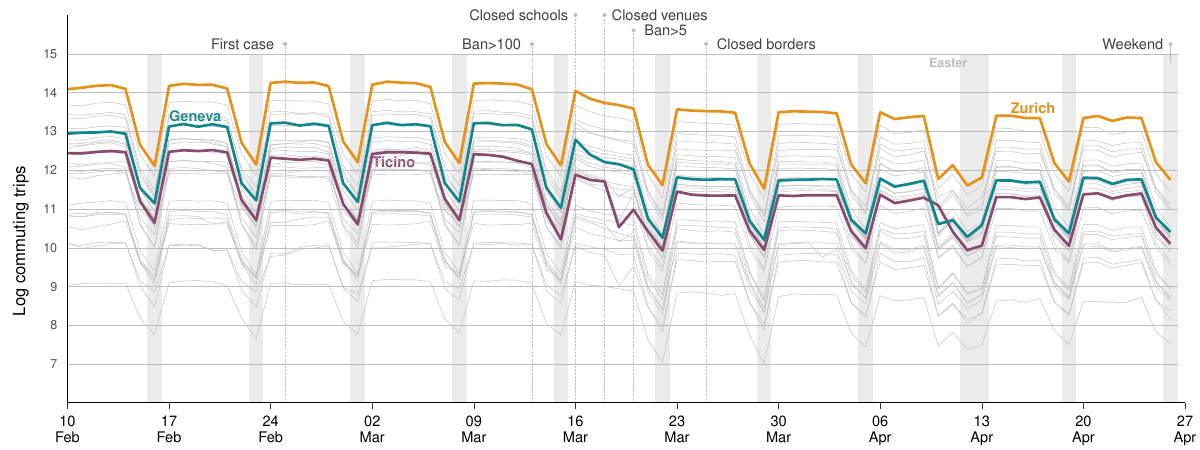}
        \end{subfigure}
    \end{minipage}
    \\
    \vspace{1cm}
    \begin{minipage}[t]{0.9\textwidth}
        \figletter{b} \\
        \begin{subfigure}{\linewidth}
            \includegraphics[width=\linewidth]{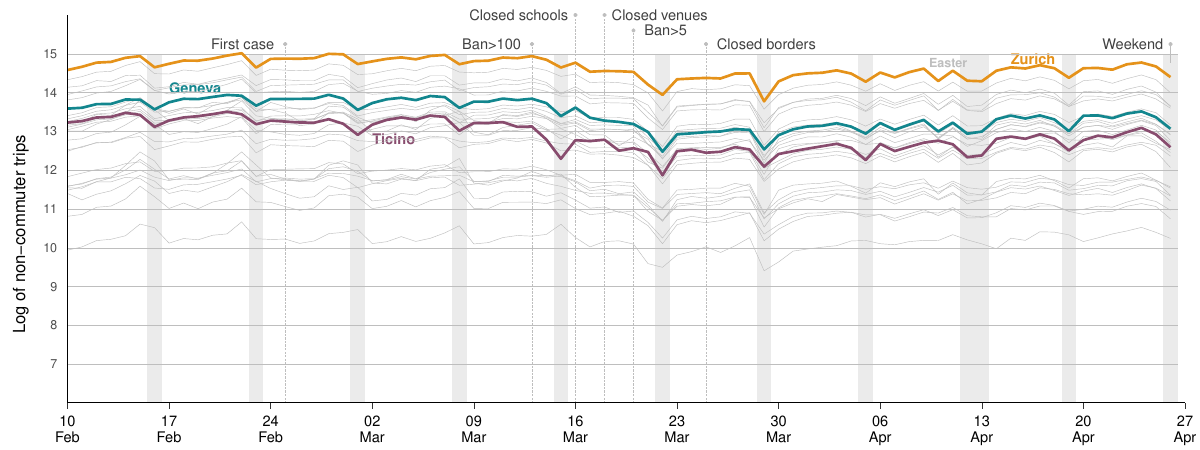}
        \end{subfigure}
    \end{minipage}
\caption{Parallel trends in log mobility by purpose across cantons: \textbf{(a)} log commuting trips, and \textbf{(b)} log non-commuting trips.}
\label{fig:parallel_trends_reason}
\end{figure}

\begin{figure}[H]
    \centering
            \includegraphics[width=0.9\textwidth]{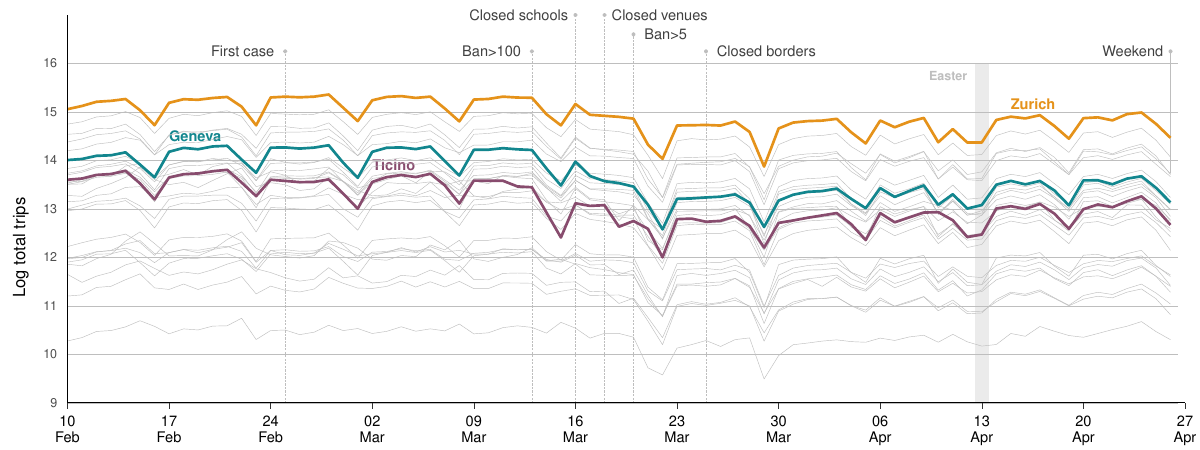}
\caption{Parallel trends in log total mobility across cantons.}
\label{fig:parallel_trends_total}
\end{figure}

\Crefrange{fig:parallel_trends_mode}{fig:parallel_trends_total} show the time series of log trip counts by mode, purpose, and in total. The figures show that for no mobility variable (\eg, mobility by mode, purpose, or in total) and in no canton did log trips counts change substantially before a policy measure was implemented. Thus, policy measures do not seem to have been implemented in response to an increase in mobility, and people do not seem to have restricted their mobility in anticipation of policy measures. Hence, assumption (1a) appears to be credible. To further increase the credibility of the assumption, our regression model controls for unobserved differences between cantons and weekdays, the time that has passed since the canton's first reported case, and differences between canton population sizes. Regarding assumption (1b), \Crefrange{fig:parallel_trends_mode}{fig:parallel_trends_total} also show that mobility by mode, purpose, and in total was approximately parallel in log-levels across cantons over the whole study period. Importantly, there is no apparent trend in log mobility for any canton in the period before the policy measures. It thereby seems plausible that mobility would also be parallel in log-levels across cantons over the whole study period even if the policy measures would not have been implemented. Hence, assumption (1b) appears to be credible.

\subsection{Credibility of Assumption 2}
\label{supp:check_assumption_2}

The second assumption required for the DiD analysis is differences between cantons were stable over time. Technically, this means that the cross-canton distributions of factors that determine differences between cantons should be similar over time \cite{Wing.2018}. Then, the compositions of cantons relative to each other in terms of such factors does not change, and they provide valid control and treatment groups for each other throughout the study period. As a result, the reductions in mobility can, with greater credibility, be attributed to the implementation of the policy measures alone. Importantly, only factors whose change over time is not due to the policy measures shall be controlled for. The following example explains why.

Say that we would have a covariate that measured the attitudes towards social distancing per day among the inhabitants of each canton, for instance, via the daily average of canton inhabitants' individual attitudes. Imagine that people's attitudes towards social distancing are associated with their mobility behavior and that these attitudes change when policy measures are implemented. Then controlling for this covariate in the regression model would estimate the part of the effect of policy measures on mobility that is left after the effect of attitudes on mobility is removed. Clearly, this is not the estimate we are interested in. What we want to estimate the effect of policy measures on mobility irrespective of the behavioral changes that explains the change in mobility. In this hypothetical example, the change in attitude is also an outcome of the policy measures, and thus, we do not wish to ``control away'' its impact on mobility. 

More generally, controlling for any covariate that is also an outcome of the ''treatment'' (here: policy measures) will induce post-treatment bias in the estimated effect of the treatment on the outcome. To avoid introducing such bias in our results, our regression model does not include any time-varying variables related to people's behavior. Instead, we only control for time-invariant differences (heterogeneity) between cantons and time-varying factors independent of policy measures in our regression model. The canton random effect and the canton population size offset control for the former whereas the weekday fixed effect and trend variable control for the latter. For any other factors we do not include in our regression, we must check if they remaining approximately constant over the study period. We do so in the following.

\begin{figure}[H]
    \centering
    \begin{minipage}[t]{\linewidth}
        \figletter{a} \\
        \begin{subfigure}{\linewidth}
            \includegraphics[width=\linewidth]{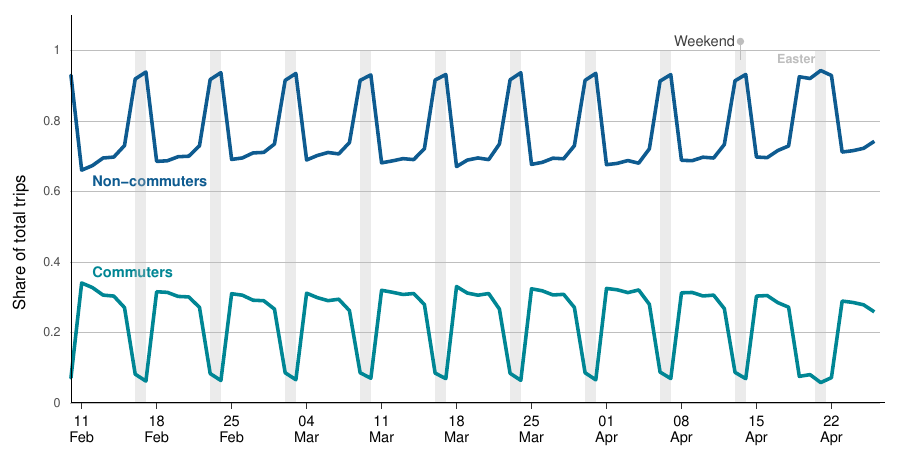}
        \end{subfigure}
    \end{minipage}
    \\
    \vspace{1cm}
    \begin{minipage}[t]{\linewidth}
        \figletter{b} \\
        \begin{subfigure}{\linewidth}
            \includegraphics[width=\linewidth]{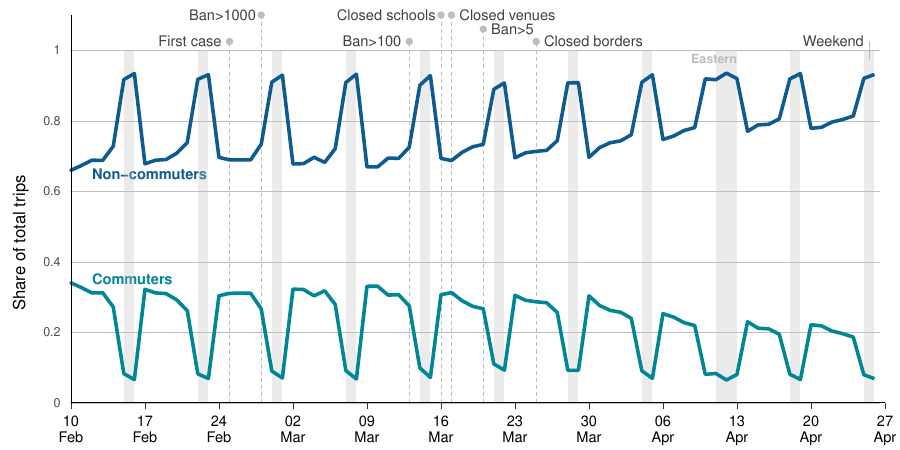}
        \end{subfigure}
    \end{minipage}
\caption{Constant composition of mobility over time in terms of the share of total trips by purpose: \textbf{(a)} 2019, and \textbf{(b)} 2020.}
\label{fig:share_trips}
\end{figure}

One factor that could change over time and is possibly independent of the policy measures is the composition of total trips in terms of the share of trips by purpose. If the share of trips by, \eg, commuters change drastically after policy measures are implemented, then trips by commuters in cantons in which policy measures are not yet implemented do not provide a valid control group for cantons in which policy measures have been implemented. This is sometimes referred to as that the ``treatment'' puts groups on differential trends \cite{Goodman.2018}. \Cref{fig:share_trips} shows time series of the share of total trips by purpose over the study period in 2020 and the same period in 2019. Overall, the differences between the shares remain almost constant over time within years, and are practically equal between years. There is a gradual change towards relatively more non-commuting trips during the last weeks of the study period in 2020 (\Cref{fig:share_trips}b). Comparing \Cref{fig:1}c and \Cref{fig:parallel_trends_reason}b reveals that this is due to that commuting trips stay relatively constant while non-commuting trips increase slightly. A potential explanation for this result is that gradual improvements in weather during spring 2020 led to more non-commuting trips but not more commuting trips as schools and workplaces were closed. 

Another factor that can change over time independent of policy measures is canton population sizes. The reasoning is that the onset of the epidemic, but not the policy measures, would lead people to move out of, \eg, metropolitan areas to the country side or even abroad. Here, substantial changes in canton population sizes could affect our results in two ways. First, a change in the distribution of population sizes across cantons would imply a change in the composition of the study population. Second, a substantial change in the population size of a given canton could explain both changes in mobility in the canton and why the policy measures where implemented at the observed date for the canton, thereby confounding the relationship between policy measures and mobility. Hence, we investigate the change in canton population sizes over the study period. 

We obtain official statistics on the canton population sizes per quarter of 2020 (\Cref{Table: population size quarter}). The table shows that changes in canton population sizes between Q1 and Q2 of 2020 was of similar magnitude to the changes between 2019 and Q1 or between Q2 and Q3 in 2020. Hence, the change during the study period was similar of similar magnitude to other points in time. Note that the quarterly population statistics include changes due to people moving into and out of cantons. 

Because canton population may change on a monthly basis within quarters, we further investigate if the canton populations changed on a monthly basis. As official statistics on the Swiss population per month of 2020 is not yet publicly available, we calculate the population sizes per canton and month with the following procedure:

To obtain the population per canton in January, 2020, we take the official 2019 population sizes per canton, add the number of births in January per canton and subtract the number of deaths in January per canton. The population per canton in February, 2020, is then calculated by adding births and subtracting deaths per canton in February from the number we got for January. The procedure is repeated analogously for March and April. The calculation does not account for potential changes in population sizes due to people moving in and out of cantons as no such monthly data is yet publicly available. However, and as previously noted, such changes are captured in the quarterly statistics. Nonetheless, given the official recommendations to limit long-distance travel between cantons and the lockdown of many European countries at the time, it seems unlikely that a substantial number of people would move into or out of cantons during the study period, and that changes in population sizes due to such behavior would differ substantially between months.

The results of our calculation shows that the canton population sizes barely changed during the study period relative to the time periods before and afterwards (\Cref{Table: population size}). This, together with that our data covers the whole of Switzerland, that the Swisscom telecommunication infrastructure covers 99.9\% of the geographic area in Switzerland \cite{MIP.2020}, and that the use of mobile device is widespread in Switzerland, indicate that the population whose mobility was tracked remained close to constant over the study period.

All statistics are obtained from the Swiss Federal Statistical Office, BFS (population per canton in 2019: \url{https://www.bfs.admin.ch/bfs/en/home/statistics/catalogues-databases/data.assetdetail.14087625.html}, births per month and canton in 2020: \url{https://www.bfs.admin.ch/bfsstatic/dam/assets/15464262/master}, and deaths per month and canton in 2020: \url{https://www.bfs.admin.ch/bfsstatic/dam/assets/15464250/master}, population per quarter and canton in 2020: \url{https://www.bfs.admin.ch/bfs/de/home/statistiken/bevoelkerung.assetdetail.14941425.html}).

\renewcommand{\arraystretch}{0.75}
\begin{table}[H]
\begin{center}
\begin{tabular}{l r r r r r}
\toprule
                                & \multicolumn{5}{c}{Population size}          \\
    \cmidrule(lr){2-6}
    Canton code	& 2019	        & Jan 2020	& Feb 2020	& Mar 2020	& Apr 2020  \\
    \midrule
    AG	        & 690 445	    & 690 400	& 690 603	& 690 738	& 690 877  \\
    AI	        & 16 192	    & 16 158	& 16 205	& 16 198	& 16 199  \\
    AR	        & 55 541	    & 55 614	& 55 535    & 55 543	& 55 547  \\
    BE	        & 1 045 579	    & 1 045 580	& 1 045 387	& 1 045 307	& 1 045 267  \\
    BL	        & 291 063	    & 291 055	& 291 016	& 290 941	& 290 882  \\
    BS	        & 199 213	    & 199 218	& 199 174	& 199 071	& 199 020  \\
    FR	        & 323 499	    & 323 465	& 323 674	& 323 724	& 323 762  \\
    GE	        & 507 442	    & 507 442	& 507 613	& 507 646	& 507 592  \\
    GL	        & 40 904	    & 40 995	& 40 919	& 40 911	& 40 900  \\
    GR	        & 206 336	    & 206 404	& 206 276	& 206 244	& 206 257  \\
    JU	        & 73 804	    & 73 891	& 73 780	& 73 755	& 73 763  \\
    LU	        & 415 919	    & 415 929	& 416 027	& 416 107	& 416 188  \\
    NE	        & 177 790	    & 177 781	& 177 765	& 177 722	& 177 686  \\
    NW	        & 43 280	    & 43 299	& 43 298	& 43 299	& 43 306  \\
    OW	        & 38 269	    & 38 280	& 38 258	& 38 257	& 38 259  \\
    SG	        & 513 553	    & 513 653	& 513 681	& 513 681	& 513 783  \\
    SH	        & 83 013	    & 83 024	& 83 012	& 82 998	& 82 988  \\
    SO	        & 277 160	    & 277 128	& 277 142	& 277 131	& 277 100  \\
    SZ	        & 161 175	    & 161 184	& 161 200	& 161 211	& 161 235  \\
    TG	        & 281 393	    & 281 570	& 281 493	& 281 521	& 281 567  \\
    TI	        & 354 264	    & 354 283	& 354 158	& 353 905	& 353 621  \\
    UR	        & 37 037	    & 37 031	& 37 037	& 37 040	& 37 026  \\
    VD	        & 815 261	    & 815 257	& 815 590	& 815 627	& 815 561  \\
    VS	        & 354 148	    & 354 256	& 354 208	& 354 179	& 354 096  \\
    ZG	        & 129 458	    & 129 432	& 129 510	& 129 562	& 129 569  \\
    ZH	        & 1 550 173	    & 1 550 208	& 1 550 810	& 1 551 155	& 1 551 461  \\
    \toprule
\end{tabular}
\caption{Population sizes per canton and month prior to and during the study period (February~10--April~26, 2020). The 2019 numbers are official statistics whereas the monthly 2020 numbers are calculated using official birth and death statistics.}
\label{Table: population size}
\end{center}
\end{table}

\renewcommand{\arraystretch}{0.75}
\begin{table}[H]
\begin{center}
\begin{tabular}{l r r r r}
    \toprule
                    & \multicolumn{4}{c}{Population size}          \\
    \cmidrule(lr){2-5}
    Canton code     & 2019 	    & Q1 2020       & Q2 2020       & Q3 2020       \\
    \midrule
    AG	&           690 445	    & 691 554	    & 693 676	    & 696 639       \\
    AR	&           55 541	    & 55 471	    & 55 426	    & 55 478       \\
    AI	&           16 192	    & 16 207	    & 16 240	    & 16 289       \\
    BL	&           291 063	    & 290 966	    & 291 444	    & 292 244       \\
    BS	&           199 213	    & 199 730	    & 199 393	    & 199 980       \\
    BE	&           1 045 579	& 1 045 678	    & 1 046 496	    & 1 048 512       \\
    FR	&           323 499	    & 324 128	    & 324 836	    & 325 818       \\
    GE	&           507 442	    & 507 535	    & 507 544	    & 508 527       \\
    GL	&           40 904	    & 41 022	    & 41 081	    & 41 138       \\
    GR	&           206 336	    & 204 427	    & 203 393	    & 203 885       \\
    JU	&           73 804	    & 73 687	    & 73 714	    & 73 923       \\
    LU	&           415 919	    & 416 791	    & 417 184	    & 418 659       \\
    NE	&           177 790	    & 177 575	    & 177 311	    & 177 499       \\
    NW	&           43 280	    & 43 215	    & 43 287	    & 43 449       \\
    OW	&           38 269	    & 38 211	    & 38 190	    & 38 312       \\
    SH	&           83 013	    & 83 061	    & 83 250	    & 83 602       \\
    SZ	&           161 175	    & 160 915	    & 161 274	    & 162 029       \\
    SO	&           277 160	    & 277 383	    & 277 834	    & 279 050       \\
    SG	&           513 553	    & 514 408	    & 515 179	    & 517 439       \\
    TI	&           354 264	    & 353 354	    & 352 926	    & 353 502       \\
    TG	&           281 393 	& 281 745	    & 283 300	    & 284 771       \\
    UR	&           37 037	    & 37 017	    & 36 951	    & 37 009       \\
    VD	&           815 261	    & 816 233	    & 816 356	    & 820 672       \\
    VS	&           354 148	    & 353 529	    & 352 017	    & 353 591       \\
    ZG	&           129 458	    & 129 101	    & 129 201	    & 129 609       \\
    ZH	&           1 550 173	& 1 552 721	    & 1 553 722	    & 1 559 699       \\

   \toprule
\end{tabular}
\caption{Population sizes per canton and quarter prior to, during, and after the study period (February~10--April~26, 2020).}
\label{Table: population size quarter}
\end{center}
\end{table}

\newpage
\section{Robustness checks}
\label{supp:robustness_checks}

The previous section demonstrated the credibility of the identifying assumptions for the DiD analysis. This suggests that it is possible to obtain valid estimates of the effect of the policy measures on mobility and the predicted change in reported cases associated with reductions in mobility, given that the regression models are correctly specified. Thus, we now test the robustness of our estimates to different specifications of the regression models.

\subsection{Alternative specifications of time effects}
\label{supp:robustness_checks_time}

Because the data cover daily observations over a relatively long time period, there are many possible ways of controlling for different time effects. We therefore perform a robustness check of our estimates by re-fitting the models with two alternative specifications of time effects. 

\Cref{Table: alt model NPI-mobility} summarises how the main and alternative model specifications differ. Like the main models, both alternative specifications include weekday fixed effects. The alternative specification~(a) uses trend variables of the non-transformed and squared number of days since the first case in each canton was reported (the main specification instead uses the logarithm). Compared to the main specification, it simply postulates a linear and quadratic effect of the number of days since the first case. The alternative specification~(b) adds week fixed-effects to the main specification. Thus, it also controls heterogeneity between weeks.

A prior of $\mathsf N(0,1)$ is assigned to the parameters for the linear and quadratic time trends in the second alternative specification. The week fixed-effects in the third alternative specification are assigned a weakly informative prior of $\textsf{Half-t}(3,0,2.5)$. All other priors are the same as in the main specification. 

The model with the first alternative specification of time effects needed 6000 iterations during MCMC estimation to converge (3000 for samples for warm-up and 3000 samples from its posterior distributions).

\setlength{\tabcolsep}{12pt}
\begin{table}[H]
\begin{center}
\begin{tabular}{l c c c}
  \toprule
                                        & \multicolumn{3}{c}{Model specification}          \\
   \cmidrule(lr){2-4}
   Time-effect                                      & Main 	     & (a)        & (b)        \\
   \midrule
   Weekday fixed effect 		                    & \checkmark & \checkmark & \checkmark \\ 
   Week fixed-effect 	    	                    & 		 	 &            & \checkmark \\ 
   Number of days since 1st reported case 		    &            & \checkmark &            \\
   Squared number of days since 1st reported case   & 		     & \checkmark &            \\
   Log number of days since 1st reported case	    & \checkmark &            & \checkmark \\ 
   \toprule
\end{tabular}
\caption{Specifications of time effects in the models of mobility on policy measures and reported number of new cases on lagged mobility. The symbol "\checkmark" means that a term for the effect is included in the model. }
\label{Table: alt model NPI-mobility}
\end{center}
\end{table}

\begin{figure}[H]
    \begin{minipage}[t]{\linewidth}
        \hspace{1cm}
        \figletter{a} \\
        \begin{subfigure}{\linewidth}
            \centering
            \includegraphics[width=\linewidth]{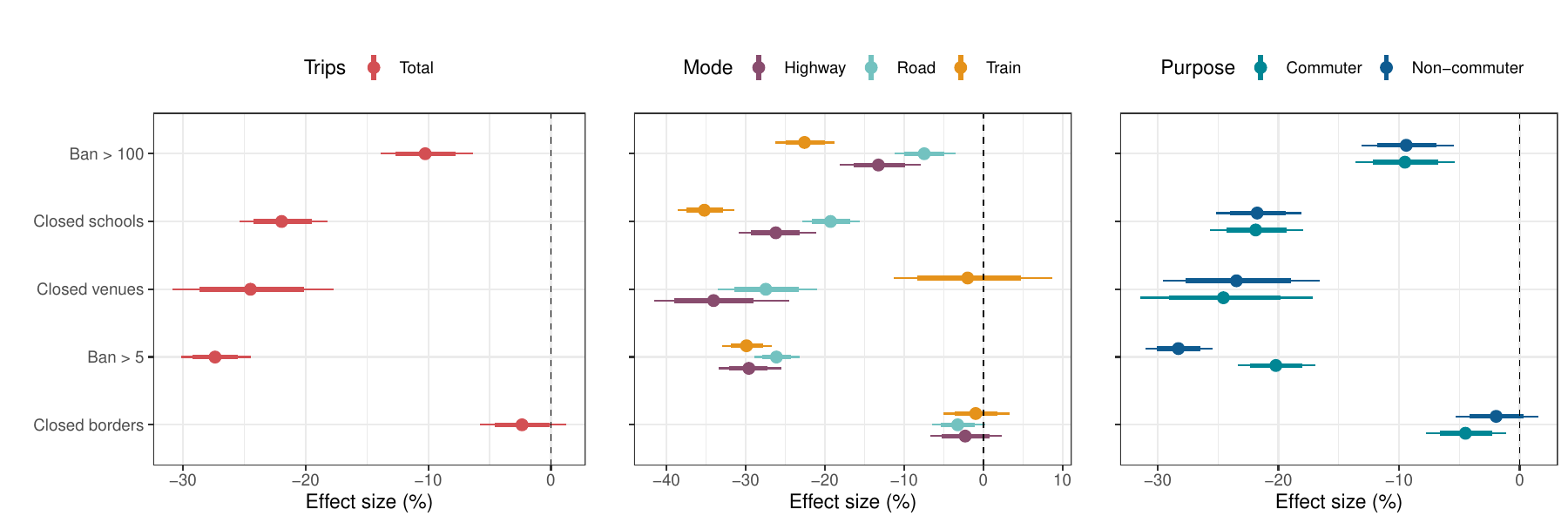}
        \end{subfigure}
    \end{minipage}
    \\
    \begin{minipage}[t]{\linewidth}
        \hspace{1cm}
        \figletter{b} \\
        \begin{subfigure}{\linewidth}
            \centering
            \includegraphics[width=\linewidth]{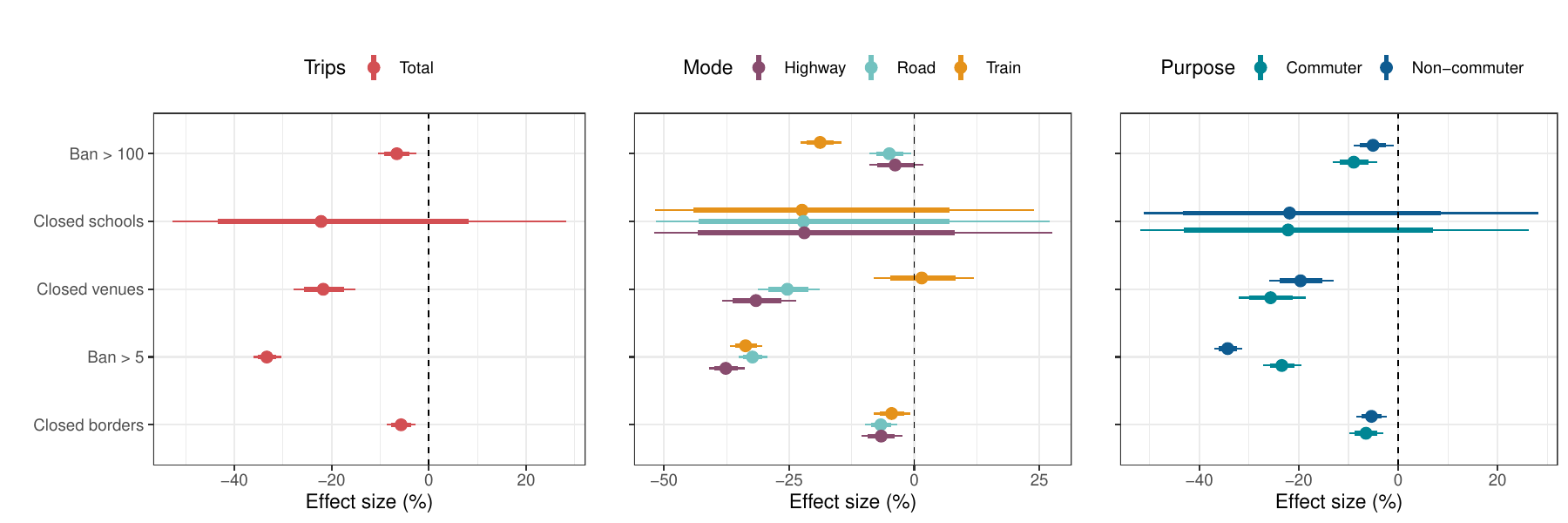}
        \end{subfigure}
    \end{minipage}
\caption{Estimated effect of policy measures on total trips, trips by mode and trip by purpose for three alternative model specifications: 
\textbf{(a)} weekday fixed effects and linear and quadratic trend of the number of days since the first reported case in each canton, and \textbf{(b)} weekday fixed effects, week fixed effects, and a trend of the log number of days since the first reported case in each canton. Posterior means are shown as dots, while 80\,\% and 95\,\% credible intervals are shown as thick and thin bars, respectively. Policy measures are arranged in the order in which they were implemented (cf. Supplement~\ref{supp:data}).}
\label{fig:robustness_checks_mobility}
\end{figure}

The models of mobility with alternative specifications of time effects give similar estimates as the main specification (\Cref{fig:robustness_checks_mobility}). Specifically, replacing the logarithmic trend variable with the corresponding linear and squared trends leaves the effects of the policy measures intact (\Cref{fig:robustness_checks_mobility}a). As expected, adding week fixed effects to the main specification reduces the estimated effects of some policy measures. It also widens the credible intervals for school closures (\Cref{fig:robustness_checks_mobility}b). Apart from this the estimates are very similar to those of the main model specification. Hence, the estimated policy effects are robust against a range of alternative controls for time effects.

\begin{figure}[H]
    \begin{minipage}[t]{\linewidth}
        \hspace{0.6cm}
        \figletter{a} \\
        \begin{subfigure}{\linewidth}
            \includegraphics[width=\linewidth]{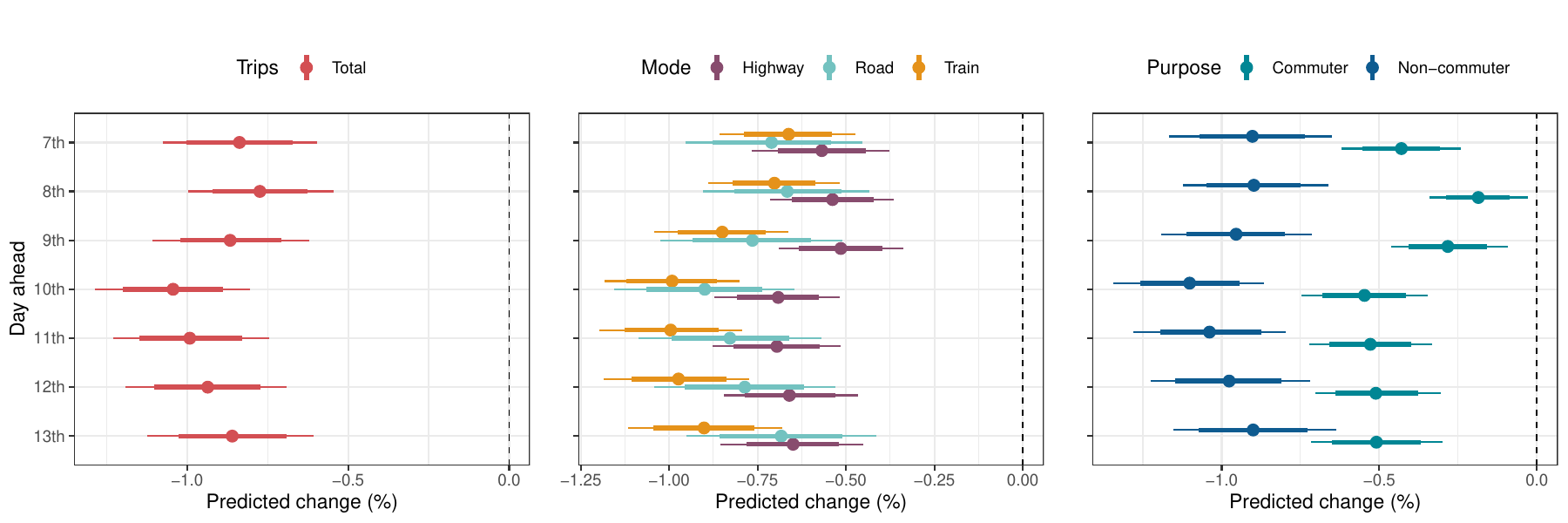}
        \end{subfigure}
    \end{minipage}
    \\
    \begin{minipage}[t]{\linewidth}
        \hspace{0.6cm}
        \figletter{b} \\
        \begin{subfigure}{\linewidth}
            \includegraphics[width=\linewidth]{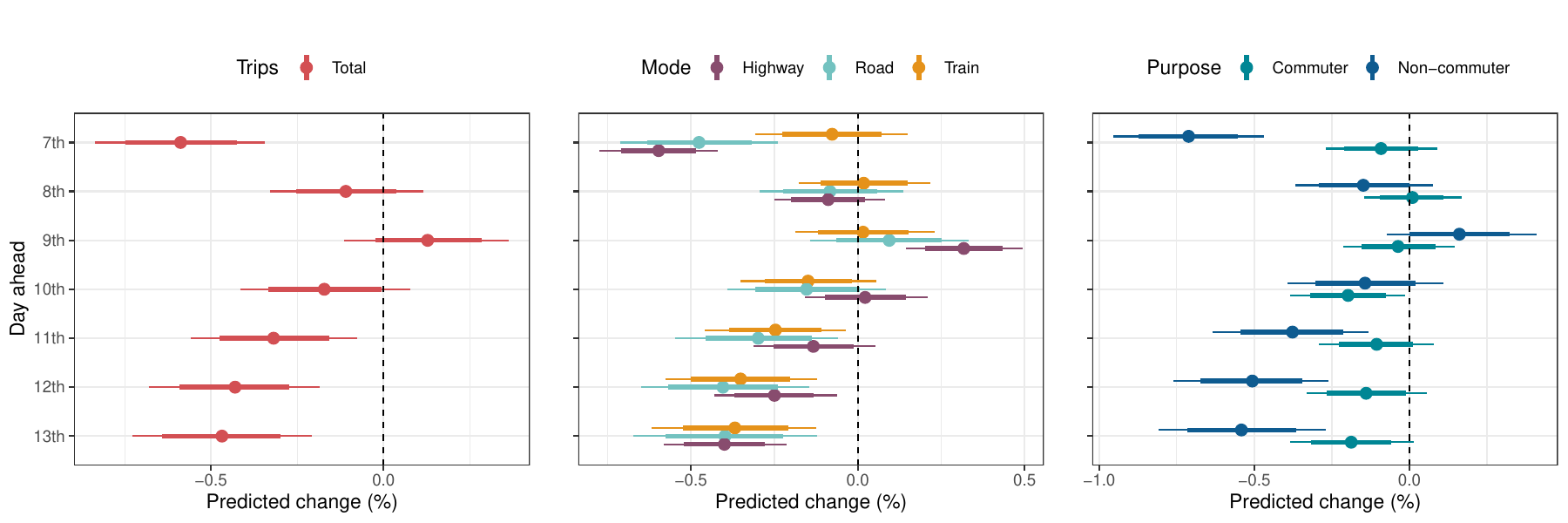}
        \end{subfigure}
    \end{minipage}
\caption{Predicted change in reported COVID-19 cases given a 1\,\% reduction of total trips, trips by mode, and trips by purpose lagged by one of 7--13 days for alternative specifications of time effects: 
\textbf{(a)} weekday fixed effects and linear and quadratic trend of the number of days since the first reported case in each canton, and \textbf{(b)} weekday fixed effects, week fixed effects, and a trend of the log number of days since the first reported case in each canton. Posterior means are shown as dots, while 80\,\% and 95\,\% credible intervals are shown as thick and thin bars, respectively. Policy measures are arranged in the order in which they were implemented (cf. Supplement~\ref{supp:data}).}
\label{fig:cases_alt_specification}
\end{figure}

The alternative specifications of time effects are also applied to the model of the reported number of new cases on lagged mobility. \Cref{fig:cases_alt_specification} shows their predictions of the percentage change in the reported number of new cases given a 1\,\% decrease in mobility lagged by 7--13 days. Replacing the logarithmic trend variable with linear and quadratic trends does not alter the estimated relationship between the mobility variables and reported case growth substantially (\Cref{fig:cases_alt_specification}a). The predicted reduction in case growth given decrease in total trips is stronger than in the main specification whereas the predicted reduction given decreases in commuter trips is slightly weaker. Adding week fixed-effects to the main specification leads to a smaller predicted change in reported case growth given reductions in mobility (\Cref{fig:cases_alt_specification}b). For some lags of the mobility variables is the predicted reduction not statistically distinguishable from zero. However, decreases in mobility still predicts reductions in reported case growth for the larger lags.

Altogether, the robustness checks show that the predicted reductions in reported cases given decreases in mobility are robust to alternative specifications of time effects.

\subsection{Controlling for changes in testing}
\label{supp:robustness_checks_tests}

The number of positive reported cases depends on the number of tests performed, which can vary over the course of the pandemic. We therefore re-estimate our models of the reported number of new cases on lagged mobility while controlling for the number of tests performed every day. This is based on the assumption that cantons might not conduct equally many tests each day. Instead, the number of tests that each cantons carry out might be determined by their population size and other relevant risk factors in their population. Since test availability determines the number of positive tests that can be detected, we extend the model of the relation between mobility and the reported case growth with the number of daily tests per canton. 

Prior to May~23, 2020, the Swiss Federal Office of Public Health (FOPH) did not report the number of daily tests conducted per canton, but only the number of daily tests in the country as a whole (\url{https://www.covid19.admin.ch/en/epidemiologic/test}). We therefore estimate the number of daily tests conducted per canton over our study period with the the number of daily tests in the whole of Switzerland standardized to the canton population shares. That is, we divide the number of daily tests in Switzerland as a whole with each cantons population share. 

Formally, let $u_{it}$ be the true unknown number of tests reported in canton $i$ in day $t$ and let $u_t = \sum^N_{i=1} u_{it}$ be the total and observed number of reported tests in the country in day $t$. We estimate the number of tests in canton $i$ in day $t$ with
\begin{equation}
    \hat u_{it}
    = u_t \times \frac{E_i}{\sum^N_{i=1} E_i},
\end{equation}
where $E_i$ is the population in canton $i$, and $\sum^N_{i=1} E_i$ is the population in the country. The estimate $\hat u_{it}$ thereby approximates the number of COVID-19 tests conducted in each canton and day under the assumption that the number of daily tests per canton is proportionate to the share of the population that lives in each canton. We include the logarithm of $\hat u_{it}$ in the model so that it estimates the percentage change in the reported number of new cases when the estimated number of daily tests increases by 1\,\%. We assign the log-transformed variable a prior of $\mathsf N(1,1)$, meaning that we expect the effect size to be 1\,\% on average, but likely not negative or larger than 2\,\%.

\begin{figure}[H]
    \begin{minipage}[t]{\linewidth}
        \begin{subfigure}{\linewidth}
            \includegraphics[width=\linewidth]{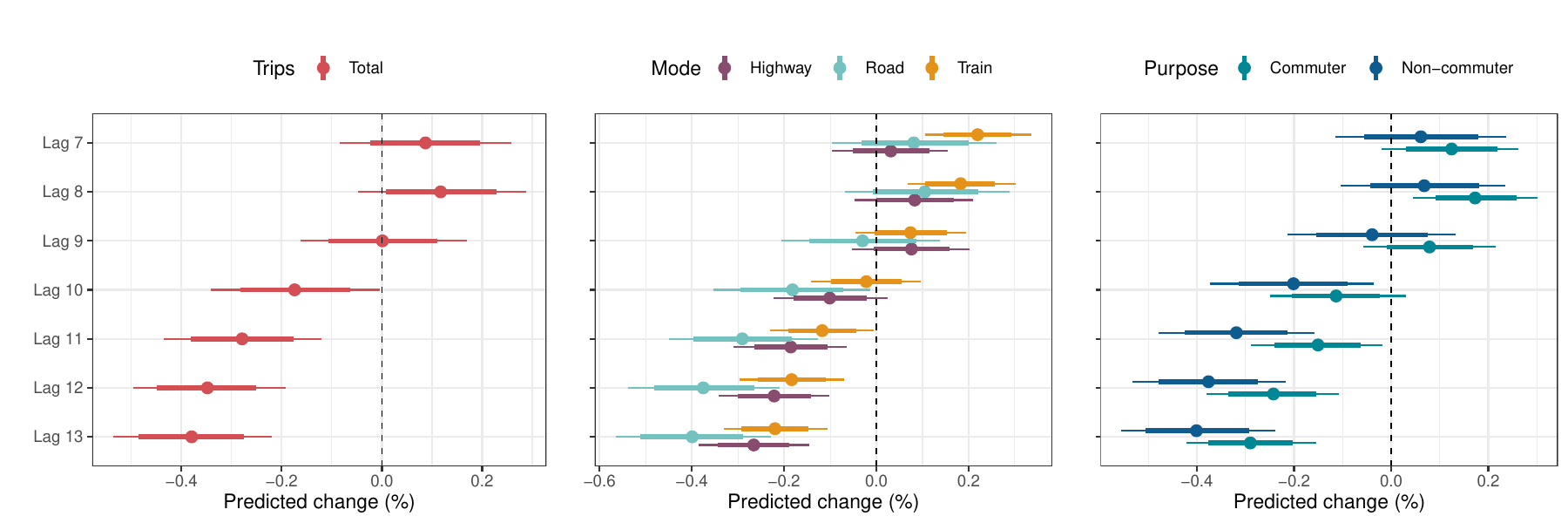}
        \end{subfigure}
    \end{minipage}
\captionsetup{justification=justified}
\caption{Predicted change in reported COVID-19 cases given a 1\,\% reduction of lagged total trips, mobility by mode, and mobility by purpose after controlling for the number of daily tests performed in each canton. Posterior means are shown as dots, while 80\,\% and 95\,\% credible intervals are shown as thick and thin bars, respectively.}
\label{fig:cases_tests}
\end{figure}

After controlling for the number of daily tests per canton, decreases in mobility lagged by 10--13 days still predict reductions in the reported number of new cases (\Cref{fig:cases_tests}). Our findings are therefore qualitatively the same for the majority of lags. The inconclusive effects for lags of 7--9 days may be due to ``post-treatment bias'' in the coefficient on lagged mobility, referring to that the regression controls for a covariate (here: number of tests) that occurs after the main covariate of interest (here: lagged mobility).

The number of daily tests performed may not be an appropriate measure of the success of a testing program as it ignores how many infected people are detected. Thus, and as an alternative robustness check, we instead include the share of positive test results (called the positivity rate or ``percent positive'') in our model. The positivity rate is a measure of testing intensity relative to the daily number of reported cases. A low positivity rate means a large number of tests relative the number of number of reported cases, which implies comprehensive testing. A high positivity rate, in turn, indicates that there may be many infected cases not yet reported, and that more comprehensive testing is needed. Compared to the number of tests performed, the positivity rate thereby inherently indicates how "difficult" it is to detect infected people, and thus provides policy makers with more information on when to enforce policy measures. Hence, this robustness check may be interpreted as controlling for testing over the study period relative to the stage of the epidemic.

We estimate the positivity rate for each canton $i$ and day $t$ by dividing the observed number of daily tests in the country, $u_t$, with the corresponding number of positive tests, and then multiplying this rate with 100 to obtain it on the percentage scale. This estimate would recover the true positivity rate if, for all days, the share of positive tests is equal across cantons. While this may not hold in practice, the lack of data on testing per canton would requires us to use additional complex modeling to estimate canton-varying positivity rates, and this is outside the scope of this paper. 

We re-estimate the model for the relationship between lagged mobility and reported cases now included the positivity rate. We assign its coefficient a weakly informative prior of $\mathsf U(0,10)$, meaning that a one percentage point increase in the share of positive tests is expected to be associated with an increase in the reported number of cases of zero to 10 percent.

\begin{figure}[H]
    \begin{minipage}[t]{\linewidth}
        \begin{subfigure}{\linewidth}
            \includegraphics[width=\linewidth]{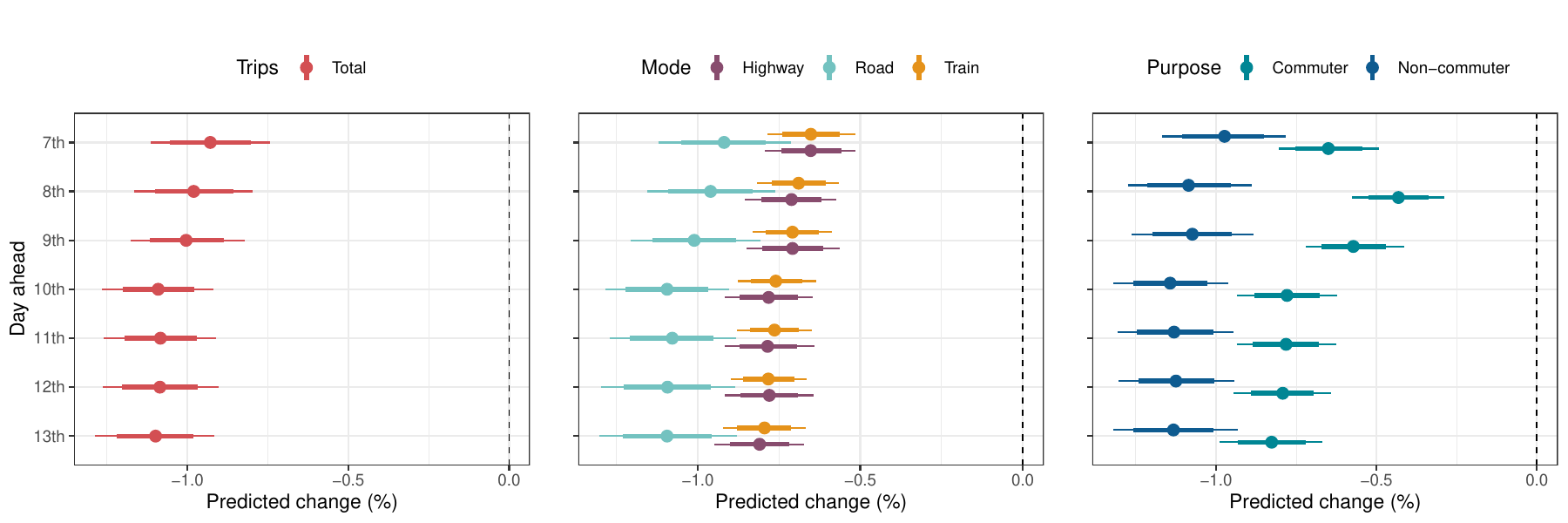}
        \end{subfigure}
    \end{minipage}
\captionsetup{justification=justified}
\caption{Predicted change in reported COVID-19 cases given a 1\,\% reduction of lagged total trips, mobility by mode, and mobility by purpose after controlling for the share of positive tests. Posterior means are shown as dots, while 80\,\% and 95\,\% credible intervals are shown as thick and thin bars, respectively.}
\label{fig:cases_positivity}
\end{figure}

Controlling for test positivity rate instead of the number of tests leads to almost identical estimates for mobility as in the main model that does not control for testing (\Cref{fig:cases_positivity}). Hence, when controlling for the positivity rate, we find no evidence of that testing confounded our model estimates.

\subsection{Predicting hospitalizations and deaths}
\label{supp:deaths_hospitalizations}

As a third approach to account for changes in testing capacity over the study period, we replace the dependent variable of our regression model with daily cantonal hospitalizations and deaths attributed to COVID-19. This avoids the problem of potential changes in testing over time as hospitalizations or deaths are registered regardless of testing capacity. We obtain official statistics on hospitalizations and deaths attributed to COVID-19 from the Federal Office of Public Health of the Swiss Confederation; BAG (\url{https://www.covid19.admin.ch/en/overview}). 

The Federal Office of Public Health of the Swiss Confederation (BAG) also publishes estimates of the effective reproductive number, but we do not consider it as the dependent variable for the following reason. First, the published data do not cover our whole study period and have a different number of missing observations across cantons. Second, using the effective reproductive number as the dependent variable in a regression would ignore that the effective reproductive number is itself an estimate. The credible intervals for the predictive effect of lagged mobility would not capture this additional uncertainty, and hence, be too narrow. We thus restrict this additional robustness check to hospitalizations and deaths as the dependent variable. 

For estimation, we increase the lag on mobility to 10--20 days to accommodate the additional delay between a reported case and hospitalizations or death. The priors on the parameters are the same as in the main model of the relation between lagged mobility and reported cases. This is justified by that the parameters reflect multiplicative changes in the dependent variable, and as such, are invariant to the actual values of the dependent variable. 

\begin{figure}[H]
    \centering
    \begin{minipage}[t]{\linewidth}
        \figletter{a} \\
        \begin{subfigure}{\linewidth}
            \includegraphics[width=\linewidth]{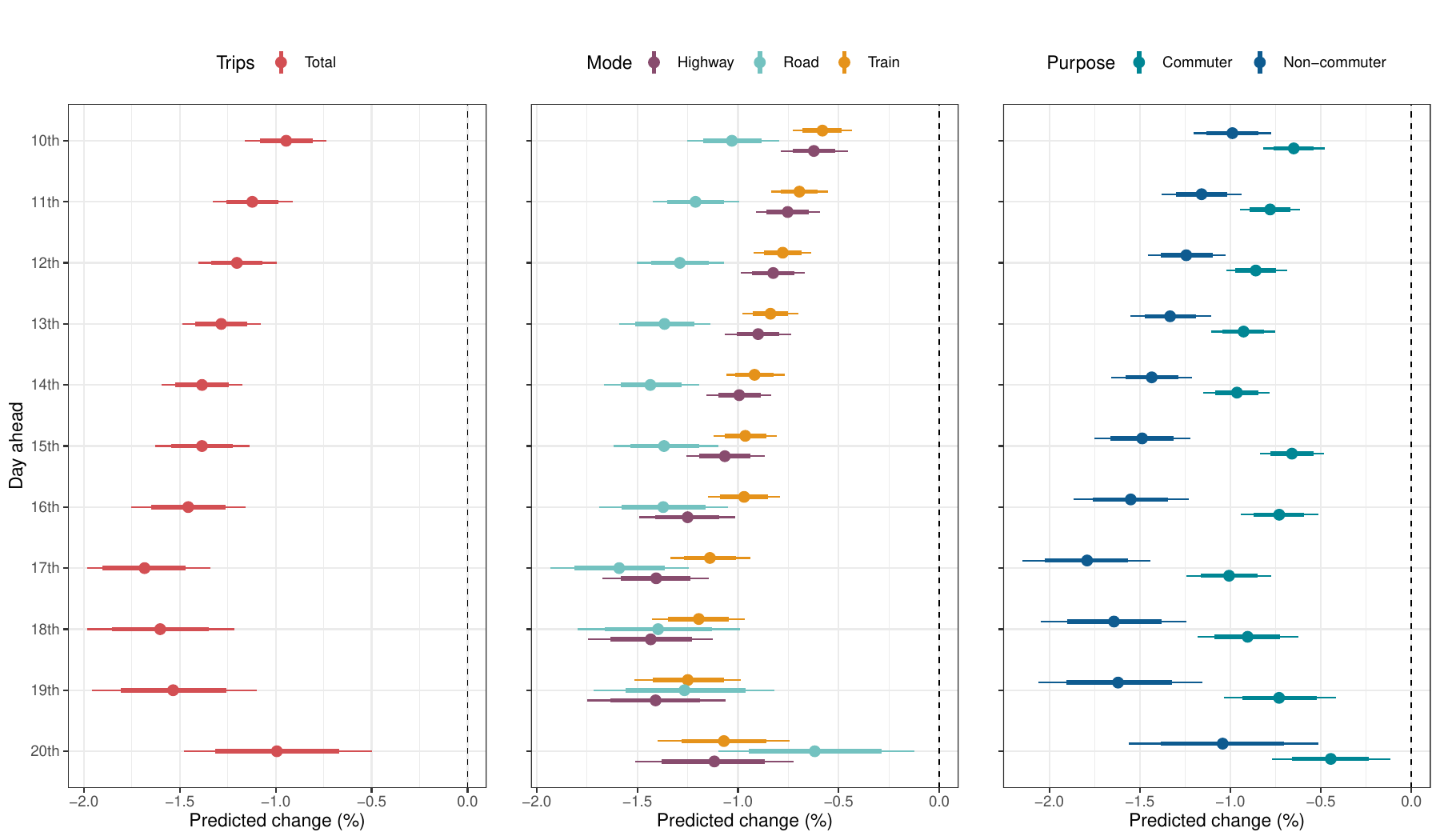}
        \end{subfigure}
    \end{minipage}
    \\
    \begin{minipage}[t]{\linewidth}
        \figletter{b} \\
        \begin{subfigure}{\linewidth}
            \includegraphics[width=\linewidth]{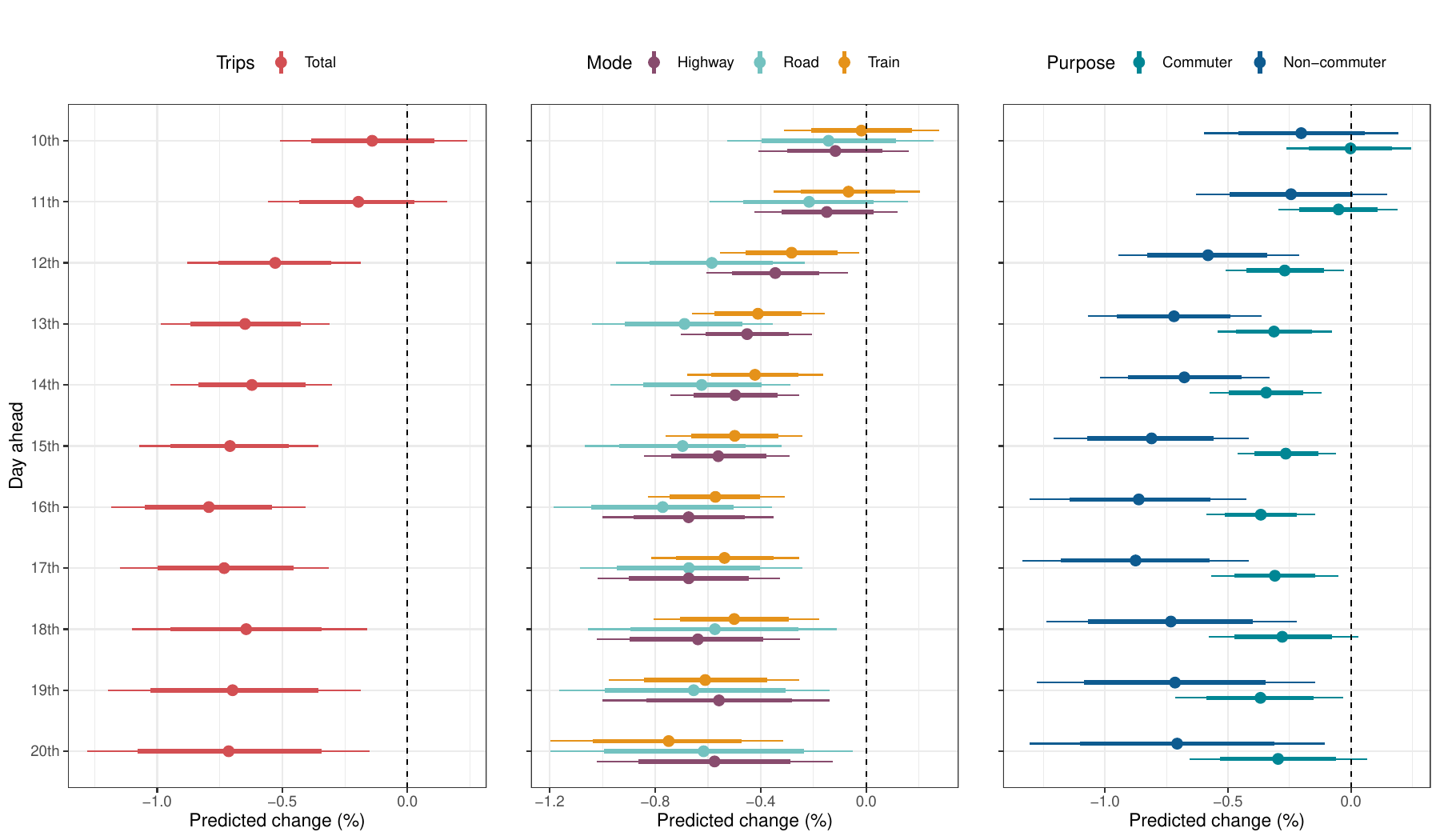}
        \end{subfigure}
    \end{minipage}
\captionsetup{justification=justified}
\caption{Predicted change given a 1\,\% reduction of lagged total trips, mobility by mode, and mobility by purpose in: \textbf{(a)} hospitalizations attributed to COVID-19, and \textbf{(b)} deaths attributed to COVID-19. Posterior means are shown as dots, while 80\,\% and 95\,\% credible intervals are shown as thick and thin bars, respectively.}
\label{fig:deaths_hosp}
\end{figure}

The predicted change in hospitalizations given a 1\,\% reduction of lagged trips is somewhat larger than the corresponding estimates with reported cases as dependent variable (\Cref{fig:deaths_hosp}). Conversely, the predicted change in deaths given a 1\,\% reduction of lagged trips are a bit smaller. Overall, reductions in mobility predicts reductions in both hospitalizations and deaths 10--20 days later for both total trips and all categories of trips by mode and purpose. However, for deaths, the credible intervals for lags of 10 and 11 days cover zero. This could reflect that the average time from infection during trips to deaths takes longer than 11 days. Again, we find mobility estimates that are consistent with those in the main paper. In particular, mobility is a predictor of both hospitalizations and deaths, thereby confirming the robustness of our work.

\clearpage
\section{Model diagnostics}
\label{supp:model_diagnostics}

This section presents an analysis of model diagnostics. For reasons of brevity, we only show diagnostics for the models with the total trips variable. The diagnostics for the models of trips by each mode and purpose also showed good fit.

The \textsf{R} package \texttt{loo} \cite{Vehtari.2020} version 2.3.1 is used to estimate the Pareto tail shape parameter and, based on it, check for influential observations. The other diagnostic plots are obtained with the \textsf{R} package \texttt{bayesplot} \cite{Gabry.2020b, Gabry.2019} version 1.7.2.

\subsection{Checking posterior predictive ability}
\begin{figure}[H]
        \centering
        \includegraphics[width=\linewidth]{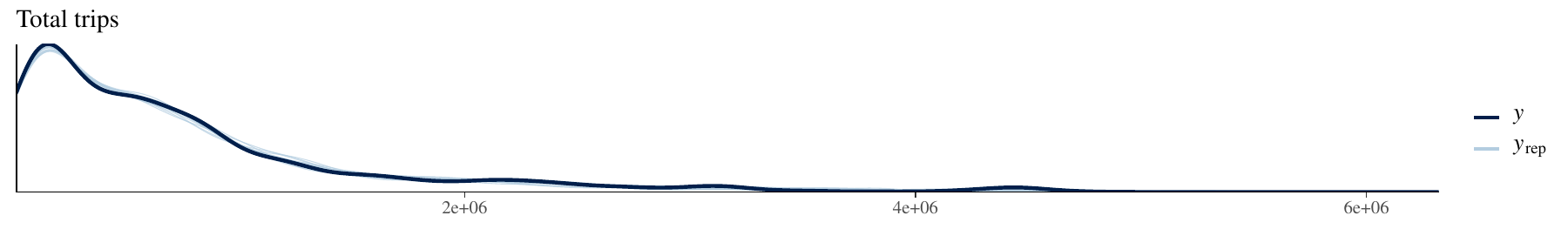}
    \caption{Posterior predictive plot for the model of the effect of policy measures on total trips. $y$ and $y_{\text{rep}}$ are kernel density estimates of the observed response values and 10 simulated replicates, respectively.}
\label{fig:traffic_pp_check_total}
\end{figure}

\begin{figure}[H]
        \centering
        \includegraphics[width=\linewidth]{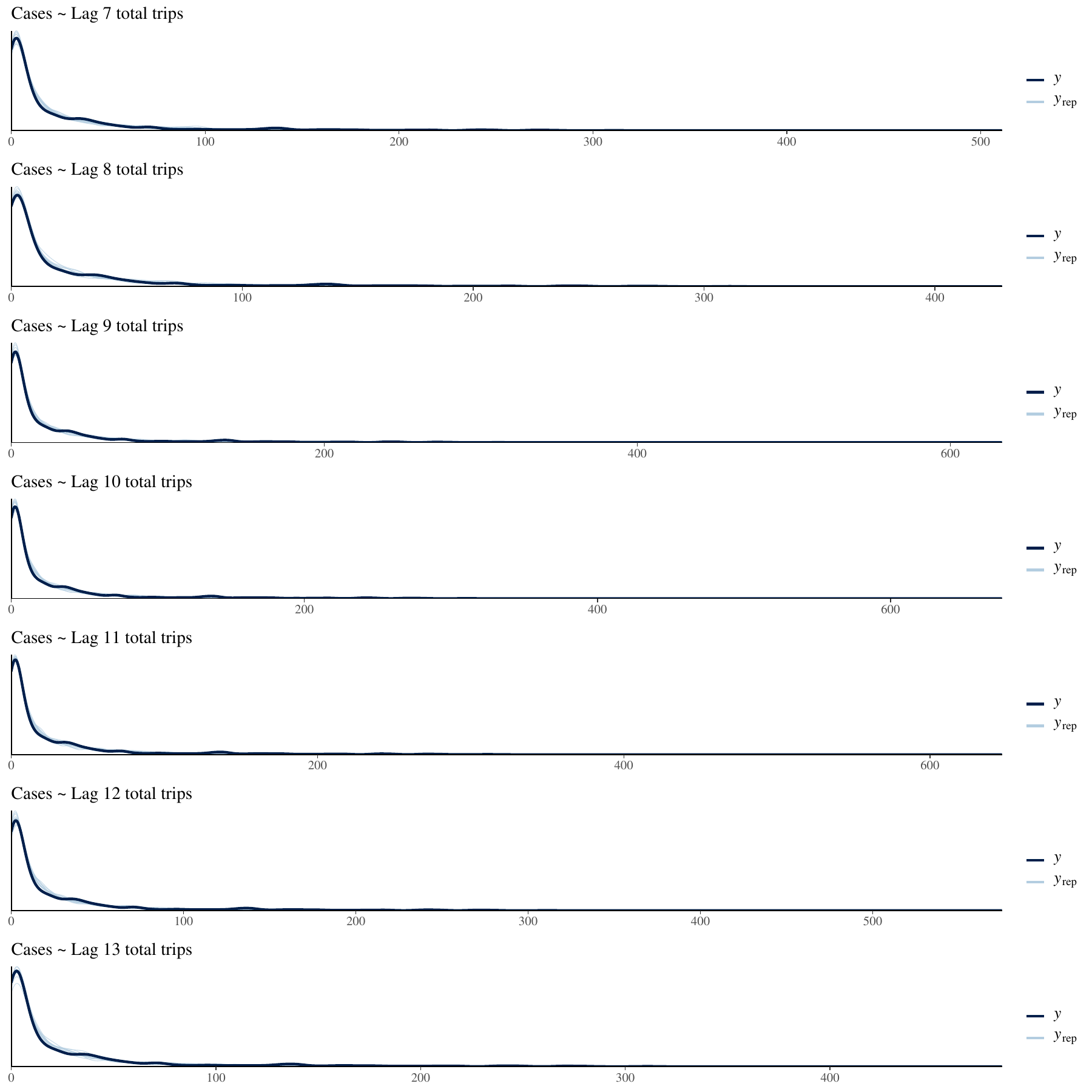}
    \caption{Posterior predictive plot of the model for the effect of each lag of total trips on the reported number of new cases. $y$ and $y_{\text{rep}}$ are kernel density estimates of the observed response values and 10 simulated replicates, respectively.}
\label{fig:cases_pp_check_total}
\end{figure}

We assess model fit through posterior predictive checks. For each model, we simulate 10 draws from its posterior predictive distributions with the same values of the explanatory variables as those used to fit the model, resulting in simulated replicates of the response variable. Consistent with the notation in \cite{Gelman.2013} we denote these $y_{rep}$. \Cref{fig:traffic_pp_check_total} and \Cref{fig:cases_pp_check_total} shows kernel density estimates of the replicated responses and the observed responses for the models of mobility and cases. The replicated response densities fit the observed response density well, indicating good model fit.

\subsection{Checking for divergent transitions}
\begin{figure}[H]
        \centering
        \includegraphics[width=\linewidth]{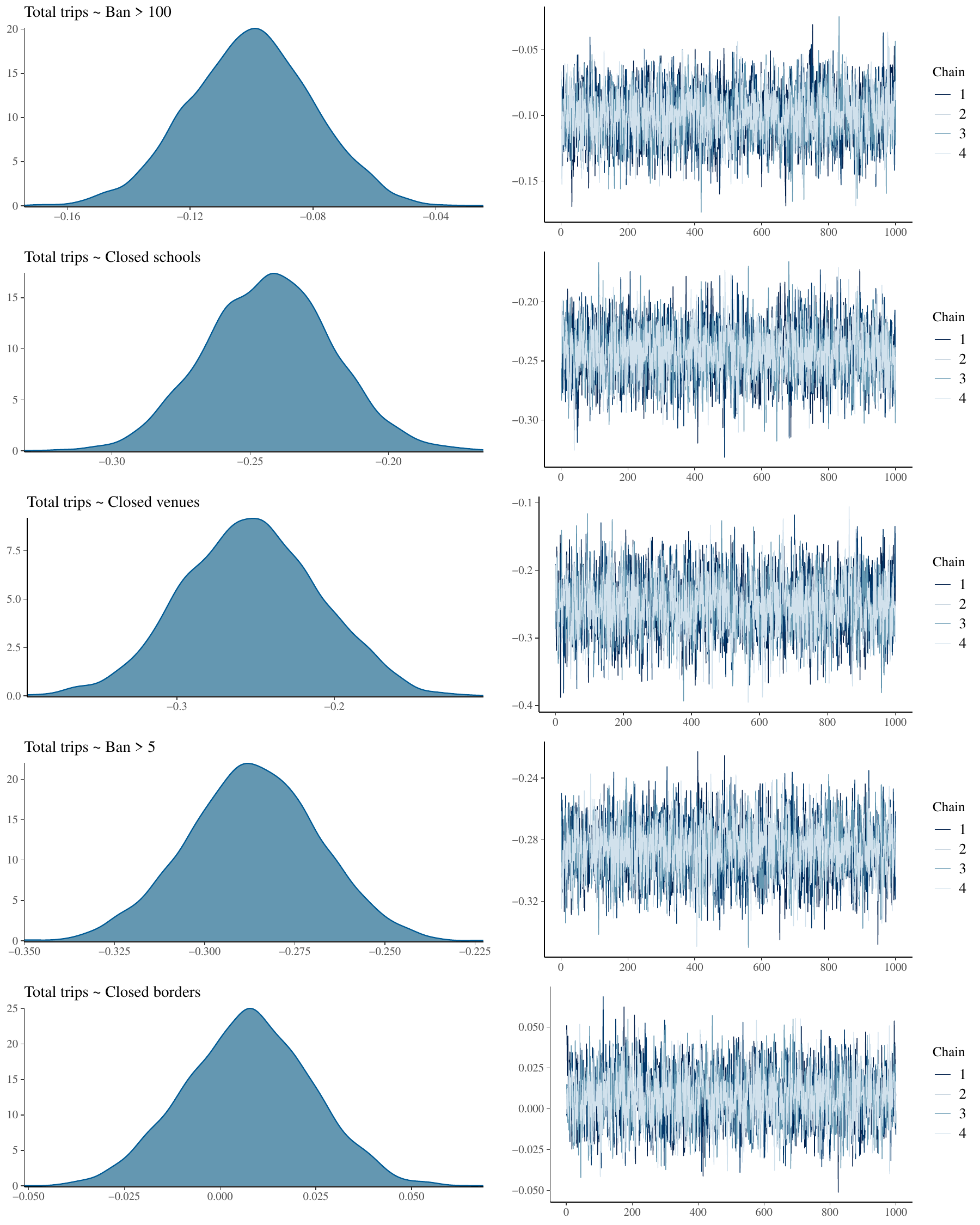}
    \caption{Kernel density and time series plots of the MCMC samples of the parameter for the effect of each policy measure and total trips.}
\label{fig:traffic_trace_density_total}
\end{figure}

\begin{figure}[H]
        \centering
        \includegraphics[width=\linewidth]{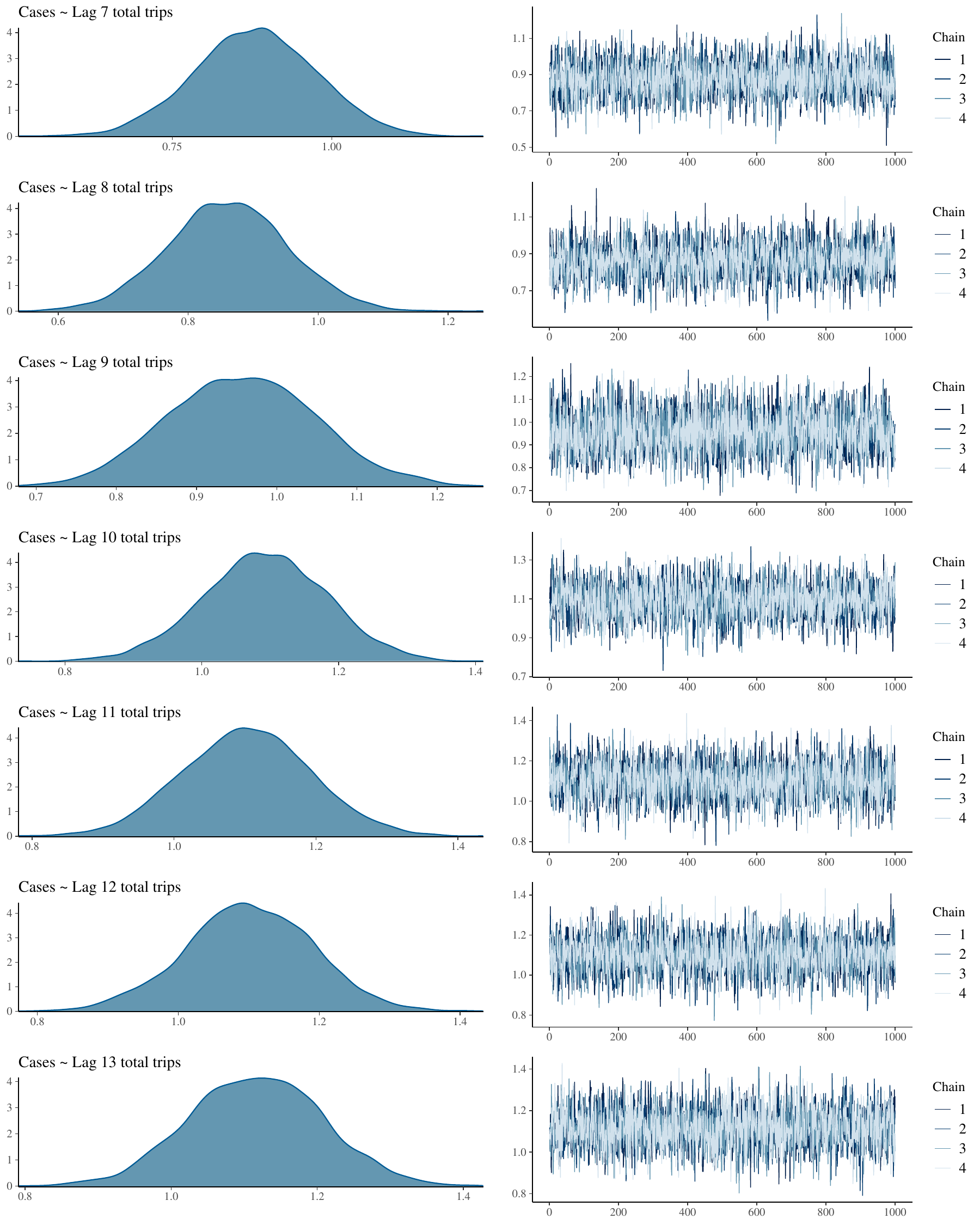}
    \caption{Kernel density and time series plots of the MCMC samples of the parameter for the effect of each lag of total trips on the reported number of new cases.}
\label{fig:cases_trace_density_total}
\end{figure}

\begin{figure}[H]
        \centering
        \includegraphics[width=\linewidth]{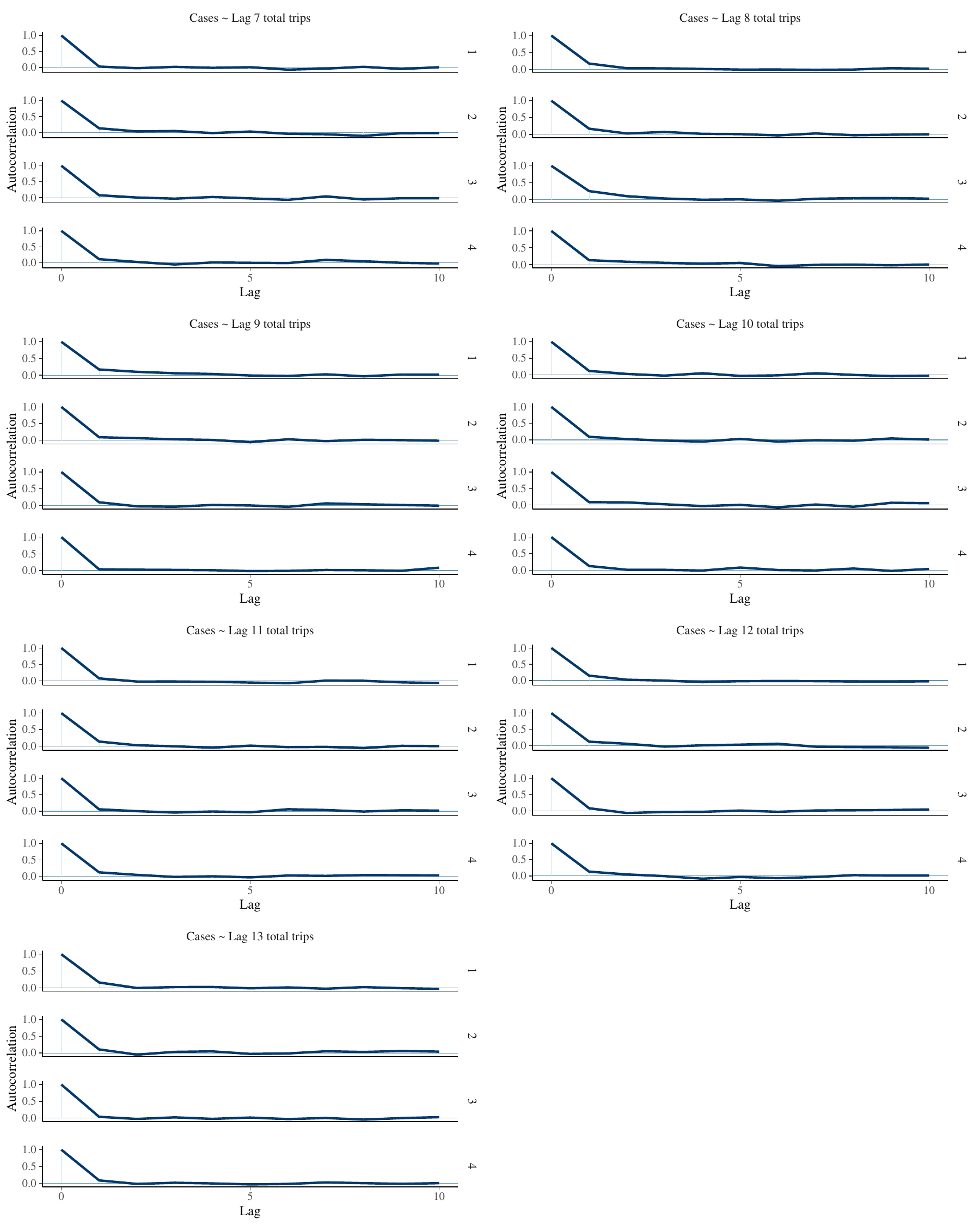}
    \caption{Autocorrelation plots of MCMC samples for each chain of the parameter for the effect of each lag of total trips on the reported number of new cases.}
\label{fig:cases_acf_total}
\end{figure}

\Cref{fig:traffic_trace_density_total} shows kernel density and time series of the MCMC samples of the parameters for the mobility model. \Cref{fig:cases_acf_total} shows the autocorrelation among the MCMC samples of the same parameters. \Cref{fig:cases_trace_density_total} and \Cref{fig:cases_acf_total} shows the same information for the models with cases as dependent variables. For brevity we omit the corresponding plots for control variables, intercepts and random and fixed effects. For each lag of total trips, the posterior density of the parameter is approximately Gaussian and that their chains have good mixing with no signs of divergence or substantial autocorrelation.

\subsection{Checking the effective sample size and convergence of Markov chains}
\begin{figure}[H]
        \centering
        \includegraphics[width=\linewidth]{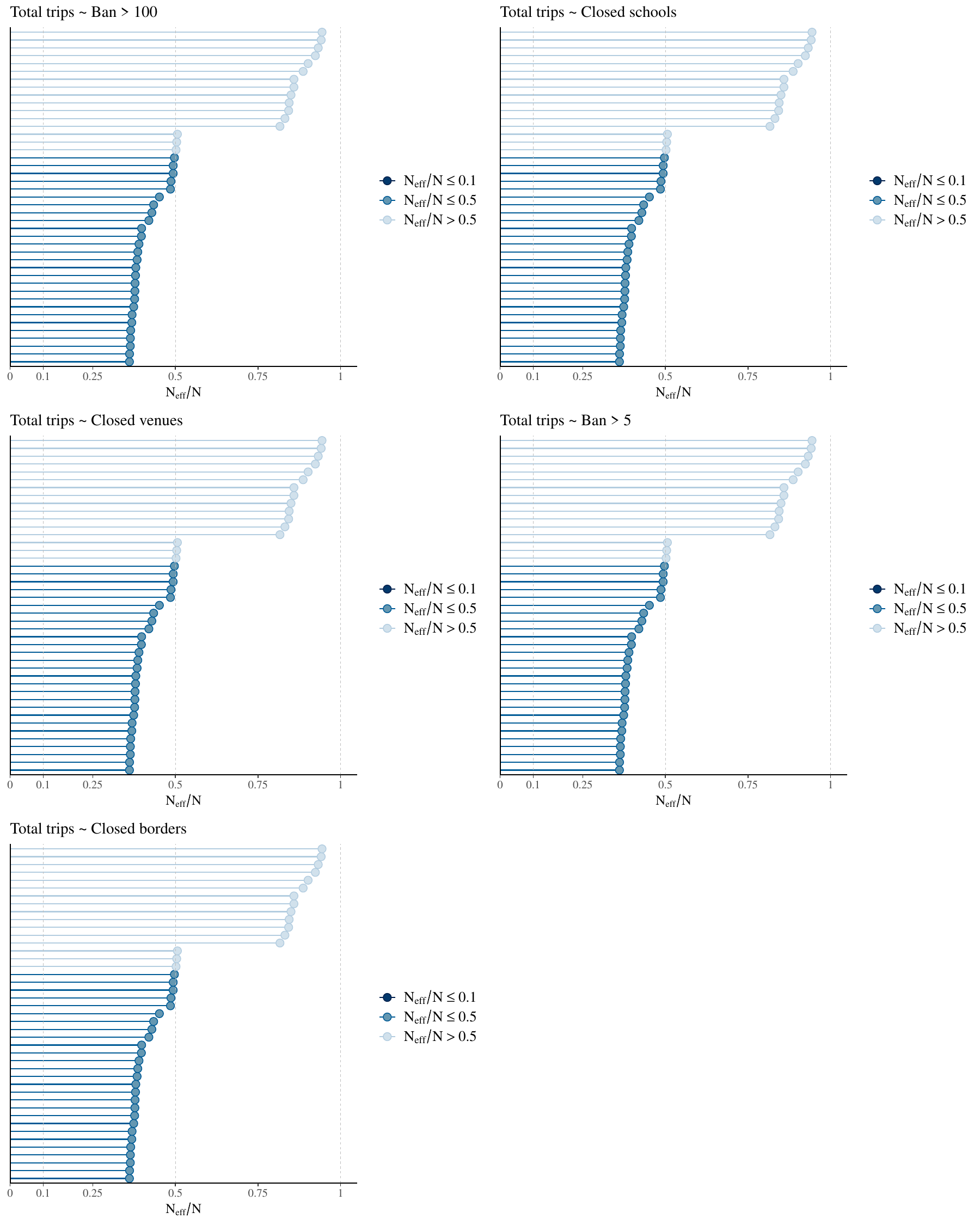}
    \caption{Ratios of effective sample size to total sample size for MCMC samples of the parameter for the effect of each policy measure on total trips.}
\label{fig:traffic_neff_total}
\end{figure}

\begin{figure}[H]
        \centering
        \includegraphics[width=\linewidth]{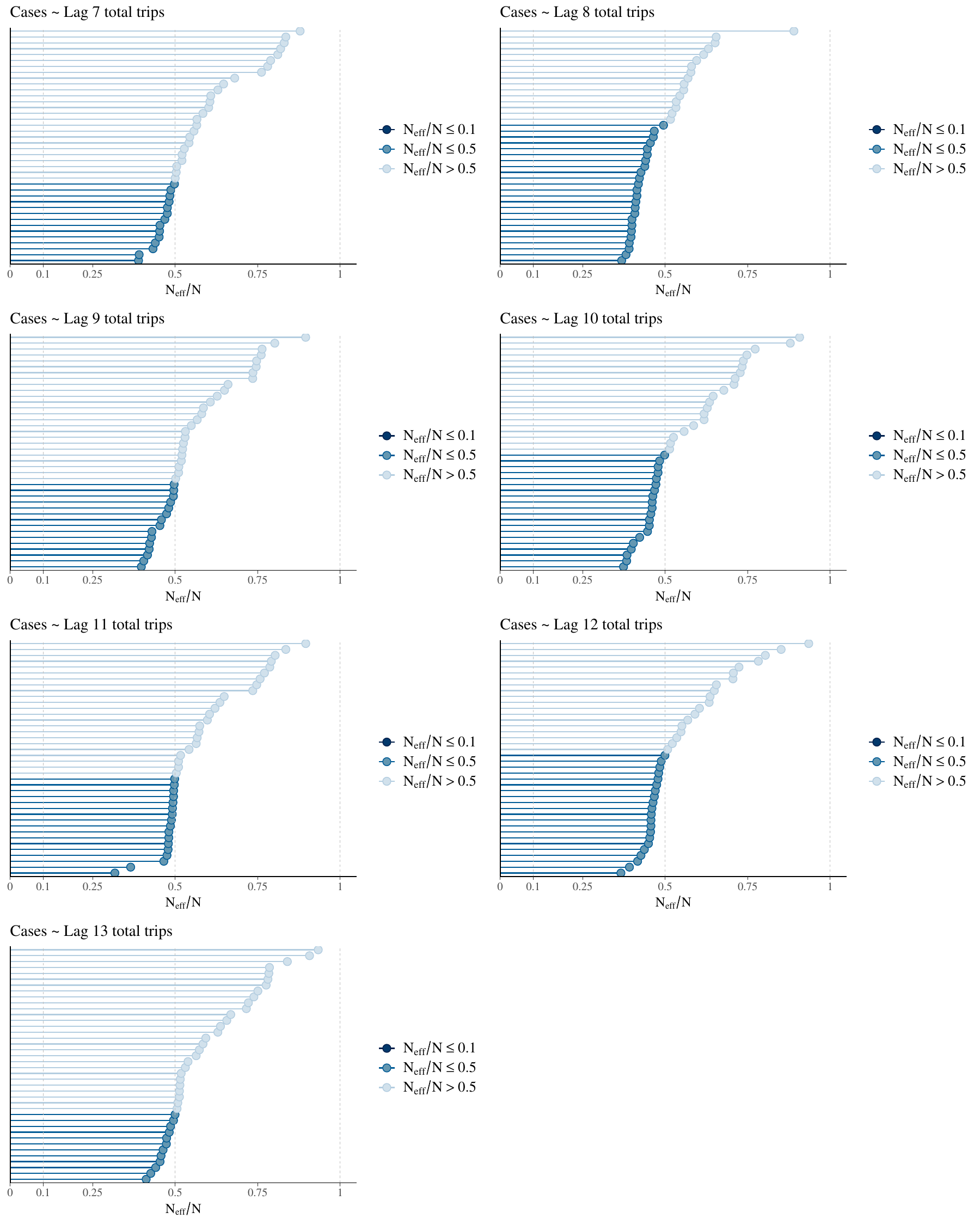}
    \caption{Ratios of effective sample size to total sample size for MCMC samples of the parameter for the effect of each lag of total trips on the reported number of new cases.}
\label{fig:cases_neff_total}
\end{figure}

\begin{figure}[H]
        \centering
        \includegraphics[width=\linewidth]{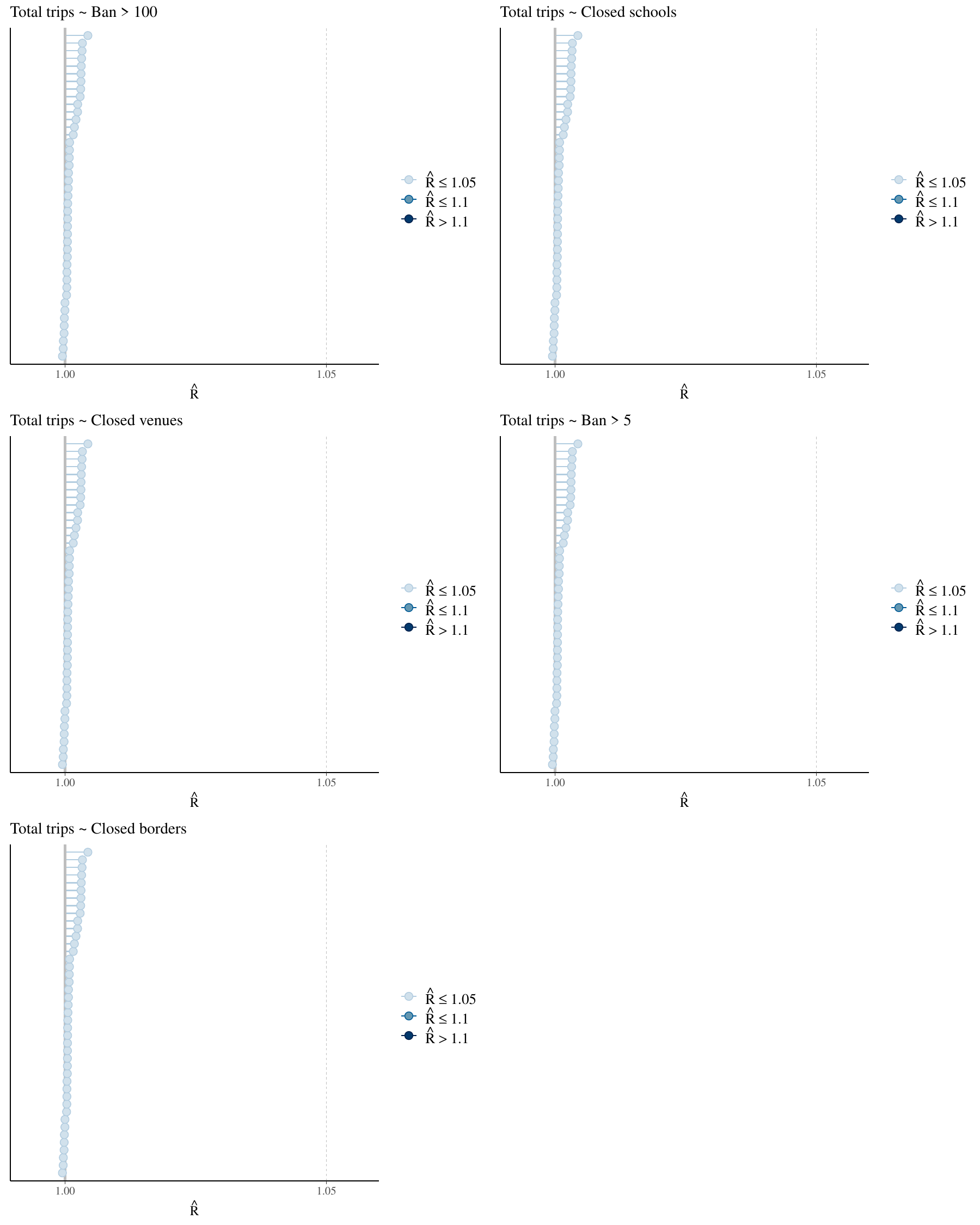}
    \caption{$\hat R$ values for MCMC samples of the parameter for the effect of each policy measure on total trips.}
\label{fig:traffic_rhat_total}
\end{figure}

\begin{figure}[H]
        \centering
        \includegraphics[width=\linewidth]{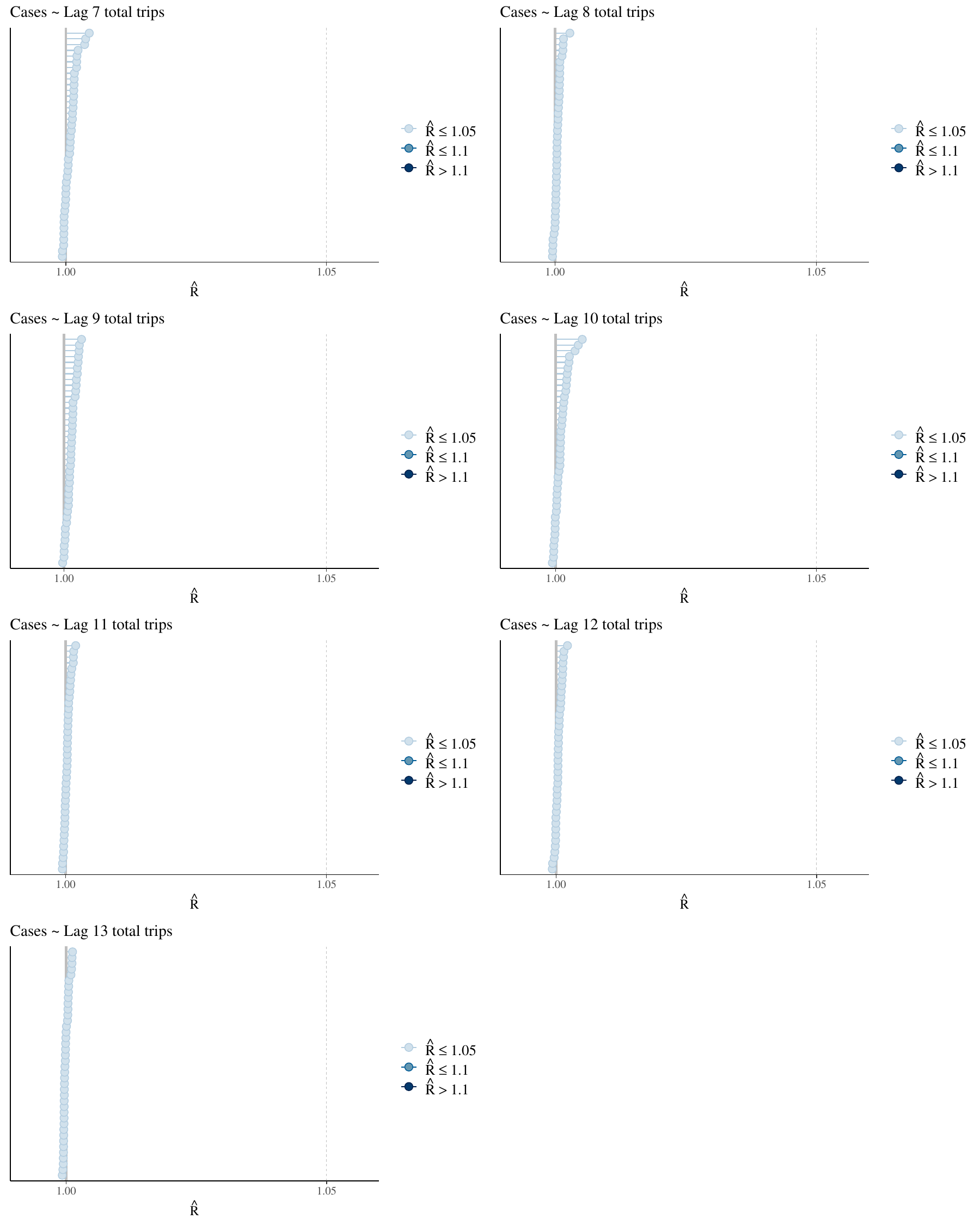}
    \caption{$\hat R$ values for MCMC samples of the parameter for the effect of each lag of total trips on the reported number of new cases.}
\label{fig:cases_rhat_total}
\end{figure}

\Cref{fig:traffic_neff_total} and \Cref{fig:traffic_rhat_total} shows ratios of effective sample size to total sample size ($\hat{n}_{\text{eff}} / N$) and $\hat R$ values, respectively, for the MCMC samples of all parameters in the regression models. The corresponding plots for the models of cases are available in \Cref{fig:cases_neff_total} and \Cref{fig:cases_rhat_total}. The ratio of effective sample size to total sample size is between around 0.4 and 1 for most parameters, indicating a sufficient number of independent draws from the posterior distributions \cite{Gelman.2013}. The $\hat R$ values are close to 1 for all parameters and models. Hence, this indicates convergence of the Markov chains.

\subsection{Checking for overdispersion}
\label{supp:model_diagnostics_overdispersion}

The estimated overdispersion parameter (for the models with total trips) is as follows:
\begin{itemize}
\item The first model (for estimating the reduction in human mobility due to policy measures) has an estimated overdispersion parameter of 51.33 (95\,\% CrI: 47.02--55.88). 
\item The second model (for estimating the relationship between mobility and cases) has an estimated overdispersion parameter of 3.43 (95\,\% CrI: 2.92--3.99) at a lag of 7 days. The parameter estimate amounts to 3.95 (95\,\% CrI: 3.34--4.65) at a lag of 13 days. 
\item In the mediation analysis, the estimated overdispersion parameter of the mediation model amounts to 54.40 (95\,\% CrI: 49.49--59.54) for a lag of 7 days. The estimate increases to 61.83 (95\,\% CrI: 56.14--67.71) for a lag of 13 days. For the outcome model, the estimated overdispersion parameter increases from 4.29 (95\,\% CrI: 3.60--5.06) at a lag 7 to an estimate of 5.42 (95\,\% CrI: 4.48--6.49) at lag 13. 
\end{itemize}
Overall, there is substantial overdispersion in the dependent variables. Because this, the use of a negative binomial distribution for the dependent variables is recommended.

\subsection{Checking for influential observations}

\begin{figure}[H]
        \centering
        \includegraphics[width=\linewidth]{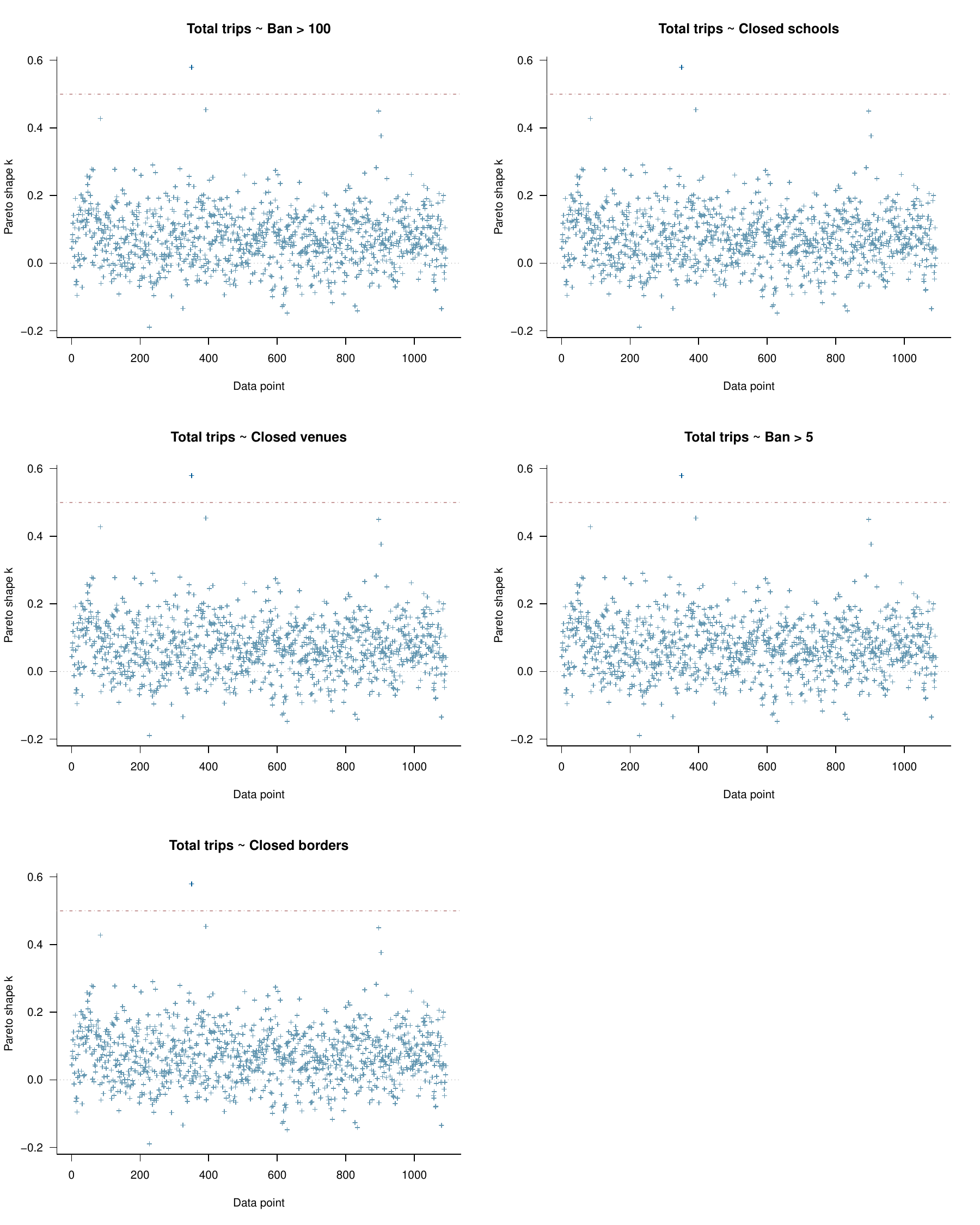}
    \caption{Estimated Pareto tail shape parameter $k$ against the observation indices for the model of the effect of policy measures on total trips.}
\label{fig:traffic_pareto_total}
\end{figure}

\begin{figure}[H]
        \centering
        \includegraphics[width=\linewidth]{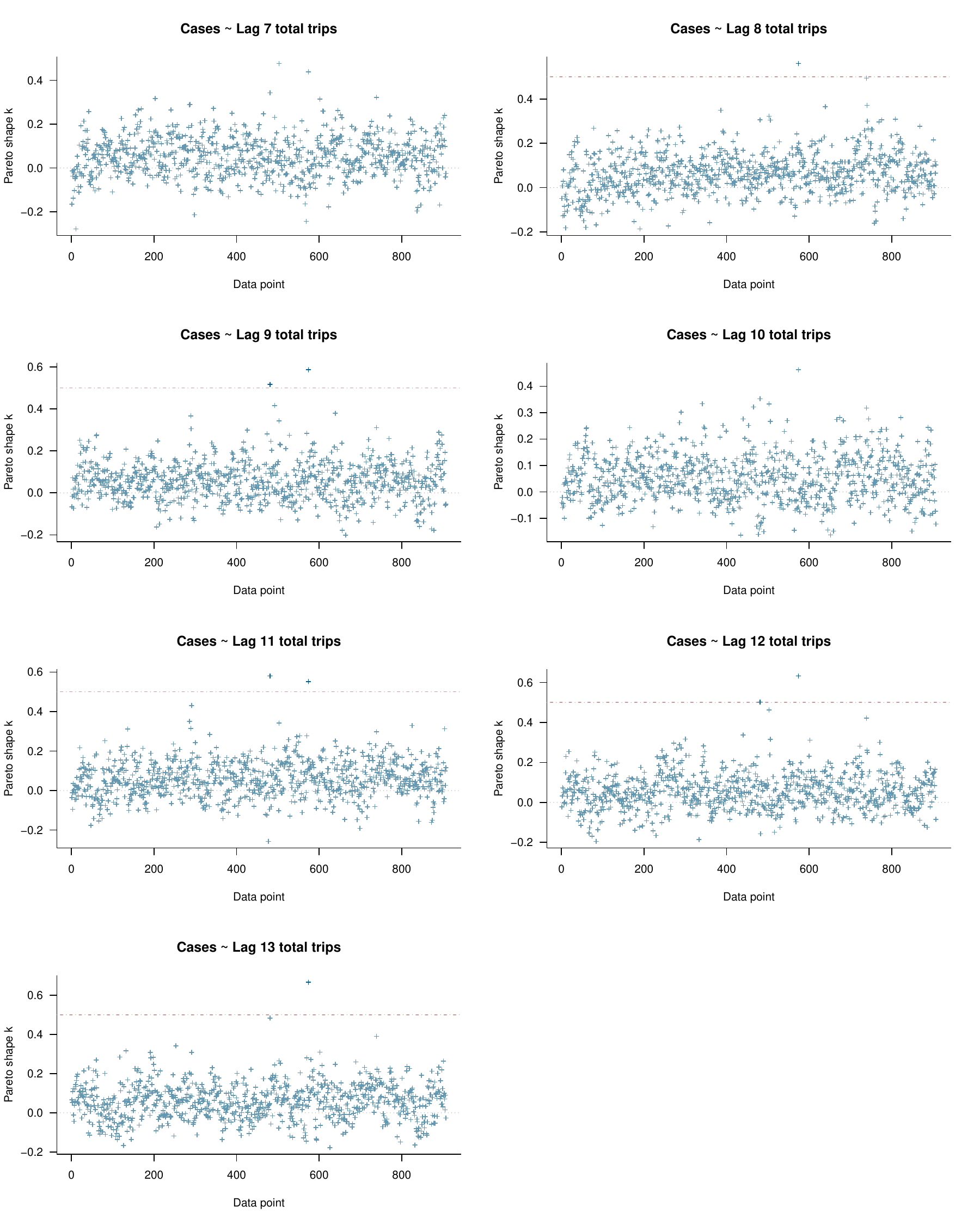}
    \caption{Estimated Pareto tail shape parameter $k$ against the observation indices for the models of the effect of each lag of total trips on the reported number of new cases.}
\label{fig:cases_pareto_total}
\end{figure}

Influential observations can have a negative effect on model fit. To check for the presence of influential observations, we plot the Pareto tail shape parameter $k$ against the observation indices for each of the models. The results for the model of total trips are shown in \Cref{fig:traffic_pareto_total} while the results for the model of cases on total trips are available in \Cref{fig:cases_pareto_total}. A value of $k$ less than 0.5 is generally considered good whereas a value between 0.5--0.7 is considered okay \cite{Vehtari.2017, Vehtari.2019}. No model has many observations with an estimated value of $k$ above 0.5.

\subsection{Checking for correlations between parameters}
\label{supp:param_corr}

\begin{figure}[H]
        \centering
        \includegraphics[width=\linewidth]{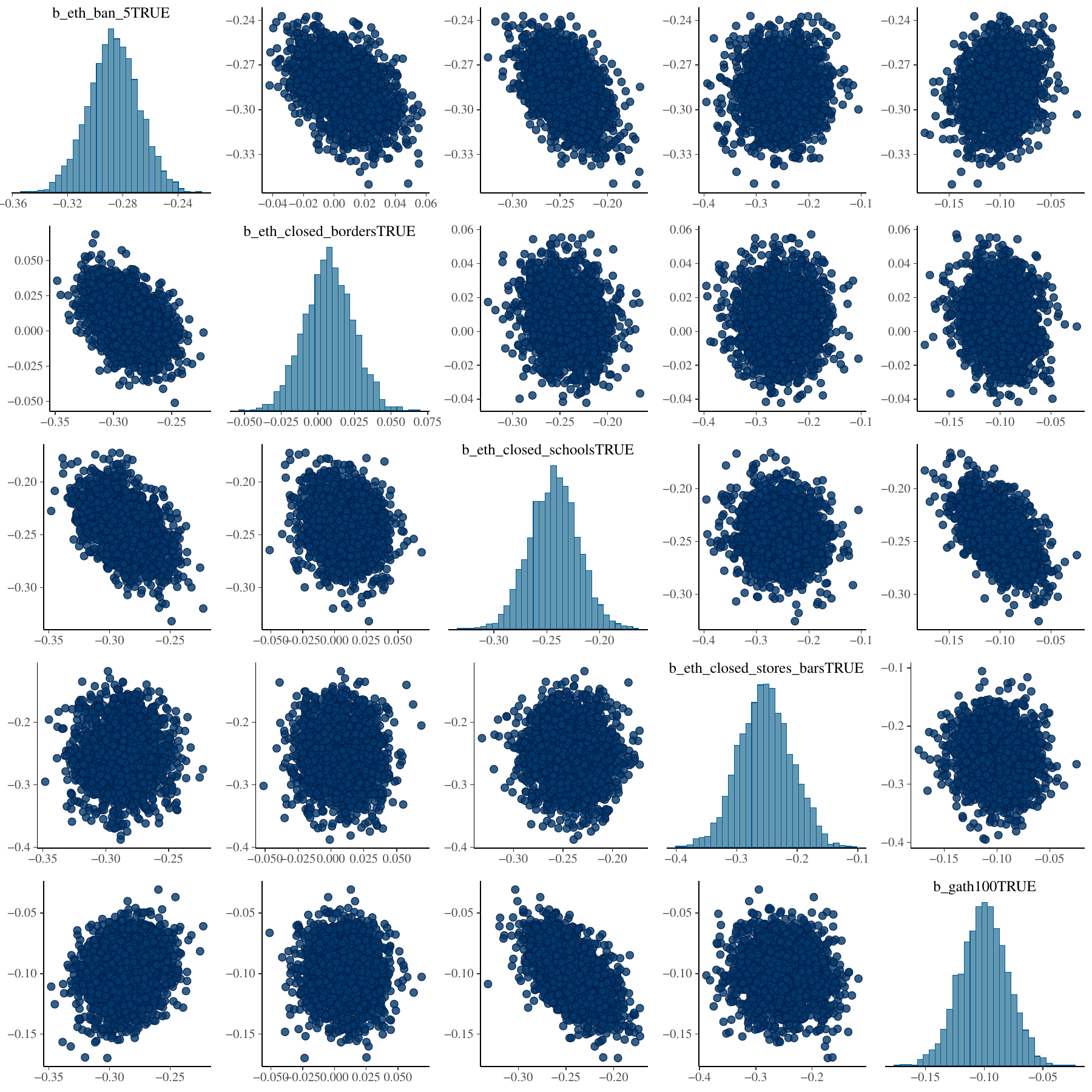}
    \caption{Pairwise bivariate posterior distributions of the policy measure parameters in the model of total trips.}
\label{fig:traffic_bivardist_total}
\end{figure}

A similar timing of the policy measures across cantons could make it difficult to distinguish their individual effects. To investigate this issue, \Cref{fig:traffic_bivardist_total} depicts scatterplots of the pairwise bivariate posterior distributions of the policy measure parameters. There is a negative correlation between the posterior samples of the parameters for some of the NPI pairs, reflecting a difficulty to distinguish the effects of the two policy measures if they are introduced close in time. It seems to be somewhat difficult to distinguish between the effects of the ban on gatherings and school closures, as they were implemented close in time in many cantons (\Cref{tbl:NPI_diff}). The same holds for the bans of gatherings of more than five people and border closures. In contrast, the effects of border closures, school closures, and venue closures can be distinguished, as judged by the even spread of their pairwise scatters (\Cref{fig:traffic_bivardist_total}).

\newpage

\begin{figure}[H]
        \centering
        \includegraphics[width=\linewidth]{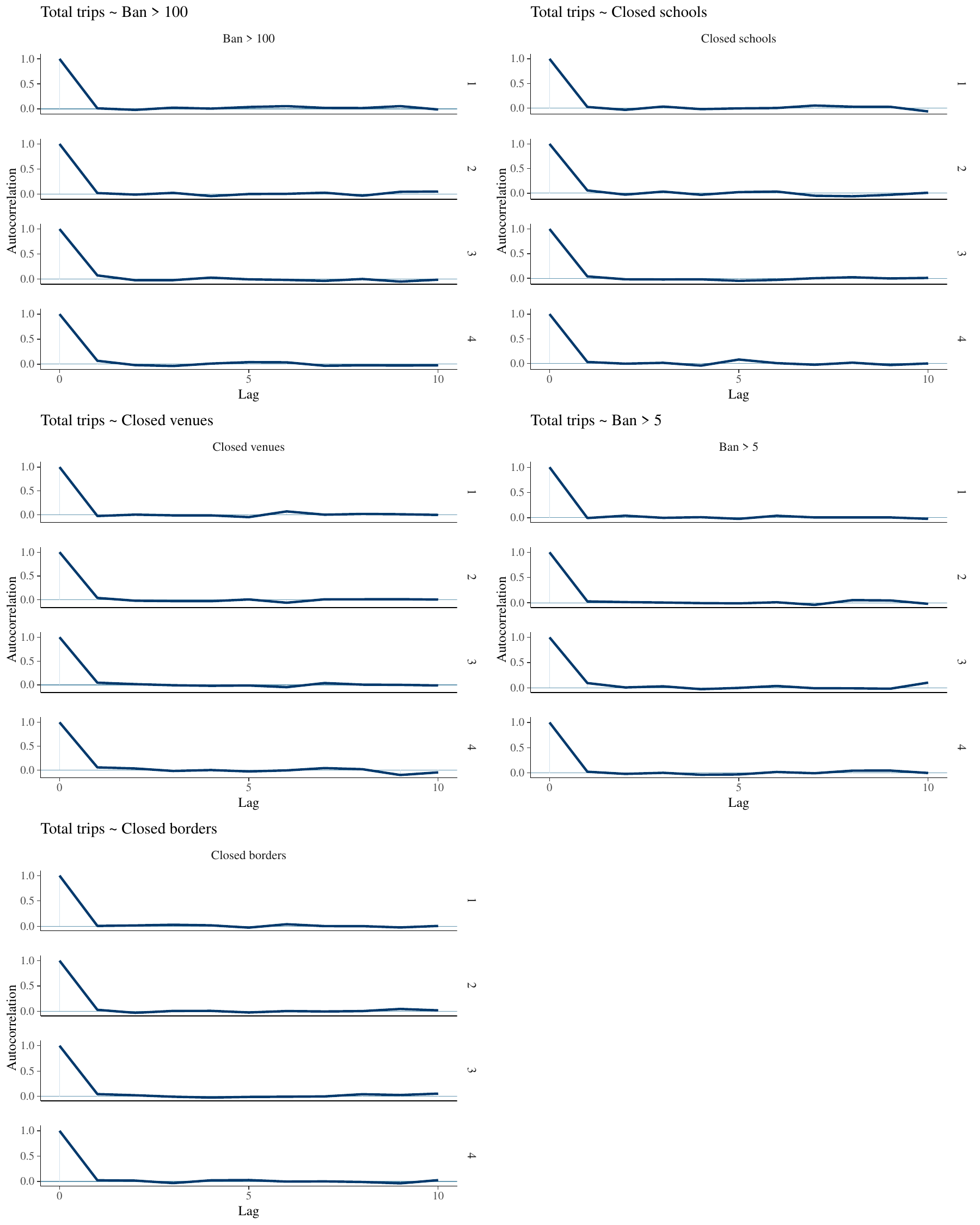}
    \caption{Autocorrelation plots of MCMC samples for each chain of the parameter for the effect of each policy measure on total trips.}
\label{fig:traffic_acf_total}
\end{figure}

\clearpage
\section{Modeling extensions}
\label{supp:modeling_extensions}

This section includes extensions of the models featured in the paper. Estimation details are in Supplement~\ref{supp:estimation_extended}.

\subsection{Accounting for the spatial dependence in mobility among neighboring cantons}
\label{supp:extension_spatial}

Our main models treat mobility as being independent across cantons. Modeling cross-sectional units as independent is a conventional approach in panel and longitudinal data analysis. However, one would reasonably expect that the level of mobility is similar for neighboring cantons but not so similar for geographically distant cantons. If a dependence in mobility between cantons is not modeled, it will enter the error term, violating the assumption that error terms are independent across cantons. For this purpose, we extend the mobility model with a spatial random effect that captures the similarity in mobility among neighboring cantons prior to any policy measure.   

Let $\phi^{(k)}_i$ be a spatial random effect for canton $i$ that has an additive effect on mobility variable $k$ on the log scale. The model for the effect of the policy measure on mobility variable $k$ is then given by
\begin{align}
    \label{eq:mobility_model_spatial}
    \log \mathbb E[M_{itk}\,|\, \eta^{(k)}_{it}, E_i]
    &=
    \log E_i +
    \alpha^{(k)} + 
    \theta^{(k)}_i + 
    \phi^{(k)}_i + 
    \delta^{(k)}_{w(t)} +
    \gamma^{(k)} \log z_{it} +
    \gamma^{(k)}_B \widebar{\log z_i} +
    \sum^L_{l=1} \beta^{(k)}_l
    d_{itl}.
\end{align}
The spatially-extended model may be viewed as a longitudinal version of the Besag-York-Mollié (BYM) model \cite{Besag.1991} commonly used for disease mapping applications in epidemiology. The term ``disease mapping'' refers to statistical methods that estimates the spatial correlation between observations and where interest lies in the distribution of a disease over geographical units. The BYM model, in turn, is a Bayesian hierarchical Poisson regression model with two types of random effects: (1)~random effects that capture spatial correlation between units, and (2)~random effects that captures unobserved heterogeneity between units \cite{Wakefield.2000}. Our spatially extended model adapts the original BYM model with the use of a negative binomial distribution to account for the overdispersion in the trip counts. It also adapts the original BYM model by treating the random effects for unobserved heterogeneity as Mundlak-style correlated random effects. This enables us to get correct inference even if the date that each cantons first case is reported depends on unobserved canton-specific factors. 

As the original BYM model, we use an intrinsic conditional auto-regressive (ICAR) model as a prior on the spatial random effect $\phi_i$. The ICAR model of the conditional distribution of each $\phi_i$ is
\begin{equation}\label{eq:ICAR_prior}
    \phi_i \,|\, \{\phi_j \colon j \in \mathcal N_i \}
    \sim
    \mathsf N 
    \left( 
        \frac{ 1 } { | \mathcal N_i | } 
        \sum_{j \in \mathcal N_i} \phi_j, \
        \frac{1}{ \tau | \mathcal N_i | }
    \right), 
\end{equation}
where $\mathcal N_i$ is the set of neighbors of canton $i$, meaning the cantons that share a border with the canton, and $|\mathcal N_i|$ denotes the cardinality of $\mathcal N_i$, that is, its number of neighbors. For practical implementation, the sets of neighbors can be constructed by defining a symmetric $N \times N$ adjacency matrix whose element in row $i$ and column $j$ equals 1 if cantons $i$ and $j$ share a common border, and 0 otherwise.

The ICAR model states that each $\phi_i$ is conditionally Gaussian with mean equal to the average of its neighbors spatial random effects and a variance that decreases with more neighbors. The term $\tau$ is a hyperparameter for the variance of the composite random effect $\upsilon_i = \theta_i + \phi_i$, where $\theta_i$ is the random effect for unobserved heterogeneity. Based on derivations in \cite{morris.2019b}, the joint distribution of $\bm \phi = (\phi_1, \ldots,\phi_N)$ is proportional to a particular pairwise difference form:
\begin{equation}\label{eq:pairwise_diff}
    p(\bm \phi|\tau)
    \,\propto
    \exp
    \left(
        -\frac{\tau}{2} 
        \sum_{i,j : j \in \mathcal N_i}
        \left(
            \phi_i - \phi_j
        \right)^2
    \right)
    =
    \exp
    \left(
        -\frac{\tau}{2} 
       \bm \phi^\top \bm Q \bm \phi
    \right).
\end{equation}
Here, $\bm Q$ is the precision matrix, \eg, the inverse spatial covariance matrix, that has entries
\begin{align*}
    Q_{i,j}
    =
    \begin{cases}
        |\mathcal N_i|,  \quad &\text{if } i = j, \\
        -1,              \quad &\text{if } j \in \mathcal N_i, \\
        0,               \quad &\text{else}.
    \end{cases}
\end{align*}
The pairwise difference form~\eqref{eq:pairwise_diff} shows that the ICAR model penalises large differences in the values of neighboring cantons spatial random effects. Hence, finding the values of $\{\phi_i\}^N_{i=1}$ that minimize these differences leads to local spatial smoothing \cite{morris.2019a}. However, the distributions in~\eqref{eq:ICAR_prior} and~\eqref{eq:pairwise_diff} are improper priors since they define the value of each cantons spatial effect relative to its neighbors. As a consequence, the priors do not identify the overall mean among $\{\phi_i\}^N_{i=1}$. To solve this, the soft constraint $N^{-1}\sum^N_{i=1} \phi_i \sim \mathsf N(0, 0.001)$ is used. 

Another problem is that the random effects $\phi_i$ and $\theta_i$ cannot be separately identified. Thus, heterogeneity across cantons that should be attributed to $\bm \theta$ will be modeled as spatial correlation by $\bm \phi$ even when no spatial dependence is present. To solve this issue, we use the reformulated BYM2 model \cite{Riebler.2016}. It involves a scaling \cite{Simpson.2017} and reparameterization \cite{Dean.2001} of $\bm \upsilon = \bm \theta + \bm \phi$, explained next.

Let $\phi^*_i \coloneqq \phi_i/\sqrt \kappa$ be the spatial random effect for canton $i$ after scaling. The scaling factor $\kappa$ is calculated from the neighborhood structure over all cantons and ensures that $\mathbb V[\phi^*_i] \approx \mathbb V[\theta_i] \approx 1$ for each canton $i$. The scaled reparameterization of $\upsilon_i = \theta_i + \phi_i$ is
\begin{equation}\label{eq:scaled_reparam}
    \frac{1}{\sqrt{\tau}}
    \left(
        \big(\sqrt{ 1 - \varphi  }\big) \theta_i +
        \sqrt{ \varphi } \phi^*_i
    \right).
\end{equation}
The corresponding covariance matrix of $\bm \upsilon = (\upsilon_1,\ldots,\upsilon_N)$ is
\begin{equation}
    \mathbb V[ \bm \upsilon\,|\,\tau, \varphi]
    =
    \frac{1}{\tau}
    \Big( 
        (1 - \varphi)\bm I 
         + \varphi \bm Q^-_*
    \Big).
\end{equation}
Here, $\bm Q^-_*$ is the generalized inverse of the scaled precision matrix $\bm Q_*$, $\varphi \in [0,1]$ is a mixing parameter, and $1/\tau$ is the marginal variance of log relative mobility that is explained by the combined random effect $\bm \upsilon$. The fraction of this variance explained by the canton random effect $\bm \theta$ and the scaled spatial random effect $\bm \phi^*$ are $(1-\varphi)$ and $\varphi$, respectively \cite{Simpson.2017}. 

Summarizing, the random effect $\bm \theta$ for between-canton heterogeneity has prior distribution $\mathsf {MVN}(\bm 0, \bm I)$ and the spatial random effects $\bm \phi$ has prior marginal distribution $\mathsf {MVN}(\bm 0, \bm Q^-_*)$ and prior conditional distribution given by \eqref{eq:ICAR_prior} for each $\phi_i$. See Riebler et al. \cite{Riebler.2016} for details. Replacing $\upsilon_i = \theta_i + \phi_i$ in~\eqref{eq:mobility_model_spatial} with \eqref{eq:scaled_reparam} gives the explicit equation for the model of mobility variable $k$ with the BYM2 spatial random effect.

\begin{figure}[H]
    \centering
    \includegraphics[width=0.5\linewidth]{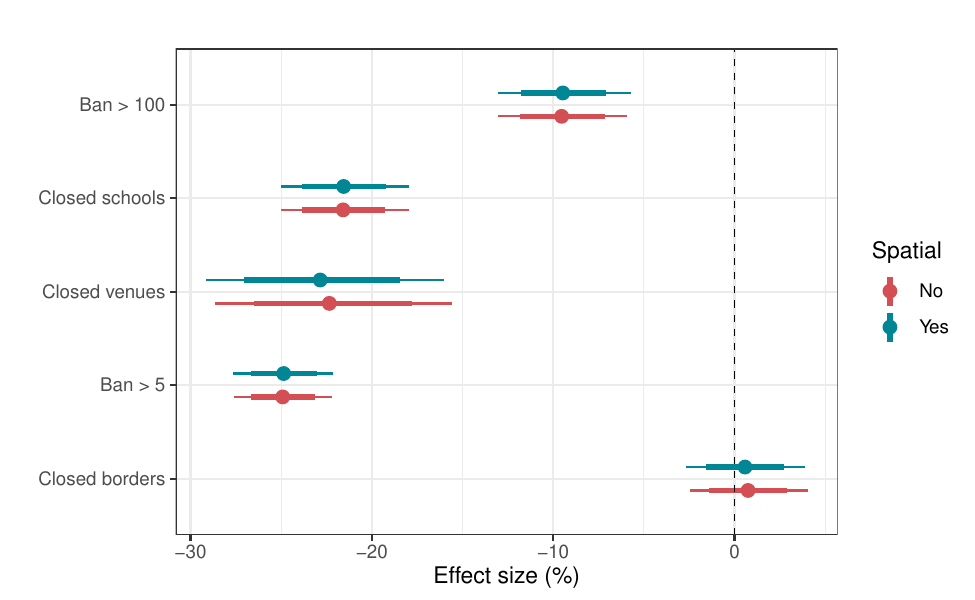}
\caption{Estimated effect of policy measures on total trips with and without a BYM2 ICAR random effect for spatial dependence between neighboring cantons. Posterior means are shown as dots, while 80\,\% and 95\,\% credible intervals are shown as thick and thin bars, respectively. Policy measures are arranged in the order in which they were implemented (cf. Supplement~\ref{supp:data}).}
\label{fig: spatial}
\end{figure}

Accounting for potential spatial dependence in mobility among neighboring cantons does not alter the estimates of the policy measures (\Cref{fig: spatial}). In other words, the spatial dependence in baseline mobility is low.  We therefore do not include the BYM2 spatial random effect in our main models.

\subsection{Accounting for the dependence between the mobility variables}
\label{supp:extension_multivariate}

The variables for trips by mode or purpose are different subsets of the total trips. As a result, the trip variables are dependent on each other. One way in which this dependence could arise is through the variation in baseline trip counts across cantons prior to any reported COVID-19 case and policy measure. As an example, we would expect the cross-canton variation in baseline commuting trips to be correlated with the cross-canton variation in baseline non-commuting trips. Since each regression equation's canton random effect $\bm \theta^{(k)} = (\theta^{(k)}_1, \theta^{(k)}_2,\ldots, \theta^{(k)}_N)$ gives the cross-canton variation in mobility variable $k$ at baseline, we can estimate the dependence between the mobility variables at baseline with the correlation coefficient between the canton random effects. To do so, we model $\bm \theta^{(1)}, \bm \theta^{(2)}, \ldots, \bm \theta^{(K)}$ as jointly multivariate Gaussian distributed:
\begin{equation}\label{eq:multivarate_gaussian_mobility}
    \begin{bmatrix}
           \bm \theta^{(1)} \\
           \bm \theta^{(2)} \\
           \vdots \\
           \bm \theta^{(K)}
    \end{bmatrix}
    \sim
    \mathsf {MVN}
    (
        \bm 0, \bm \Omega_M
    ),
    \quad
    \bm \Omega_M
    =
    \begin{bmatrix}
           \sigma^2_{\bm\theta^{(1)}} \bm I & \sigma_{\bm\theta^{(2)} \bm\theta^{(1)}} \bm I & \hdots & \sigma_{\bm\theta^{(K)}} \sigma_{\bm\theta^{(1)}} \bm I \\
           \sigma_{\bm\theta^{(1)} \bm\theta^{(2)}} \bm I & \sigma^2_{\bm\theta^{(2)}} \bm I & \hdots & \sigma_{\bm\theta^{(K)}} \sigma_{\bm\theta^{(2)}} \bm I \\
           \vdots  & \vdots & \ddots & \vdots \\ 
           \sigma_{\bm\theta^{(1)} \bm\theta^{(K)}} \bm I & \sigma_{\bm\theta^{(2)} \bm\theta^{(K)}} \bm I & \hdots & \sigma^2_{\bm\theta^{(K)}} \bm I
    \end{bmatrix}
    = \bm \Sigma_M \otimes \bm I,
\end{equation}
where $\sigma^2_{\bm\theta^{(k)}}$ is the variance of $\bm\theta^{(k)} = (\theta^{(k)}_1,\theta^{(k)}_2,\ldots, \theta^{(k)}_N)$ of regression equation $k=1,2,\ldots,K$, the term $\sigma_{\bm\theta^{(k)} \bm\theta^{(p)}}$ is the corresponding covariance between $\bm \theta^{(k)}$ and $\bm \theta^{(p)}$ from regressions equations $k$ and $p$, respectively, and the $K \times K$ covariance matrix $\bm \Sigma_M$ contains these variance and covariance terms. Moreover, $\bm I$ is an $N \times N$ identity matrix and the symbol $\otimes$ denotes the Kronecker product.

We then obtain the correlations between all pairs of canton random effects for each pair of $k,p = 1,2,\ldots,K$, $k \neq p$, by computing the Pearson coefficient
\begin{equation}
    \rho_{\bm\theta^{(k)} \bm\theta^{(p)}}
    =
    \frac{ \sigma_{\bm\theta^{(k)} \bm\theta^{(p)}} }
    { \sqrt{ \sigma^2_{ \bm\theta^{(k)} } \sigma^2_{ \bm\theta^{(p)} } } }.
\end{equation}
The correlation coefficients are collected as off-diagonal elements in the $K \times K$ correlation matrix $\bm R_M$ corresponding to $\bm \Sigma_M$ and can be interpreted as summaries of the dependence between the mobility variables' baseline relative trip counts across cantons in the absence of any policy measure and reported COVID-19 case. By modeling this dependence, we obtain more precise estimates of uncertainty in parameter values than estimating each mobility variable's regression equation separately. The approach is analogous to that of seemingly unrelated regressions \cite{Zellner.1962}, except that each regression model is a negative binomial generalized mixed model with a log-link. 

\begin{figure}[H]
     \begin{minipage}[t]{\linewidth}
            \hspace{1.4cm}
        \figletter{a}
    \end{minipage}
    \begin{minipage}[t]{\linewidth}
        \centering
        \includegraphics[width=\linewidth]{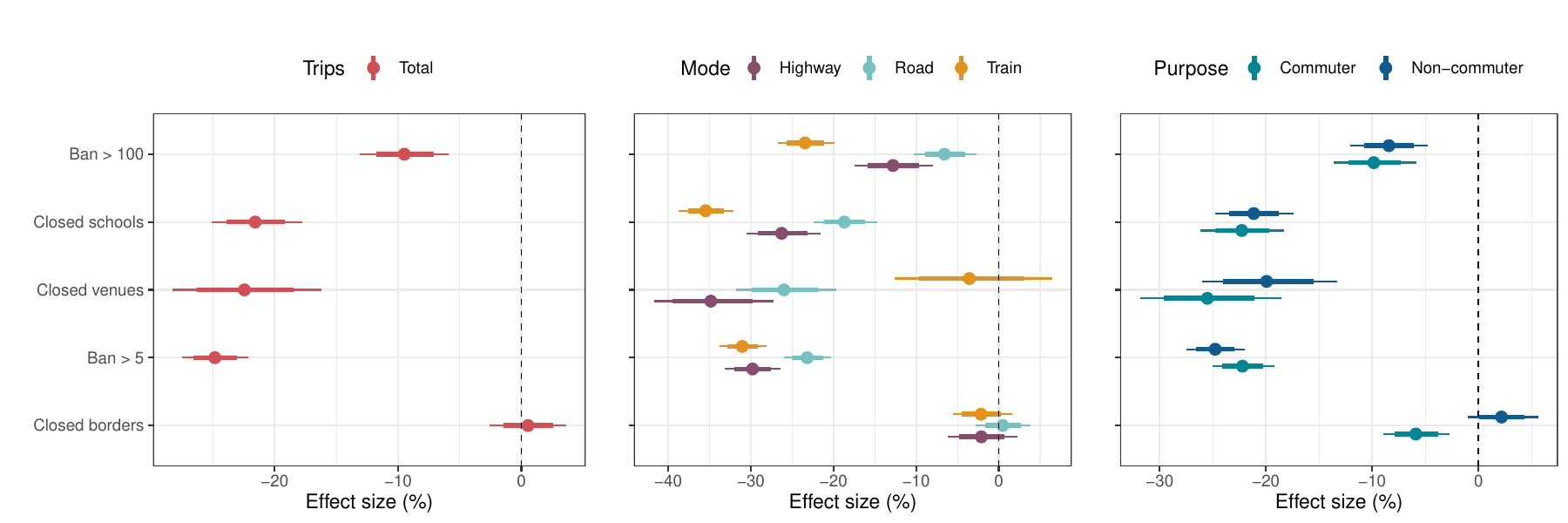}
    \end{minipage}
    \\
    \begin{minipage}[t]{\linewidth}
            \hspace{1.4cm}
        \figletter{b}
    \end{minipage}
    \begin{minipage}[t]{\linewidth}
        \centering
        \includegraphics[width=\linewidth]{Plots/traffic_analysis/did_traffic_nb_logtrend.pdf}
    \end{minipage}
\caption{Estimated effects of policy measures on total trips, trips by mode, trips by purpose from $(\mathbf{a})$ the joint multivariate model over all mobility variables, and $(\mathbf{b})$ the separate regressions per mobility variable. The joint multivariate model estimates the correlation between the baseline levels of the mobility variables across cantons by modeling the canton random effects as drawn from a multivariate Gaussian distribution.
Posterior means are shown as dots, while 80\,\% and 95\,\% credible intervals are shown as thick and thin bars, respectively. Policy measures are arranged in the order in which they were implemented (cf. Supplement~\ref{supp:data}).}
\label{fig: multivariate mobility-NPI}
\end{figure}

\Cref{fig: multivariate mobility-NPI} shows estimates of the effect of the policy measures on mobility from the multivariate model (\Cref{fig: multivariate mobility-NPI}a) against the estimates from the separate regressions per mobility variables for which the correlation between their canton random effects is not modeled (\Cref{fig: multivariate mobility-NPI}b). Comparing the multivariate and separate regressions, we find that the point estimates given by the posterior means are practically identical. There is a small tendency of the credible intervals in the multivariate model to be tighter. Note that estimating the multivariate model requires considerably more computational runtime. 

\clearpage
\section{Mediation analysis of the role of mobility in the effect of policy measures on reported cases}
\label{supp:mediation_analysis}

We perform a mediation analysis to decompose the effects of policy measures on reported COVID-19 cases into (1)~their direct effects, meaning the part of their effects that is due to behavioral adaptions not related to mobility, and (2)~their indirect effects, meaning the part of their effects that is explained by changes in mobility. The mediation analysis is performed by combining our two regression models into a structural equation model. Estimation details are provided in Supplement~\ref{supp:estimation_extended}.

\subsection{Mediation model}
\label{supp:mediation_model}

The structural equation model consists of the mediation model
\begin{align}
    \label{eq:mediator_model}
    \begin{aligned}
    \log \mathbb E[M_{i,t-s,k} \,|\, \eta^{(k,s,M)}_{i,t-s}, E_i]
    &= \mu(M_{i,t-s,k}) \\
    &=
    \log E_i +
    \alpha^{(k,s,M)}_i + 
    \delta^{(k,s,M)}_{w(t-s)} +
    \gamma^{(k,s,M)} \log z_{i,t-s} +
    \sum^L_{l=1} \beta^{(k,s)}_l
    d_{i,t-s,l}\\
    \alpha^{(k,s,M)}_i
    &=
    \alpha^{(k,s,M)} + \theta^{(k,s,M)}_i + \gamma^{(k,s,M)}_B \widebar{\log z_i} 
    \end{aligned}
\end{align}
and the outcome model
\begin{align}
    \label{eq:outcome_model}
    \begin{aligned}
    \log \mathbb E[Y_{it}\,|\, \eta^{(k,s,Y)}_{it}, E_i]
    &= \mu^{(k,s)}(Y_{it}) \\
    &=
    \log E_i +
    \alpha^{(k,s,Y)}_i + 
    \delta^{(k,s,Y)}_{w(t)} + 
    \gamma^{(k,s,Y)} \log z_{it} +
    \psi_{ks} \log m_{i,t-s,k} +
    \sum^L_{l=1} \lambda^{(k,s)}_l
    d_{i,t-s,l} \\
    \alpha^{(k,s,Y)}_i
    &=
    \alpha^{(k,s,Y)} + \theta^{(k,s,Y)}_i + \gamma^{(k,s,Y)}_B \widebar{\log z_i} + \psi_{ks,B} \widebar{\log m_{ik}}.
    \end{aligned}
\end{align}
For notation, the bar again denotes the average value of an expression. The superscripts $M$ and $Y$ are added to indicate that the parameter values are generally different for the mediation model and outcome model with the same mobility variable $k$ lagged by $s$ days. The conditional variance of $M_{i,t-s,k}$ and $Y_{it}$ is given by
\begin{align}
    \label{eq:mediation_model_variance_}
    \mathbb V[M_{i,t-s,k}\,|\,\eta^{(k,s,M)}_{i,t-s}, E_i, \zeta^{(k,s,M)}]
    &=
    \mu(M_{i,t-s,k})
    \left(
        1 +
        \frac{ \mu(M_{i,t-s,k}) }{\zeta^{(k,s,M)}}
    \right)
\end{align}
and
\begin{align}
    \label{eq:outcome_model_variance}
    \mathbb V[M_{i,t-s,k}\,|\,\eta^{(k,s,Y)}_{it}, E_i, \zeta^{(k,s,Y)}]
    &=
    \mu^{(k,s)}(Y_{it})
    \left(
        1 +
        \frac{ \mu^{(k,s)}(Y_{it}) }{\zeta^{(k,s,Y)}}
    \right).
\end{align}
Here, $\zeta^{(k,s,M)}$ and $\zeta^{(k,s,Y)}$ is the overdispersion parameter from the mediation model and outcome model, respectively, in which mobility variable $k$ is lagged by $s$ days. 

The mediation model is identical to the model that links policy measures to mobility, but has lagged values of all time-varying variables since their effects on the reported number of new cases are delayed. The outcome model is the same as the model that links mobility to reported cases, but now also  includes the policy measures. The reason for this is that the structural equation model shall estimate (1) the effects of mobility conditional on the policy measures, and (2) the effects of the policy measures conditional on mobility. Here, (1) gives the mediation effect of mobility that enables us to estimate the indirect effect of policy measures via mobility, and (2) gives the direct effect of policy measures that explained by other behavioral adaptions. The added superscripts $m$ and $y$ signifies that the parameters generally take different values in the mediation model and outcome model. 

To facilitate the interpretation of our results, we state how the parameters of interest shall be interpreted in the mediation analysis: 
\begin{itemize}
    \item $\beta^{(k,s)}_l$ is the effect of policy measure $l$ on the log expected number of trips on mobility variable $k$.
    \item $\lambda^{(k,s)}_l$ is the direct effect of policy measure $l$ lagged by $s$ days on the log expected number of reported cases, conditional on mobility variable $k$ lagged by $s$ days and the other policy measures lagged by $s$ days . Formally, this is the change that would occur in the reported number of new cases if (lagged) policy measure $l$ would have been implemented but mobility would not have changed as a consequence of policy measure $l$.
    \item $\psi_{ks}$ is the expected percentage change in the reported number of new cases as the number of trips on mobility variable $k$ lagged by $s$ days increases by 1\,\%, conditional on all policy measures lagged by $s$ days being held fixed.
    \item $\beta^{(k,s)}_l \times \psi_{ks}$ is the indirect effect of policy measure $l$ via mobility variable $k$, both lagged by $s$ days, on the log expected reported number of new cases. The indirect effect gives the expected change that would occur in the reported number of new cases if policy measure $l$ lagged by $s$ days would always have been implemented but mobility variable $k$ lagged by $s$ days would have changed as if the policy measure would vary as in the data.
    \item $\lambda^{(k,s)}_l + \beta^{(k,s)}_l \times \psi_{ks}$ is the total effect of policy measure $l$ lagged by $s$ days on the log expected number of reported cases. It is the effect of policy measure $l$ with the mobility variable $k$ changing as a consequence of implementing the policy measure. Hence, the total effect gives the change in the reported number of new cases due to any behavioral adaption.
\end{itemize}

As shown in the above, we compute the indirect effect using the product method \cite{Baron.1986} and sum the direct and indirect effect to obtain the total effect. This is applicable to linear Bayesian multilevel models with random intercepts and no mediator-treatment interaction \cite{Yuan.2009} (\ie, no treatment-moderated mediation), as the models of this study. 

We check the assumption of no treatment-mediator interaction by, for every lag of 7--13 days, re-estimating the outcome model \eqref{eq:outcome_model} with an interaction term between each policy measure and the mobility variable. We find that the parameters of the interaction terms have point estimates close to zero with some of the credible intervals covering zero. We take this as sufficient evidence of no treatment-mediator interaction and, therefore, proceed with the outcome model without such terms. 

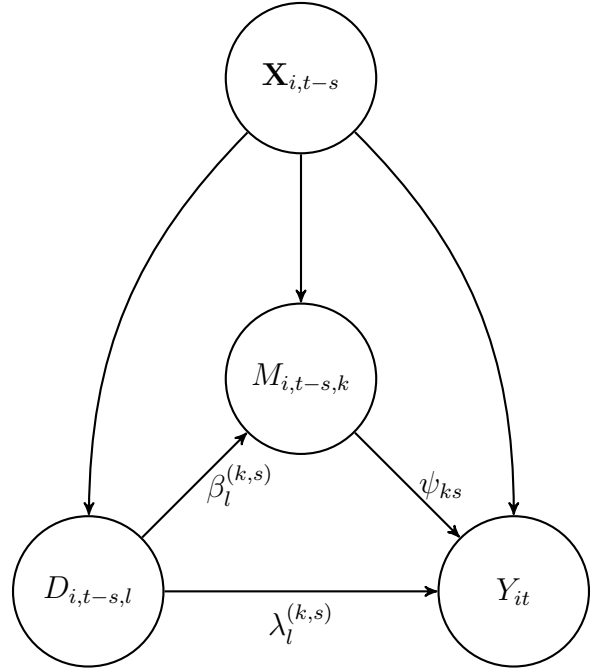
\begin{figure}[H]
\begin{center}

\begin{tikzpicture}[
        ->, 
        >=stealth', 
        auto, 
        node distance=4cm,
        thick, 
        main node/.style={circle, draw, minimum size=2cm}
        ]
  	
	\node[main node] (1) {$M_{i,t-s,k}$};
	\node[main node] (2) [below left of=1]{$D_{i,t-s,l}$};
 	\node[main node] (3) [below right of=1] {$Y_{it}$}; 
 	\node[main node] (4) [above of=1] {$\mathbf X_{i,t-s}$}; 

  	\path[every node/.style={line width=1pt}]
	 (2) edge node [right] {$\beta^{(k,s)}_l$} (1)
	 (2) edge node [below] {$\lambda^{(k,s)}_l$} (3)
	 (1) edge node [right] {$\psi_{ks}$} (3)
	 
	 (4) edge node [right] {} (1)
	 (4) edge [bend right=22.5] node [right] {} (2)
	 (4) edge [bend left=22.5] node [right] {} (3);
\end{tikzpicture}

\end{center}
\caption{A directed acyclic graph of the structural equation model within a fixed point in time. The vector $\mathbf X_{i,t-s}$ denotes the weekday, time, and canton effects in the models.}
\label{fig:DAG}
\end{figure}

The directed acyclic graph in \Cref{fig:DAG} visualises the relationships between the variables in the structural equation model at a given day. For simplicity, we denote weekday, time and unobserved canton factors lagged by $s$ days with the vector $\mathbf X_{i,t-s}$. The structural relationships among the variables are as follows: At day $t$, the weekday, the number of days that has passed since the first case was reported in the canton, and unobserved canton-specific factors jointly determine the number of trips in the canton on each mobility variable and whether each policy measure is implemented in the canton that day. The weekday, number of days that has passed since the first reported case, and unobserved canton-specific factors also determine the reported number of new cases, but with a lag of $s$ days due to incubation periods and reporting delay. Whether a policy measure is implemented in a canton at day $t$ affects the trip count on each mobility variable in same the canton that day (path $\beta^{(k,s)}_l$). Given the implementation of policy measures at day $t$, the level of mobility affects the number of reported cases $s$ days later (path $\psi_{ks}$). Hence, each policy measure has an immediate effect on mobility, which is transferred into a delayed indirect effect on the reported number of new cases through mobility (the product of paths $\beta^{(k,s)}_l$ and $\psi_{ks}$). Each policy measures also has a direct effect on the number of reported cases via behavioral adaptions not related to mobility. This effect is also subject to a lag of $s$ days (path $\lambda^{(k,s)}_l$). 

Note that every time-varying variable in the mediation model is lagged by $s$ days, but that in the outcome model, only the mobility variable and the policy measures are subject to this lag. With this specification, we estimate how the instantaneous effect of the policy measures on mobility variable $k$ in day $t-s$ affect the number of reported new cases in day $t$ directly and indirectly via mobility.

For estimation, we allow the canton random effects $\bm \theta^{(k,M)} = (\theta^{(k,s,M)}_1,\ldots,\theta^{(k,s,M)}_N)$ and $\bm \theta^{(k,s,Y)}=(\theta^{(k,s,Y)}_1,\ldots,\theta^{(k,s,Y)}_N)$ of the mediator model and outcome model, respectively, to be correlated. We thereby account for possible correlation between the baseline relative mobility and baseline relative (reported) infection risk across cantons. We achieve this by modeling the random effects as drawn from a bivariate Gaussian distribution:
\begin{align}\label{eq:multivarate_gaussian_mediation}
    \begin{bmatrix}
           \bm \theta^{(k,s,M)} \\
           \bm \theta^{(k,s,Y)}
    \end{bmatrix}
    \sim
    \mathsf {MVN}
    (
        \bm 0, \bm \Omega_{M,Y}
    ),
    \quad
    \bm \Omega_{M,Y}
    =
    \begin{bmatrix}
           \sigma^2_{\bm \theta^{(k,s,M)}} \bm I & \sigma_{\bm\theta^{(k,s,M)}\bm\theta^{(k,s,Y)}} \bm I \\ \sigma_{\bm\theta^{(k,s,Y)}\bm\theta^{(k,s,M)}} \bm I &
           \sigma^1_{\bm\theta^{(k,s,Y)}} \bm I 
    \end{bmatrix}
    = 
    \bm \Sigma_{M,Y} \otimes \bm I.
\end{align}
Here, $\sigma^2_{\bm \theta^{(k,s,M)}}$ and $\sigma^2_{\bm \theta^{(k,s,Y)}}$ is the cross-canton variance of the random effect of the mediation model and outcome model $(k,s)$, respectively, and the off-diagonal term is the cross-canton covariance between the models' respective random effects. The $2 \times 2$ covariance matrix $\bm \Sigma_{M,Y}$ contains these variance and covariance terms. Computing the Pearson correlation coefficient
\begin{equation}
    \rho_{\bm\theta^{(k,s,M)}\bm\theta^{(k,s,Y)}}
    =
    \frac{ \sigma_{\bm\theta^{(k,s,M)}\bm\theta^{(k,s,Y)}} }
    { \sqrt{ \sigma^2_{\bm\theta^{(k,s,M)}} \sigma^2_{\bm\theta^{(k,s,Y)}} } }
\end{equation}
then gives the correlation between the random effects of mediation model and outcome model $(k,s)$. The correlation coefficient is in turn the off-diagonal element in the $2 \times 2$ correlation matrix $\bm R_{M,Y}$ that corresponds to the covariance matrix $\bm \Sigma_{M,Y}$.

We estimate the structural equation model separately for mobility and the policy measures lagged by 7--13 days to see how the direct, indirect and total effects vary across the delay of the effects. To save space, however, we only estimate the models with the total trips mobility variable as a mediator.

\subsection{Identification}
\label{supp:mediation_identification}

We now state the identification conditions and assumptions required for the mediation analysis. Imai et al. \cite{Imai.2010a, Imai.2010b} propose a ``sequential ignorability'' assumption under which the direct and indirect effects are identified if the structural equation model is linear and contains no treatment-mediator interaction. The assumption consist of two parts, which for the setting of this study, are: (1)~each policy measure was implemented independently of mobility and the reported number of new cases, and (2) among observations with the same status on the policy measures, the number of trips is independent of the potential number of reported cases. The first part of the assumption means that the control variables, the fixed effects, and the random effects in the models are sufficient to remove any confounding in the relationships between the policy measures and mobility as well as any confounding in the relationships between policy measures and the reported number of new cases. The second part of the assumption implies that, if we also condition on the policy measures, there is no confounding of the relationship between mobility and the reported number of new cases. This is a stronger assumption since it must hold for every observed combination of the policy measures in the data. A sensitivity analysis for the assumptions is presented in Supplement~\ref{supp:mediation_error_correlation}. Another assumption necessary for identifying indirect effects is that the mediator (here: each mobility variable) is measured without error. This assumption is untestable but may plausibly hold since our data cover every trip between any two postcodes made with a mobile device.

\subsection{Results of mediation analysis}
\label{supp:mediation_results}

\begin{figure}[H]
    \begin{minipage}[t]{\linewidth}
        \hspace{0.6cm}
        \figletter{a} \\
        \begin{subfigure}{\linewidth}
            \includegraphics[width=\linewidth]{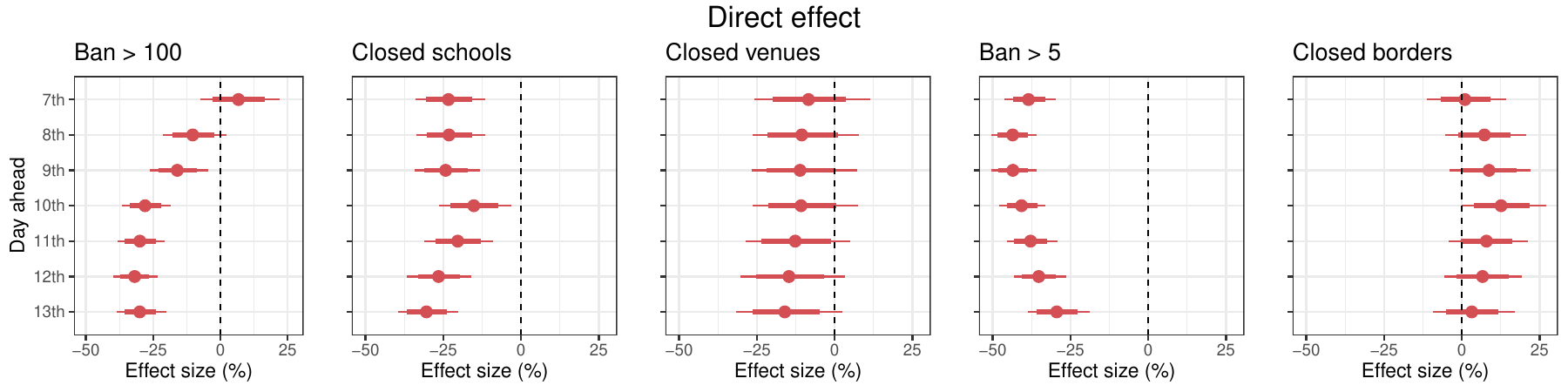}
        \end{subfigure}
    \end{minipage}
    \\
    \begin{minipage}[t]{\linewidth}
        \hspace{0.6cm}
        \figletter{b} \\
        \begin{subfigure}{\linewidth}
            \includegraphics[width=\linewidth]{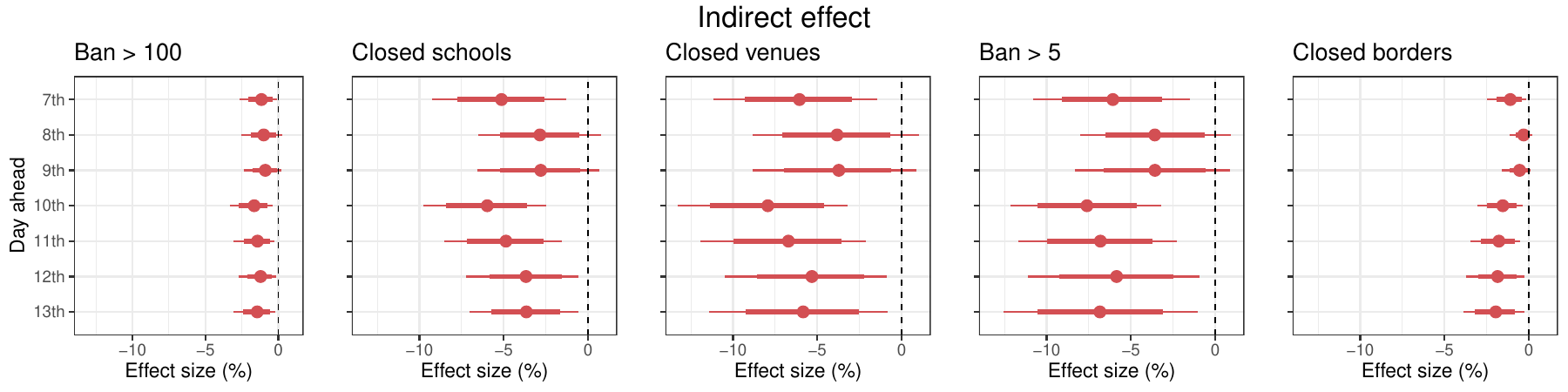}
        \end{subfigure}
    \end{minipage}
    \\
    \begin{minipage}[t]{\linewidth}
        \hspace{0.6cm}
        \figletter{c} \\
        \begin{subfigure}{\linewidth}
            \includegraphics[width=\linewidth]{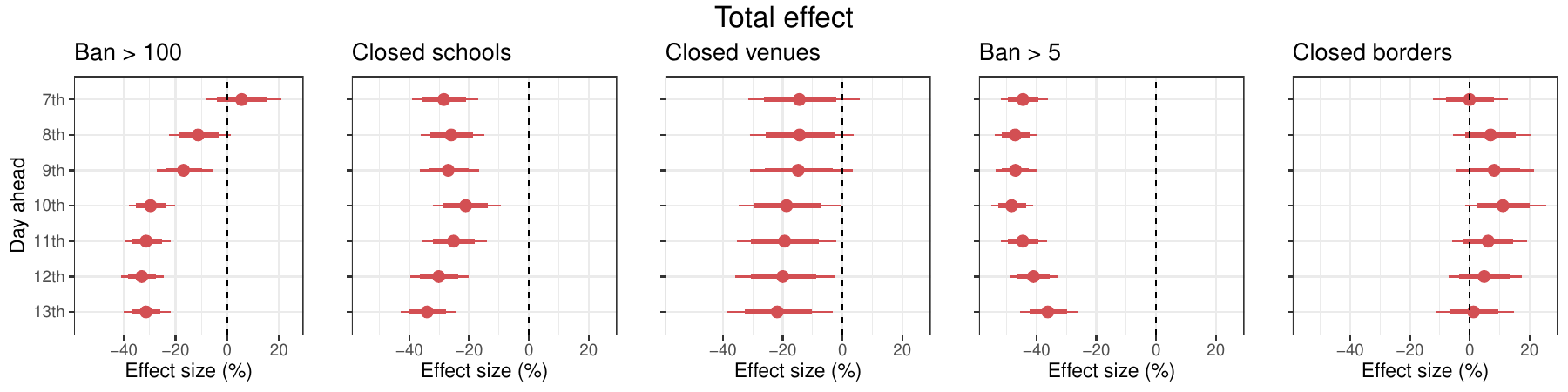}
        \end{subfigure}
    \end{minipage}
    \caption{Mobility mediates the effect of policy measures on the reported number of new cases. Estimated \textbf{(a)}~direct effect of policy measures, \textbf{(b)}~indirect effect of policy measures via total trips, and \textbf{(c)}~total effect of policy measures on the 7th to 13th day ahead. Posterior means are shown as dots, while 80\,\% and 95\,\% credible intervals are shown as thick and thin bars, respectively. Policy measures are arranged from left to right in the order in which they were implemented (cf. Supplement~\ref{supp:data}).}
\label{fig:mediation_results_2}
\end{figure}

The results from the mediation analysis with total trips as a mediator are shown in \Cref{fig:mediation_results_2}. 

\textbf{Direct effects.} The ban on gatherings of more than 5 people has the strongest estimated direct effect on the reduction in the reported number of new cases. Its estimated 95\,\% credible intervals range between a reduction of 18.8--38.7\,\% (at a lag of 13 days) to a reduction of 35.9--50.5\,\% (at a lag of 8 days). The credible interval of the direct effect of bans on gatherings of more than 100 people includes zero for lags 7--8 days. Hence, bans on larger gatherings may have a longer delayed direct effects than bans on smaller gatherings. Border closures do not appear to have directly reduced the reported number of new cases.

\textbf{Indirect effects.} The estimated indirect effects show that mobility mediates the effects of policy measures on the reported number of new cases. The largest indirect effects are found for venue closures and bans on gatherings of more than 5 people. Both policy measures are estimated to have reduced the reported case growth with around 6--8 \% at the higher order lags indirectly via mobility. Of particular interest is the estimated indirect effect for border closures. Recall that the estimated direct effects of border closures include zero. Hence, this implies that the effect of border closures occurs exclusively through mobility. This is to be expected considering that border closures may not affect other types of behavior than mobility.

\textbf{Total effects.} Combining the direct effects and indirect effects gives the total effects. Again, bans on gatherings of more than 5 people, bans on gatherings of more than 100 people, and school closures have the largest estimated total effects, in part due to their pronounced indirect effects. For any of the days that schools were closed, the estimates imply that there would on average be 21.2\,\% (95\,\% CrI: 9.48--32.3\,\%) more reported cases at the 10th day ahead and 34.1\,\% (95\,\% CrI: 24.3--42.9\,\%) more reported cases at the 13th day ahead if schools would instead have remained open. Note that for several lags, the indirect effect of venue closures makes up more than a third of its total effect. 

Overall, the mediation analysis demonstrates that the effects of social distancing policies operate -- to a large degree -- through mobility. Hence, telecommunication data provides valuable information for estimating effects of policy measures aimed at reducing mobility.

\subsection{Sensitivity analysis for the assumption of no unmeasured confounding}
\label{supp:mediation_error_correlation}

The mediation analysis rests on the sequential ignorability assumption of no unmeasured confounding in both the mediation model and the outcome model. Now, if there exists unobserved confounders that affect both the mediator and the outcome, then those variables will be part of both models' error terms. Hence, under the assumption of no unmeasured confounding, there should be no correlation between the models' respective errors. Based on this observation, Imai et al. \cite{Imai.2010b} propose a sensitivity analysis based on the correlation between the residuals in the two models.

To conduct the sensitivity analysis, we draw 4000 samples (\eg, as many as the length of the parameters' Markov chains, including the warm-up) from the models' predictive posterior distributions, each of the same size as the number of observations in the data. For each of the models, we subtract these values from the observed responses to get posterior samples of the model's predictive errors. We then compute the Pearson correlation coefficient between the mediation model's predictive errors and the outcome model's predictive errors for each of the 4000 posterior error samples. 

\begin{figure}[H]
    \begin{minipage}[t]{\linewidth}
        \centering
        \includegraphics[width=\linewidth]{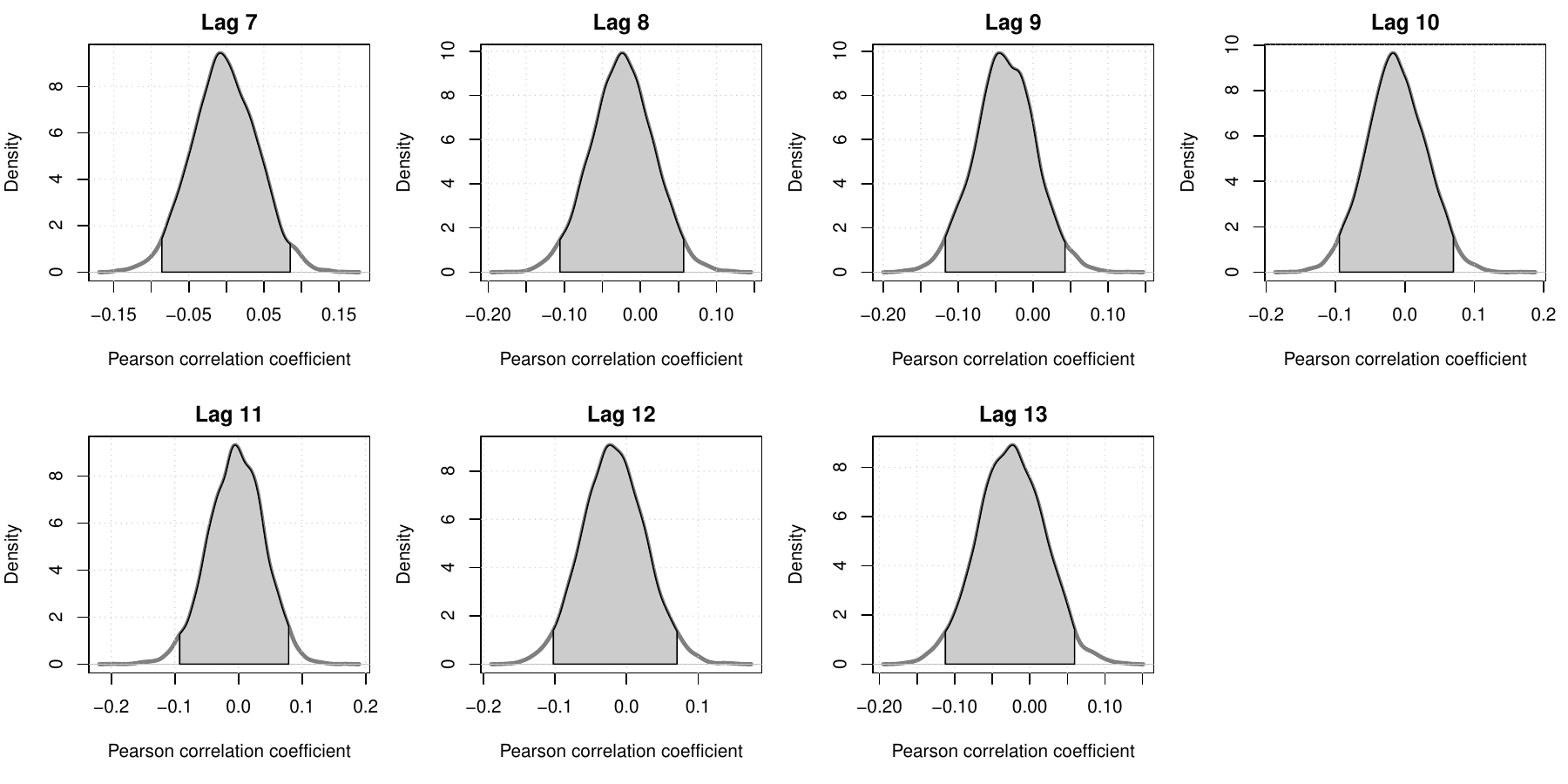}
    \end{minipage}
    \caption{Kernel density estimates of the distribution of Pearson correlation coefficients between posterior samples of the mediation model's and outcome model's respective predictive errors. The shaded regions are 95\,\% credible intervals computed as the the area between the 2.5th and 97.5th percentile Pearson correlation coefficient.}
\label{fig: error cor dist}
\end{figure}

\Cref{fig: error cor dist} shows kernel density estimates of the Pearson correlation coefficient distributions. The mean correlation coefficient is for each lag close to zero with the 95\,\% credible intervals covering zero. Hence, the sensitivity analysis suggests that there is no unmeasured confounding in the mediation analysis.

\clearpage
\section{Estimation details for modeling extensions and mediation analysis}
\label{supp:estimation_extended}

\Cref{tbl:all_priors} shows our choices of prior distributions for the parameters in the extended models. The new parameters are (1)~the parameters for the direct effects in the mediation analysis, (2)~the ICAR parameters in the spatial model, (3)~the correlation coefficients for the dependence between the canton random effects in the outcome model and mediation model, and (4)~the correlation coefficients for the dependence between the canton random effects in the multivariate mobility model. Below we explain the priors for the new parameters. The remaining priors are explained in \Cref{sec:estimation}.

As in the main paper, the prior on $\beta_l$ reflects that we expect each policy measure to reduce log expected trips by on average 25\,\%. The prior on the direct effect parameter $\lambda_l$ means that we expect each policy measure to reduce the log expected number of reported cases with on average 25\,\% conditional on mobility, with a variance such that increases in reported cases due to the policy measures are unlikely. Moreover, the prior on the mean and variance of $\psi_{ks}$ implies that we expect mobility to predict half as large reduction in cases once we condition on the policy measures, but that negative parameter estimates are still unlikely. Note that in the spatially extended model, the canton random effects are assigned a prior of $\mathsf N(0,1)$ due to the BYM2 reparametrization and scaling of the combined canton and spatial random effects. Moreover, the prior on the spatial random effects is technically on its conditional distribution given its neighboring cantons spatial random effects, not its marginal distribution as it is for simplicity written in \Cref{tbl:all_priors}. The ICAR parameters are assigned priors according to the recommendations by Morris \cite{morris.2019b} and Gelman et al. \cite{Gelman.2006}. The prior on the canton random effects correlations is, per the standard procedure in \text{Stan}, assigned on the Cholesky factor of the correlation matrix. We assign a LKJ correlation distribution with a shape parameter of 2 which makes extreme correlations less likely. All variables that are also included in the non-extended models are given the same priors as in those models, see \Cref{tbl:priors} in the paper. The ICAR model is estimated via the implementation available in the \texttt{brms} package. See \cite{morris.2019a} for details. 

\begin{table}[H]
\begin{center}
\setlength{\tabcolsep}{4pt}
\footnotesize
\begin{tabular}{l l l l}
\toprule
\text{Parameter} & \text{Description} & \text{Prior} & \text{Model} \\
\midrule
$\beta_l$        & \text{Effect of policy measure $l$} &  $\mathsf{N}(-0.25,0.25)$ & \eqref{eq:mobility_model_spatial}, \eqref{eq:mediator_model} \\
$\lambda_l$      & \text{Effect of policy measure $l$} &  $\mathsf{N}(-0.25,0.125)$ & \eqref{eq:outcome_model} \\
$\psi_{ks}$      & \text{Effect of log mobility variable $k$ with a lag of $s$} & $\mathsf{N}(0.5,0.125)$  & \eqref{eq:outcome_model} \\
$\theta_i$       & \text{Canton random effect}              & $\mathsf{N}(0, \sigma_\theta)$        & \eqref{eq:mediator_model}, \eqref{eq:outcome_model}, \eqref{eq:multivarate_gaussian_mobility} \\
$\sigma_\theta$  & \text{Standard deviation for canton random effect}              & $\mathsf {Half\text{-}t}(3, 0, 2.5)$        & \eqref{eq:mediator_model}, \eqref{eq:outcome_model}, \eqref{eq:multivarate_gaussian_mobility} \\
$\theta_i$       & \text{Canton random effect in the spatial model}              & $\mathsf{N}(0, 1)$        & \eqref{eq:mobility_model_spatial} \\
$\alpha$         & \text{Intercept}                             & $\mathsf {Half\text{-}t}(3, 1.8, 2.5)$        & All \\
$\delta_{w(t)}$  & \text{Effect of weekday $w$ compared to Monday}                 & $\mathsf{N}(0,0.5)$        & All \\
$\gamma$         & \text{Effect of log no. of days since 1st reported case}    & $\mathsf{N}(1,1)$       & All \\
$\gamma_B$       & \text{Effect of between-canton average of log no. of days since 1st reported case}    & $\mathsf N(0,5)$       & All \\
$\psi_{ks,B}$    & \text{Effect of between-canton average of log mobility with a lag of $s$} & $\mathsf{N}(0,5)$  & \eqref{eq:outcome_model} \\
$\zeta$          & \text{Overdispersion in dependent variable}        & $\mathsf {Gamma}(0.01, 0.01)$   & All \\
$\phi_i$           & \text{Spatial random effect}          & ICAR model \eqref{eq:ICAR_prior} & \eqref{eq:mobility_model_spatial}  \\
$\varphi$           & \text{ICAR mixing parameter}          & $\mathsf{Beta}(0.5, 0.5)$   & \eqref{eq:mobility_model_spatial}  \\
$\tau^{-0.5}$    & \text{ICAR standard deviation}        &  $\mathsf{Half\text{-}t}_{3, 0, 2.5}$   & \eqref{eq:mobility_model_spatial} \\
$\bm R_{M,Y}$        & \text{Mediation-outcome cross-canton random effects correlation}        & $\mathsf{LKJ}(2)$   & \eqref{eq:multivarate_gaussian_mediation}  \\
$\bm R_M$        & \text{Mobility variable cross-canton random effects correlation}        & $\mathsf{LKJ}(2)$   & \eqref{eq:multivarate_gaussian_mobility} \\
\bottomrule
\multicolumn{4}{p{16cm}}{\emph{Note:} The superscripts $(k)$ and $(k,s)$ are omitted as the same priors are assigned to each model. The column ``Description'' states what effect the associated parameter represent (except for the overdispersion parameter).}
\end{tabular}
\caption{\textbf{Choice of priors for all models}}
\label{tbl:all_priors}
\end{center}
\end{table}

\end{document}